\documentclass[pre,twocolumn,amsmath,amssymb,floatfix,superscriptaddress,nopacs]{revtex4}

\usepackage{graphicx}
\usepackage{latexsym}
\usepackage{amsmath}
\usepackage{amssymb}
\usepackage{amsfonts}
\usepackage{bm}
\usepackage{color}

\newcommand{\Fcal}{\ensuremath{{\cal F}}}

\newcommand{\Lcal}{\ensuremath{{\cal L}}}

\newcommand{\ahat}{\hat{a}}
\newcommand{\chat}{\hat{c}}
\newcommand{\ihat}{\hat{i}}

\newcommand{\khat}{\hat{k}}
\newcommand{\lhat}{\hat{l}}
\newcommand{\qhat}{\hat{q}}
\newcommand{\rhat}{\hat{r}}
\newcommand{\uhat}{\hat{u}}
\newcommand{\qhatvec}{\ensuremath{{\hat{\bf q}}}}
\newcommand{\rhatvec}{\ensuremath{{\hat{\bf r}}}}

\newcommand{\ddiff}{\ensuremath{\mathrm{d}}}

\newcommand{\bfzero}{\ensuremath{{\bf 0}}}

\newcommand{\evec}{\mbox{$\bf e$}}

\newcommand{\qvec}{\mbox{$\bf q$}}
\newcommand{\rvec}{\mbox{$\bf r$}}

\newcommand{\vvec}{\mbox{$\bf v$}}

\newcommand{\Avec}{\mbox{$\bf A$}}
\newcommand{\Bvec}{\mbox{$\bf B$}}
\newcommand{\Cvec}{\mbox{$\bf C$}}

\newcommand{\Tgen}{\star}

\newcommand{\Ra}{\ensuremath{R_{\alpha}}}
\newcommand{\Sa}{\ensuremath{S_{\alpha}}}
\newcommand{\Sb}{\ensuremath{S_{\beta}}}
\newcommand{\Aab}{\ensuremath{A_{\alpha\beta}}}
\newcommand{\Bab}{\ensuremath{B_{\alpha\beta}}}
\newcommand{\cab}{\ensuremath{c_{\alpha\beta}}}

\newcommand{\Eab}{\ensuremath{E_{\alpha\beta}}}

\newcommand{\Gab}{\ensuremath{\mathsf{G}_{\alpha\beta}}}

\newcommand{\Kab}{\ensuremath{K_{\alpha\beta}}}

\newcommand{\Rab}{\ensuremath{R_{\alpha\beta}}}

\newcommand{\Tab}{\ensuremath{T_{\alpha\beta}}}

\newcommand{\cabcd}{\ensuremath{c_{\alpha\beta\gamma\delta}}}

\newcommand{\cbacd}{c_{\beta\alpha\gamma\delta}}
\newcommand{\cabdc}{c_{\alpha\beta\delta\gamma}}
\newcommand{\cbadc}{c_{\beta\alpha\delta\gamma}}
\newcommand{\Eabcd}{\ensuremath{E_{\alpha\beta\gamma\delta}}}
\newcommand{\Gabcd}{\ensuremath{\mathsf{G}_{\alpha\beta\gamma\delta}}}

\newcommand{\Tabcd}{\ensuremath{T_{\alpha\beta\gamma\delta}}}

\newcommand{\IDab}{\delta_{\alpha\beta}}
\newcommand{\IDcd}{\delta_{\gamma\delta}}
\newcommand{\IDac}{\delta_{\alpha\gamma}}
\newcommand{\IDad}{\delta_{\alpha\delta}}
\newcommand{\IDbc}{\delta_{\beta\gamma}}
\newcommand{\IDbd}{\delta_{\beta\delta}}

\newcommand{\OPup}{{\cal U}}
\newcommand{\OPdown}{{\cal D}}

\newcommand{\aL}{a_{\mathrm{L}}}
\newcommand{\aN}{a_{\mathrm{N}}}

\newcommand{\bL}{b_{\mathrm{L}}}

\newcommand{\bN}{b_{\mathrm{N}}}

\newcommand{\cL}{c_{\mathrm{L}}}
\newcommand{\cG}{c_{\mathrm{G}}}

\newcommand{\dL}{d_{\mathrm{L}}}
\newcommand{\dG}{d_{\mathrm{G}}}

\newcommand{\iL}{i_{\mathrm{L}}}
\newcommand{\iG}{i_{\mathrm{G}}}
\newcommand{\iN}{i_{\mathrm{N}}}
\newcommand{\iM}{i_{\mathrm{M}}}
\newcommand{\iS}{i_{\mathrm{P}}}
\newcommand{\iT}{i_{\mathrm{T}}}

\newcommand{\kL}{k_{\mathrm{L}}}
\newcommand{\kN}{k_{\mathrm{N}}}

\newcommand{\kEone}{k^{\mathrm{E}}_{1}}
\newcommand{\kEtwo}{k^{\mathrm{E}}_{2}}
\newcommand{\kEL}{k^{\mathrm{E}}_{\mathrm{L}}}
\newcommand{\kEN}{k^{\mathrm{E}}_{\mathrm{N}}}

\newcommand{\kKL}{k^{\mathrm{K}}_{\mathrm{L}}}
\newcommand{\kKN}{k^{\mathrm{K}}_{\mathrm{N}}}

\newcommand{\ctild}{\tilde{c}}
\newcommand{\ftild}{\tilde{f}}
\newcommand{\ltild}{\tilde{l}}
\newcommand{\ktild}{\tilde{k}}

\newcommand{\itild}{\tilde{i}}

\newcommand{\dkL}{\delta k_{\mathrm{L}}}
\newcommand{\dkN}{\delta k_{\mathrm{N}}}

\newcommand{\gL}{g_{\mathrm{L}}}
\newcommand{\gG}{g_{\mathrm{G}}}

\newcommand{\gaimp}{g^{\mathrm{ex}}_{\alpha}}
\newcommand{\gbimp}{g^{\mathrm{ex}}_{\beta}}
\newcommand{\gagen}{g^{\mathrm{f}}_{\alpha}}

\newcommand{\qL}{q_{\mathrm{L}}}
\newcommand{\qG}{q_{\mathrm{G}}}

\newcommand{\tincr}{\delta \tau}
\newcommand{\tsamp}{\Delta \tau}
\newcommand{\tsampmax}{{\Delta \tau}_{\mathrm{max}}}

\newcommand{\Nt}{\ensuremath{N_\mathrm{t}}}
\newcommand{\Nc}{\ensuremath{N_\mathrm{c}}}
\newcommand{\Nk}{\ensuremath{N_\mathrm{k}}}

\newcommand{\agrid}{a_{\mathrm{grid}}}

\newcommand{\rhobarab}{\bar{\rho}_{\alpha\beta}}
\newcommand{\rhobarcd}{\bar{\rho}_{\gamma\delta}}
\newcommand{\cbar}{\bar{c}}
\newcommand{\cbarabcd}{\bar{c}_{\alpha\beta\gamma\delta}}
\newcommand{\cLbar}{\bar{c}_{\mathrm{L}}}
\newcommand{\cGbar}{\bar{c}_{\mathrm{G}}}

\newcommand{\gext}{g^{\mathrm{ex}}}

\newcommand{\rhohata}{\hat{\rho}_{\alpha}}
\newcommand{\rhohatb}{\hat{\rho}_{\beta}}
\newcommand{\rhohatab}{\hat{\rho}_{\alpha\beta}}

\newcommand{\rref}{\tilde{r}}

\newcommand{\epsab}{\varepsilon_{\alpha\beta}}
\newcommand{\epscd}{\varepsilon_{\gamma\delta}}
\newcommand{\epshatab}{\hat{\varepsilon}_{\alpha\beta}}
\newcommand{\epshatcd}{\hat{\varepsilon}_{\gamma\delta}}

\newcommand{\epsL}{\varepsilon_{\mathrm{L}}}
\newcommand{\epsN}{\varepsilon_{\mathrm{N}}}

\newcommand{\eRe}{e^{\prime}}
\newcommand{\eIm}{e^{\prime\prime}}

\newcommand{\sigab}{\sigma_{\alpha\beta}}

\newcommand{\sigabext}{\sigma^{\mathrm{ex}}_{\alpha\beta}}

\newcommand{\sigcdext}{\sigma^{\mathrm{ex}}_{\gamma\delta}}

\newcommand{\etaL}{\eta_{\mathrm{L}}}
\newcommand{\etaG}{\eta_{\mathrm{G}}}

\newcommand{\la}{\left<}
\newcommand{\ra}{\right>}

\newcommand{\tauL}{\tau_{\mathrm{L}}}
\newcommand{\tauT}{\tau_{\mathrm{G}}}
\newcommand{\tauLT}{\tau_{\mathrm{L,G}}}
\newcommand{\tauLzero}{\tau_{\mathrm{L}}^{\mathrm{v}}}
\newcommand{\tauTzero}{\tau_{\mathrm{G}}^{\mathrm{v}}}

\newcommand{\xicont}{\xi_{\mathrm{cont}}}
\newcommand{\ximon}{\xi_{\mathrm{mon}}}
\newcommand{\xihom}{\xi_{\mathrm{hom}}}
\newcommand{\xiiso}{\xi_{\mathrm{iso}}}

\newcommand{\Tglass}{T_{\mathrm{g}}}

\newcommand{\rotab}{\omega_{\alpha\beta}}
\newcommand{\rotabhat}{\hat{\omega}_{\alpha\beta}}

\newcommand{\rotbahat}{\hat{\omega}_{\beta\alpha}}

\newcommand{\gamab}{\gamma_{\alpha\beta}}

\begin{document}

\title{Isotropic tensor fields in amorphous solids:\\
Correlation functions of displacement and strain tensor fields}

\author{J.P.~Wittmer}
\email{joachim.wittmer@ics-cnrs.unistra.fr}
\affiliation{Institut Charles Sadron, Universit\'e de Strasbourg \& CNRS, 23 rue du Loess, 67034 Strasbourg Cedex, France}
\author{J. Baschnagel}
\affiliation{Institut Charles Sadron, Universit\'e de Strasbourg \& CNRS, 23 rue du Loess, 67034 Strasbourg Cedex, France}

\begin{abstract}
Generalizing recent work on isotropic tensor fields in isotropic and achiral condensed 
matter systems from two to arbitrary dimensions we address both mathematical 
aspects assuming perfectly isotropic systems and applications focusing on correlation 
functions of displacement and strain field components in amorphous solids
where isotropy may not hold.
Various general points are exemplified using simulated polydisperse Lennard-Jones particles.
It is shown that the strain components in reciprocal space have essentially a complex 
circularly-symmetric Gaussian distribution albeit weak non-Gaussianity effects become visible 
for large wavenumbers $q$ where also anisotropy effects become relevant. 
The dynamical strain correlation functions are strongly non-monotonic with respect to $q$ 
with a minimum roughly at the breakdown of the continuum limit. 
\end{abstract}
\date{\today}
\maketitle

\section{Introduction}
\label{intro}

\begin{figure}[t]
\centerline{\resizebox{0.8\columnwidth}{!}{\includegraphics*{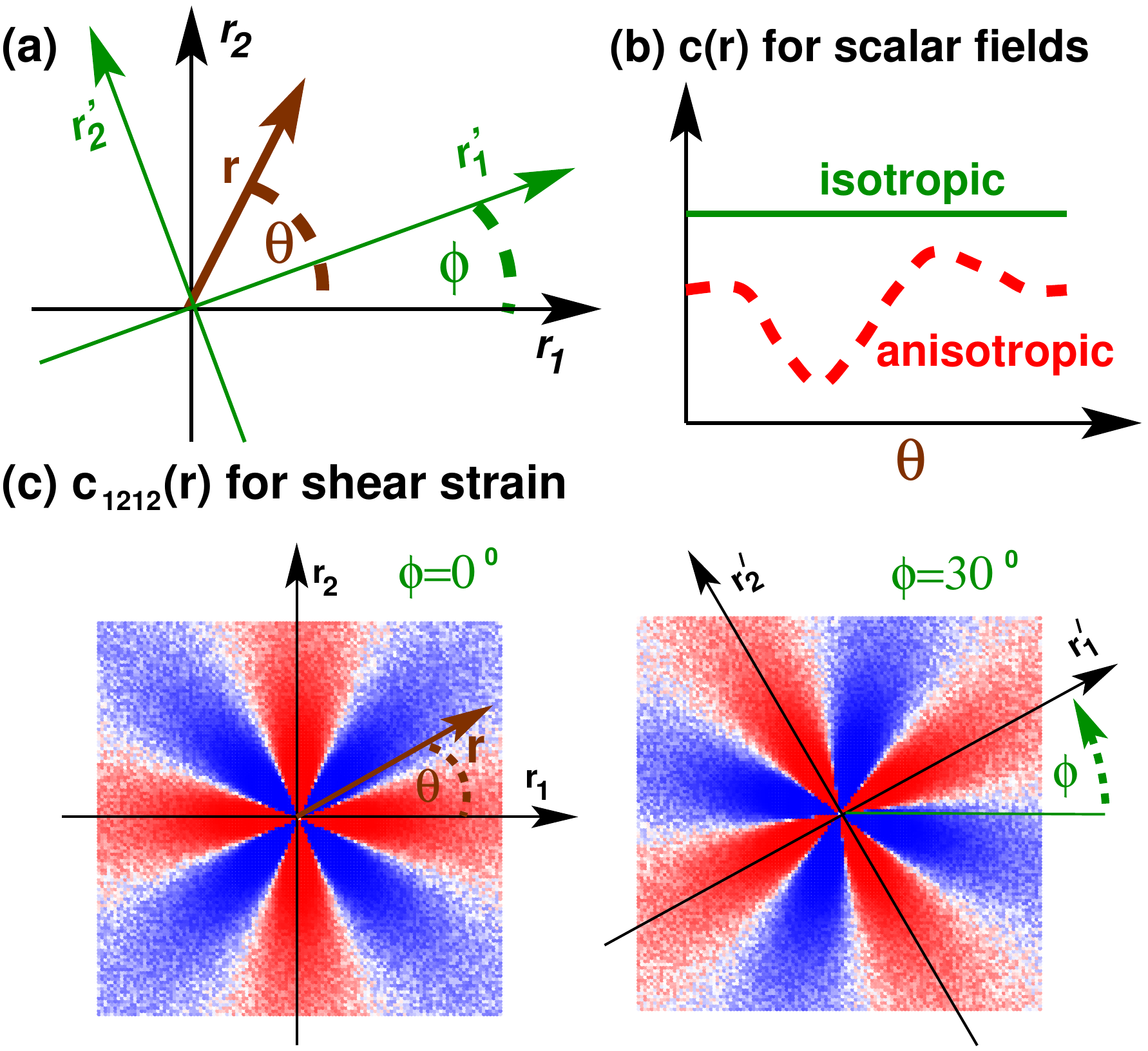}}}
\caption{Some useful notions:
{\bf (a)} 
We consider tensor fields (TFs) and their correlation functions (CFs)
under  orthogonal transformations as the shown rotation of a coordinate system
by an angle $\phi$. $\theta$ denotes the angle of the field vector $\rvec$ in the $12$-plane
of unrotated coordinates.
{\bf (b)}
CFs $c(\rvec)$ of scalar fields do not depend on the coordinate system.
For isotropic systems $c(\rvec)$ only depends on the magnitude $r=|\rvec|$ of the field 
vector $\rvec$ but not on its direction (angle $\theta$).  
{\bf (c)}
CF $c_{1212}(\rvec)$ of shear strain field $\varepsilon_{12}(\rvec)$
for an isotropic elastic body revealing an octupolar pattern. 
The CF is positive along the axes and negative along the bisection lines
of the respective axes. The pattern rotates with the coordinate frame.
}
\label{fig_intro}
\end{figure}

The characterization of structural properties of condensed matter systems 
\cite{ChaikinBook,BenoitBook,HansenBook,DhontGlasses,GoetzeBook,FerryBook,DoiEdwardsBook,RubinsteinBook}
often boils down to the experimental determination of  ``correlation functions" (CFs) 
$c(\qvec) = \la f(\qvec) f(-\qvec) \ra$ 
of fields $f(\qvec)=\Fcal[f(\rvec)]$ measured in reciprocal space 
as a function of the wavevector $\qvec$ \cite{ChaikinBook,BenoitBook}. 
($\Fcal[\ldots]$ stands here for the ``Fourier transformation" (FT) \cite{numrec}
and $\la \ldots \ra$ denotes a convenient average specified below.)
Due to the (partial) translational invariance of many systems it is
also useful in theoretical work 
\cite{ChaikinBook,ForsterBook,HansenBook,GoetzeBook,DoiEdwardsBook}
and computational studies \cite{AllenTildesleyBook}
to focus on the characterization of CFs in reciprocal space, at least as a first step.
The CFs $c(\rvec) = \Fcal^{-1}[c(\qvec)]$ in real space, as shown in Fig.~\ref{fig_intro},
may then finally be obtained by inverse FT. 
As already emphasized elsewhere 
\cite{Lemaitre15,Lemaitre17,Fuchs17,Fuchs18,lyuda18,lyuda22a,spmP5a,spmP5b}, 
it is now of importance whether the measured field $f(\qvec)$ 
is a {\em scalar field} (order $o=0$) or a component of a 
{\em ``tensor field"} (TF) of order $o>0$ 
\cite{Schouten,McConnell,Schultz_Piszachich,TadmorCMTBook}. 
Let us first focus on CFs $c(\rvec)$ of scalar fields in real space 
as sketched in the second panel of Fig.~\ref{fig_intro}.
For {\em isotropic} systems such a CF 
only depends on the magnitudes $r = |\rvec|$ or $q = |\qvec|$
of the field vectors $\rvec$ or $\qvec$ in real or reciprocal space.
If a dependency on the normalized directions $\rhatvec=\rvec/r$ or $\qhatvec=\qvec/q$
of the field vectors is observed \cite{foot_intro_carrets}, this demonstrates {\em anisotropy}.
Moreover, an observed anisotropic pattern in the material reference frame (dashed line)
does not depend on the orientation of the coordinate system. 
This is in general different if CFs of components of TFs are probed,
simply since the components of TFs depend explicitly on the coordinate system. 
This implies that components of mathematically and physically legitimate 
``isotropic tensor fields" (ITFs) may depend on the coordinate systems,
however, subject to a generic mathematical structure summarized in Sec.~\ref{ten_sum}.
As an example further discussed in Sec.~\ref{cstat},
panel {\bf (c)} of Fig.~\ref{fig_intro} shows the CF $c_{1212}(\rvec)$ of the shear strain 
component $\varepsilon_{12}(\rvec)$ for an isotropic linear elastic body \cite{LandauElasticity} 
in two dimensions \cite{spmP5b} revealing an octupolar pattern \cite{foot_wiki_multipolar}
which, moreover, turns if the coordinate frame is rotated by an angle $\phi$.
Importantly, the fourth-order strain CFs $\cabcd(\rvec)$ of ideal isotropic elastic bodies 
in $d$ dimensions can be theoretically shown to decay as 
$1/r^d$ for sufficiently large $r$.
As stressed in Refs.~\cite{Lemaitre15,Lemaitre17,spmP5a,spmP5b},
the observation of such angle-dependencies or long-range power-law decays of CFs of TFs
can thus {\em apriori} not be used as an indication of Eshelby-like plastic 
rearrangements \cite{Eshelby57,Lequeux04}.

Extending our recent studies \cite{lyuda18,lyuda22a,spmP5a,spmP5b}
the present work has five main goals:
\begin{itemize}
\item
We generalize our work on isotropic and achiral condensed matter systems
for $d=2$ \cite{spmP5a,spmP5b} to $d \ge 2$ confirming Ref.~\cite{Lemaitre15} 
that fourth-order ITFs are in general described by {\em five} invariants.
For $d=2$ this can be compressed using a simple transformation to {\em four} invariants.
\item
The invariants may be best theoretically or numerically characterized using ``Natural Rotated 
Coordinates" (NRC) in reciprocal space \cite{spmP5a,spmP5b}. 
Taking advantage of the mathematical formalism for ITFs we thus compute useful FTs relating
the invariants in real and reciprocal space for $d=2$ and $3$.
\item
As in Ref.~\cite{spmP5b} we focus on displacement and strain TFs in amorphous solid bodies.
It is shown that their static and dynamical correlations are characterized in terms of only 
{\em two independent} ``invariant CFs" (ICFs) and that these ICFs are related 
to two invariant viscoelastic material functions $L(q,t)$ and $G(q,t)$ 
being generalized moduli characterizing the longitudinal and transverse displacements in NRC
with respect to wavenumber $q$ and time $t$.
\item
Real systems cannot be isotropic on all scales \cite{Massimo21,Massimo24}.
Just as for any symmetry breaking, such anisotropies must be described 
in terms of invariants of the symmetry group assumed to be broken.
\item
As in previous work \cite{lyuda22a,spmP1,spmP5a,spmP5b} we illustrate various features 
by means of Monte Carlo (MC) simulation of ``polydisperse Lennard-Jones" (pLJ) particles. 
We characterize the static generalized moduli $L(q)$ and $G(q)$
up to $q$ corresponding to the main peak of the monomer structure factor 
and the time-dependent ICFs in the continuum $q$-limit.
Deviations from the assumed Gausianity and isotropy will be described
using proper invariants.
\end{itemize}

We begin in Sec.~\ref{ten} with a summary of general properties of ITFs
and a discussion of possibilities to pinpoint true anisotropies.
Some computational details are described in Sec.~\ref{comp}:
the computational model in Sec.~\ref{comp_algo},
averaging procedures in Sec.~\ref{comp_cklt}
and the measured instantaneous TFs in Sec.~\ref{instTFs}.
We characterize then in turn 
the static CFs of displacement and strain TFs in amorphous solids (cf.~Sec.~\ref{cstat})
and the Gaussianity (cf.~Sec.~\ref{gauss}) and the isotropy (cf.~Sec.~\ref{aniso})
of these fields.
Section~\ref{linres} discusses the linear response of TFs:
``response fields" (RFs), ``source fields" (SFs) and
``Green and growth function fields" (GFs) in Sec.~\ref{GFs},
the linear relation between GFs and CFs from the ``Fluctuation-Dissipation-Theorem" (FDT) 
\cite{ChaikinBook,DoiEdwardsBook} in Sec.~\ref{fdt},
the displacement response in amorphous bodies due to a small imposed force density
(cf.~Sec.~\ref{resp}) and the generalized Boltzmann superposition relation
for force density/stress and displacement/strain fields.
We show then in Sec.~\ref{dyn} how the time-dependent CFs
are related to $L(q,t)$ and $G(q,t)$.
Our work is summarized in Sec.~\ref{conc}.
We provide additional information on
FTs and ``Laplace-Carson transformations" (LTs) in Appendix~\ref{trans}, 
ITFs for isotropic and achiral systems in Appendix~\ref{tenmore},
CFs of the instantaneous rotation TF in Appendix~\ref{rotate},
technical consequences of the assumed stationarity in Appendix~\ref{corr}
and the large-time behavior of strain ICFs in Appendix~\ref{tauterm}.

\section{Isotropic tensor fields}
\label{ten}

\subsection{Some properties in reciprocal space}
\label{ten_sum}

We summarize here properties of ITFs used below. 
More details may be found in Appendix~\ref{ten} and in the literature
\cite{Schouten,McConnell,Schultz_Piszachich,TadmorCMTBook,spmP5a,spmP5b}.
Using the standard indicial notation \cite{TadmorCMTBook} and Cartesian coordinates 
with orthonormal basis \cite{Schultz_Piszachich} it is assumed that all second-order TFs 
are symmetric and that the minor and major index symmetries \cite{TadmorCMTBook}
hold for all fourth-order TFs.
Moreover, it is supposed that all second- and fourth-order ITFs are {\em even} 
with respect to the field vector as required for CFs of {\em achiral} systems. 
Most importantly, it is assumed that the TFs are {\em isotropic},
i.e. the isotropy condition \cite{Schultz_Piszachich,spmP5a}
\begin{equation}
T^{\Tgen}_{\alpha_1\ldots\alpha_o}(\qvec) = T_{\alpha_1\ldots\alpha_o}(\qvec^{\Tgen})
\label{eq_ten_sum_iso}
\end{equation}
holds for {\em any} orthogonal transformation (marked by ``$\Tgen$") 
of the coordinate system.
Isotropy, Eq.~(\ref{eq_ten_sum_iso}), necessarily implies homogeneity
\cite{Schultz_Piszachich}. We thus do not need to assume homogeneity explicitly.
As discussed in Appendix~\ref{ten_field}, the above assumptions imply that
\begin{eqnarray}
T_{\alpha}(\qvec) & = & l_1(q) \ \qhat_{\alpha} \label{eq_ten_sum_o1} \\
T_{\alpha\beta}(\qvec) & = & 
k_1(q) \ \delta_{\alpha\beta} + k_2(q) \ \qhat_{\alpha} \qhat_{\beta} 
\label{eq_ten_sum_o2} \\
T_{\alpha\beta\gamma\delta}(\qvec) & = &
i_1(q) \ \delta_{\alpha\beta} \delta_{\gamma\delta} \label{eq_ten_sum_o4} \\
& + & i_2(q) \left(
\delta_{\alpha\gamma} \delta_{\beta\delta} + \delta_{\alpha\delta} \delta_{\beta\gamma}
\right) \nonumber \\
& + & i_3(q) \left(
\qhat_{\alpha} \qhat_{\beta}\delta_{\gamma\delta} + \qhat_{\gamma} \qhat_{\delta}\delta_{\alpha\beta} 
\right) \nonumber \\
& + & i_4(q) \ \qhat_{\alpha} \qhat_{\beta} \qhat_{\gamma} \qhat_{\delta} \nonumber\\
& + & 
i_5(q)  \left( 
\qhat_{\alpha} \qhat_{\gamma} \delta_{\beta\delta}+  
\qhat_{\alpha} \qhat_{\delta} \delta_{\beta\gamma}+\right. \nonumber \\
& & \hspace*{.90cm}\left. \qhat_{\beta}\qhat_{\gamma} \delta_{\alpha\delta}+  
\qhat_{\beta}\qhat_{\delta} \delta_{\alpha\gamma}  
\right) \nonumber
\end{eqnarray}
for $d \ge 2$ in terms of one invariant scalar $l_1(q)$ for first-order fields,
two invariants $k_n(q)$ for the second-order fields and 
five invariants $i_n(q)$ for the fourth-order fields.
(The last relation does not hold for the CFs of the
rotation TF discussed in Appendix~\ref{rotate}.)
For $d=2$ it is possible to rewrite Eq.~(\ref{eq_ten_sum_o4})
more compactly by means of the transformation
\begin{eqnarray}
i_1(q) & \to & i_1(q) - 2 i_5(q), \label{eq_ten_sum_d2_in_transform} \\
i_2(q) & \to & i_2(q) +   i_5(q), \nonumber\\
i_3(q) & \to & i_3(q) + 2 i_5(q), \nonumber\\
i_4(q) & \to & i_4(q) \mbox{ and } \nonumber\\
i_5(q) & \to & 0  \nonumber
\end{eqnarray}
in terms of only four invariants $i_{n\le 4}(q)$. 
See the last paragraph of Appendix~\ref{ten_field} for details.
Following Refs.~\cite{lyuda18,spmP5a,spmP5b} let us rotate the coordinate system 
such that the $1$-axis points into the direction of $\qvec$, i.e.
$q_{\alpha}^{\circ} = q \delta_{1\alpha}$ with $q = |\qvec|$.
We mark coordinates in these ``Natural Rotated Coordinates" (NRC) by ``$\circ$" and define
the invariants
\begin{eqnarray}
\kL(q)  & \equiv & T^{\circ}_{11}(\qvec), \iL(q) \equiv T^{\circ}_{1111}(\qvec), \nonumber \\ 
\kN(q)  & \equiv & T^{\circ}_{22}(\qvec), \iN(q) \equiv T^{\circ}_{2222}(\qvec), \nonumber \\
\iM(q)  & \equiv & T^{\circ}_{1122}(\qvec), \nonumber \\
\iG(q)  & \equiv & T^{\circ}_{1212}(\qvec) \mbox{ and } \nonumber \\
\iS(q)  & \equiv & T^{\circ}_{2233}(\qvec) 
\label{eq_ten_sum_NRC_invariants} 
\end{eqnarray}
for second- and fourth-order ITFs in NRC.
Since the system is isotropic these functions depend on the scalar $q$ but not on
the wavevector direction $\qhatvec$, i.e. 
they are {\em invariant} under rotation and they do not change 
if one of the coordinate axes is inversed.
Both sets of invariants are linearly related by
\begin{eqnarray}
\kL(q) & = & k_1(q)+k_2(q), \kN(q) = k_1(q), \nonumber  \\ 
\iL(q) & = &  i_1(q) + 2i_2(q) + 2 i_3(q) + i_4(q) + 4 i_5(q),\nonumber \\
\iG(q) & = &  i_2(q) + i_5(q),                                \nonumber \\
\iM(q) & = &  i_1(q) + i_3(q),                                \nonumber \\
\iN(q) & = &  i_1(q) + 2i_2(q) \mbox{ and }                   \nonumber \\
\iS(q) & = &  i_1(q).               \label{eq_ten_sum_d3}
\end{eqnarray}
Using the product theorem of ITFs (cf.~Appendix~\ref{ten_product}) one may
construct ITFs by taking outer products of ITFs of lower order.
Let us introduce the linear operator
\begin{eqnarray}
\Tabcd(\qvec) & = & \OPup[\Tab(\qvec)] \mbox{ with } \nonumber \\
\OPup[\Tab(\qvec)]
& \equiv &
\frac{1}{4} \left[
\qhat_{\alpha}\qhat_{\delta} T_{\beta\gamma}(\qvec) +
\qhat_{\alpha}\qhat_{\gamma} T_{\beta\delta}(\qvec) \right.  \nonumber \\
        & + & \left. \hspace*{0.38cm}
\qhat_{\beta}\qhat_{\delta} T_{\alpha\gamma}(\qvec) +
\qhat_{\beta}\qhat_{\gamma} T_{\alpha\delta}(\qvec)\right] 
\label{eq_ten_sum_OPup_def}
\end{eqnarray}
constructing a fourth-order ITF from a given second-order ITF  $\Tab(\qvec)$.
The invariants of $\Tab(\qvec)$ and $\Tabcd(\qvec)$ in NRC are related by
\begin{eqnarray}
\iL(q) & = & \kL(q) \mbox{ and } \iG(q) = \kN(q)/4 \mbox{ while } \nonumber \\
\iM(q) & = & \iN(q) = \iS(q) = \iT(q) = 0.
\label{eq_ten_sum_OPup_invariants}
\end{eqnarray}
Similarly, it is possible to construct from a higher order ITF
by contraction with another ITF a lower order ITF. We shall thus 
use below the linear operator
\begin{equation}
\Tab(\qvec) \equiv \OPdown[\Tabcd(\qvec)] \equiv 
\qhat_{\gamma} \qhat_{\delta} T_{\alpha\gamma\beta\delta}(\qvec), 
\label{eq_ten_sum_OPdown_def}
\end{equation}
generating a second-order ITF by taking twice the inner product
of a fourth-order ITF $\Tabcd(\qvec)$ with $\qhat_{\alpha}$.
As one readily verifies the invariants of $\Tab(\qvec)$ are  
\begin{equation}
\kL(q) = \iL(q) \mbox{ and } \kN(q) = \iG(q).
\label{eq_ten_sum_OPdown_invariants}
\end{equation}
We note finally that assuming $\Aab(\qvec)$ to be an ITF one may 
define an associated inverse ITF $\Bab(\qvec)$ by
\begin{equation}
A_{\alpha\gamma}(\qvec) B_{\gamma\beta}(\qvec) = \IDab. 
\label{eq_ten_sum_inv}
\end{equation}
With $\aL(q)$ and $\aN(q)$ denoting the two invariants of $\Aab(\qvec)$ in NRC
(assumed to be finite)
and $\bL(q)$ and $\bN(q)$ the corresponding invariants of $\Bab(\qvec)$ this implies
\begin{equation}
\bL(q) = 1/\aL(q) \mbox{ and } \bN(q) = 1/\aN(q).
\label{eq_ten_sum_inv_invariants}
\end{equation}

\begin{table*}[t]
\begin{center}
\begin{tabular}{|c||c|c|c||c|c|c|}
\cline{2-7}
\multicolumn{1}{c}{$ $} &
\multicolumn{3}{|c||}{$d=2$} &
\multicolumn{3}{c|}{$d=3$}\\ \hline
invariant     &
$\eta=0$      & 
$\eta=1$      &     
$\eta=2$      &     
$\eta=0$      & 
$\eta=1$      &     
$\eta=2$      \\ \hline     
$\ltild_1(r)/i$&
$4 \lhat_1$    &
$4 \lhat_1$    &
$4 \lhat_1$    &
$\frac{8}{\pi} \lhat_1$&
$2 \lhat_1 $  &
$\frac{4}{\pi} \lhat_1 $\\ 
$\ktild_1(r)$              &
$4 \khat_2$                &
$4 [\khat_1+\khat_2]$    &
\mbox{ log. div. }               &
$2 \khat_2$                &
$\frac{4}{\pi} [\khat_1+\khat_2] $ &
$2 \khat_1 + \khat_2$     \\ 
$\ktild_2(r)$              &
$-8\khat_2$                &
$-4 \khat_2$               &
$-2 \khat_2$               &	
$-6\khat_2$                &
$-\frac{8}{\pi} \khat_2$   &
$- \khat_2$                \\ 
$\itild_1(r)  $                              &
$2[4\ihat_3+\ihat_4]$                        &
$\frac{4}{3} [3\ihat_1+6\ihat_3+\ihat_4]$    &
\mbox{ log. div. }	                     &
$1 [4\ihat_3+ \ihat_4]$                      &
$\frac{4}{3\pi} [3\ihat_1+6\ihat_3+\ihat_4]$ &
$\frac{1}{4} [8\ihat_1+8\ihat_3+\ihat_4]$    \\ 
$\itild_2(r)  $                              &
$2[4\ihat_5+\ihat_4]$                        &
$\frac{4}{3}[3\ihat_2+6\ihat_5+\ihat_4]$     &
\mbox{ log. div. }	                     &
$1 [4\ihat_5+\ihat_4]$                       &
$\frac{4}{3\pi}[3\ihat_2+6\ihat_5+\ihat_4]$  &
$\frac{1}{4} [8\ihat_1+8\ihat_5+\ihat_4]$    \\ 
$\itild_3(r)  $                      &
$-4[2\ihat_3+\ihat_4]$               &
$-\frac{4}{3}[3\ihat_3+\ihat_4]$     &
$-\frac{1}{2}[4\ihat_3+\ihat_4]$     &
$-3[2\ihat_3+\ihat_4]$               &
$-\frac{8}{3\pi}[3\ihat_3+\ihat_4]$  &
$-\frac{1}{4}[4\ihat_3+\ihat_4]$     \\ 
$\itild_5(r)  $                      &
$-4[2\ihat_5+\ihat_4]$               &
$-\frac{4}{3}[3\ihat_5+\ihat_4]$     &
$-\frac{1}{2}[4\ihat_5+\ihat_4]$     &
$-3[2\ihat_5+\ihat_4]$               &
$-\frac{8}{3\pi}[3\ihat_5+\ihat_4]$  &
$-\frac{1}{4}[4\ihat_5+\ihat_4]$     \\
$\itild_4(r)  $                      &
$16\ihat_4$                          &
$1 [4\ihat_4]$                       &
$\ihat_4$			     &
$15\ihat_4$                          &
$\frac{8}{3\pi} [4 \ihat_4]$         &
$\frac{3}{4}\ihat_4$	             \\ 
\hline
\end{tabular}
\caption[]{Invariants for first-, second- and fourth-order ICFs in real space for 
spatial dimensions $d=2$ and $d=3$ and power-law exponents $\eta=0$, $1$ and $2$.
All invariants are given in units of $8\pi r^{d-\eta}$.
Invariants of ITF of odd order $o$ are imaginary as indicated in the first column.
$\delta(r)$-singularities are not indicated.
Note that $\eta=0$ corresponds to constant invariants in reciprocal space.
To simplify the comparison of different dimensions several entries have been factorized
such that the brackets $[\ldots]$ for different $d$ but same $\eta$ are identical.
``log. div." in the fourth column marks logarithmic divergences.
} 
\label{tab_eta}
\end{center}
\end{table*}

\subsection{Inverse Fourier transformations for ITFs}
\label{ten_q2r}

We have formulated above all properties in reciprocal space.
TFs in real and reciprocal space are related by 
\begin{equation}
T_{\alpha\ldots}(\qvec) = \Fcal[T_{\alpha\ldots}(\rvec)].
\label{eq_ten_q2r_FT}
\end{equation}
Importantly, the FT of any ITF must also be an ITF, i.e. 
\begin{equation}
T^{\Tgen}_{\alpha\ldots}(\rvec) = T_{\alpha\ldots}(\rvec^{\Tgen}) 
\stackrel{\Fcal}{\Leftrightarrow} 
T^{\Tgen}_{\alpha\ldots}(\qvec) = T_{\alpha\ldots}(\qvec^{\Tgen})
\label{eq_ten_q2r_ITFs}
\end{equation}
for the isotropy conditions in, respectively, real and in reciprocal space.
This general relation holds due the linearity of the FT.
According to Eq.~(\ref{eq_ten_q2r_FT}) we have
\begin{eqnarray}
T_{\alpha}(\rvec) & = & \ltild_1(r) \ \rhat_{\alpha},\label{eq_ten_q2r_FT_o1} \\
T_{\alpha\beta}(\rvec) & = & 
\ktild_1(r) \ \delta_{\alpha\beta} + \ktild_2(r) \ \rhat_{\alpha}\rhat_{\beta},
\label{eq_ten_q2r_FT_o2} \\
T_{\alpha\beta\gamma\delta}(\rvec) & = & 
\itild_1(r) \ \delta_{\alpha\beta} \delta_{\gamma\delta} \label{eq_ten_q2r_FT_o4} \\
& + & \itild_2(r) \ \left(
\delta_{\alpha\gamma} \delta_{\beta\delta} + \delta_{\alpha\delta} \delta_{\beta\gamma}
\right) \nonumber \\
& + & \itild_3(r) \ \left(
\rhat_{\alpha} \rhat_{\beta}\delta_{\gamma\delta} + 
\rhat_{\gamma}\rhat_{\delta}\delta_{\alpha\beta} 
\right) \nonumber \\
& + & \itild_4(r) \ \rhat_{\alpha} \rhat_{\beta} \rhat_{\gamma} \rhat_{\delta} \nonumber\\
& + & 
\itild_5(r)  \left( 
\rhat_{\alpha}\rhat_{\gamma} \delta_{\beta\delta}+  
\rhat_{\alpha}\rhat_{\delta} \delta_{\beta\gamma}+\right. \nonumber \\
& & \hspace*{.90cm}\left. \rhat_{\beta}\rhat_{\gamma} \delta_{\alpha\delta}+  
\rhat_{\beta}\rhat_{\delta} \delta_{\alpha\gamma}  
\right) \nonumber
\end{eqnarray}
for $r>0$.
As for the ITFs in reciprocal space, the TF of order $o=2$ is symmetric and
the major and minor index symmetries hold for ITF of order $o=4$.
We have used here a representation for $r>0$ in terms of the components
$\rhat_{\alpha}$ of the normalized vector $\rhatvec=\rvec/r$ in real space.
If an ITF of a certain order and index symmetry is given in reciprocal space, 
the same holds in real space and the components and the invariants of each TF 
in real and reciprocal space are related by Eq.~(\ref{eq_ten_q2r_ITFs}).
The task is thus to get the invariants in real space from the invariants
in reciprocal space and {\em visa versa}.
We assume that all invariants of a field of order $o$
in reciprocal space are proportional to the {\em same} power law $s_{\eta}(q)=1/Vq^{\eta}$
characterized by an exponent $\eta \ge 0$, i.e.
$l_n(q) = \lhat_n s_{\eta}(q)$,
$k_n(q) = \khat_n s_{\eta}(q)$ and
$i_n(q) = \ihat_n s_{\eta}(q)$
with $\lhat_n$, $\khat_n$ and $\ihat_n$ being constants.
We focus on $\eta=0, 1$ and $2$.
Generalizing the procedure given in Refs.~\cite{spmP5a,spmP5b} for $d=2$ and $\eta=0$ 
the inverse FT may be computed in three steps:
\begin{itemize}
\item
One first considers the inverse FT of
$f(\qvec) = s_{\eta}(q) Y(\qhatvec)$ with $Y(\qhatvec)$ being either
a planar or a spherical harmonics.
This yields $f(\rvec) = \ftild(r) Y(\rhatvec)$
with the same function $Y(\cdot)$ as in reciprocal space and a scalar $\ftild(r)$
depending on $\eta$. 
\item
One expresses then
$\qhat_{\alpha}$, $\qhat_{\alpha}\qhat_{\beta}, \ldots$ in terms of the $Y(\qhatvec)$
and, similarly $\rhat_{\alpha}$, $\rhat_{\alpha}\rhat_{\beta}, \ldots$ in terms of the $Y(\rhatvec)$.
This gives the inverse FT of all additive terms of the ITFs in reciprocal space.
\item
A final summation over different contributions
yields the real space invariants given in Table~\ref{tab_eta}.
\end{itemize}
The fourth-order invariants for $d=2$ may be further compressed using 
Eq.~(\ref{eq_ten_sum_d2_in_transform}) consistently with Ref.~\cite{spmP5b}.

\subsection{Numerical test of isotropy hypothesis}
\label{ten_aniso}

We have assumed above that the stated symmetries hold
for {\em all} field vectors $\rvec$ or $\qvec$.
Obviously, this cannot be the case for experimentally or numerically obtained TFs.
Deviations do in practice occur at least in the low-$q$ (large-$r$) and 
the large-$q$ (small-$r$) limits.
Low-$q$ deviations commonly arise due to anisotropic 
boundary conditions \cite{foot_lowqextrapol}, e.g., the use of a standard square periodic 
simulation box in computer simulations \cite{AllenTildesleyBook}, 
large-$q$ deviations simply due to the grid symmetry and the finite grid lattice 
constant $\agrid$ in real space used for the data sampling \cite{spmP5a}.
Moreover, due to the finite size $\ximon$ of the particles and the ensuing packing 
constraints at high densities no real condensed matter system can be perfectly 
isotropic for large $q$. 
We remind that inhomogeneity necessarily implies anisotropy.
Let us assume that the system is homogeneous for small wavenumbers $q \ll 1/\xihom$ 
with $\xihom$ being set by the typical size of the
local heterogeneities. 
This implies that Eq.~(\ref{eq_ten_sum_iso}) can only hold for 
$1/L \ll q \ll 1/\xiiso$ with $L$ being the linear system size 
and $\xiiso \ge \xihom$ characterizing the size of local anisotropies.
As shown recently for TFs in real space \cite{Massimo21,Massimo24},
to test the isotropy hypothesis and to quantify possible anisotropic effects one needs 
to measure true invariants under arbitrary orthogonal coordinate transformations.
One thus has
\begin{itemize}
\item
to measure a sufficiently large number of components of the TF,
\item
to fit the invariants $k_n(q)$ or $i_n(q)$
according to the generic mathematical structure of ITFs and 
\item
to decide according to a scalar $\chi^2$-test \cite{numrec}
whether the ``isotropy hypothesis" holds.
\end{itemize}
A related alternative is to do this analysis entirely in NRC.
Let us illustrate this for a second-order TF $\Tab(\qvec)$ in $d=2$.
Instead of the two invariants $k_1(q)$ and $k_2(q)$,
cf. Eq.~(\ref{eq_ten_sum_o2}), one measures in NRC the components 
$T^{\circ}_{11}(\qvec)$ and $T^{\circ}_{22}(\qvec)$ 
parallel (longitudinal) and perpendicular (normal) to $\qvec$.
For a perfectly isotropic system these components only depend on $q$. 
In practice, even for reasonable isotropic systems some $\qhatvec$-dependence
is always present due to, e.g., thermal fluctuations. 
It is thus justified to compute $\kL(q)$ and $\kN(q)$,
cf.~Eq.~(\ref{eq_ten_sum_NRC_invariants}), using 
\begin{eqnarray}
\kL(q) = k_1(q)+k_2(q) & = & 
\la T^{\circ}_{11}(\qvec) \ra_{\qhatvec} \mbox{ and } \nonumber \\
\kN(q) = k_1(q) & = & 
\la T^{\circ}_{22}(\qvec) \ra_{\qhatvec} 
\label{eq_ten_aniso_A}
\end{eqnarray} 
by averaging over all wavevectors in a $q$-bin.
Possible anisotropies may be characterized using the moments
\begin{eqnarray}
\dkL(q) & \equiv & 
\la \left( T^{\circ}_{11}(\qvec) - \kL(q) \right)^m \ra_{\qhatvec}^{1/m} \nonumber \mbox{ and } \\ 
\dkN(q) & \equiv & 
\la \left( T^{\circ}_{22}(\qvec) - \kN(q) \right)^m \ra_{\qhatvec}^{1/m} 
\label{eq_ten_aniso_B}
\end{eqnarray}
(with $m=2,3,\ldots$) which must vanish for perfectly isotropic systems.
An example is discussed in Sec.~\ref{aniso}.

\section{Computational details}
\label{comp}

\subsection{Polydisperse Lennard-Jones particles}
\label{comp_algo}

Quite generally, computer simulations are of interest where a slow but realistic dynamical 
algorithm is mixed with a fast albeit artificial algorithm allowing to efficiently sample 
the phase space \cite{AllenTildesleyBook,LandauBinderBook}. 
This can be achieved for polydisperse glass-forming colloids by combining
``molecular dynamics" (MD) simulations or local MC hopping moves 
\cite{AllenTildesleyBook,LandauBinderBook} 
with ``swap MC moves" \cite{Berthier17} 
exchanging the diameters of two randomly chosen particles \cite{spmP1,spmP5a}.
As in previous studies \cite{spmP1,spmP5a,spmP5b},
we present numerical results obtained for two-dimensional 
``polydisperse Lennard-Jones" (pLJ) particles
quenched and tempered with switched on swap MC moves.
More information on computational details
(temperature quench, tempering, production runs, measurements of macroscopic
observables and of microscopic TFs, storage of time series, FTs of TFs using discrete grids) 
can be found in Refs.~\cite{spmP5a,spmP5b}.
Lennard-Jones units \cite{AllenTildesleyBook} are used throughout this work.  
All production runs are finally performed by switching off the swap MC moves.
All data are sampled at a temperature $T=0.2$ which is much lower than the glass transition 
temperature $\Tglass \approx 0.26$ \cite{spmP1}.
As expected for an amorphous solid, the particles are only able to move over distances 
of about $1/10$ of the typical particle size (``Lindemann criterion" \cite{DhontGlasses}).
Due to the use of an MC algorithm,
not only the particle trajectories but also collective relaxation modes reveal an overdamped dynamics
characterized by an effective friction coefficient
$\zeta$ which we shall determine in Sec.~\ref{dyn_e_simu}.
We consider systems containing up to $n=160000$ particles.
Compared to Ref.~\cite{spmP5b} we have thus increased $n$ by a factor $4$.
According to Eq.~(\ref{eq_ten_ten_o4}) 
the macroscopic elastic modulus tensor $\Eabcd$ \cite{LandauElasticity} 
may be written for isotropic systems as
\begin{equation}
\Eabcd = \lambda \delta_{\alpha\beta} \delta_{\gamma\delta}
+ \mu \left(\delta_{\alpha\gamma}\delta_{\beta\delta}+
\delta_{\alpha\delta}\delta_{\beta\gamma}\right)
\label{eq_moduli_Eabcd_iso}
\end{equation}
in terms of the two Lam\'e moduli $\lambda$ and $\mu$ \cite{LandauElasticity,TadmorCMTBook}.
Using the stress-fluctuation formalism described elsewhere \cite{Hoover69,spmP1}
we have determined $\lambda \approx 38$ and $\mu \approx 14$ for $T=0.2$. 

\subsection{Different types of averages}
\label{comp_cklt}

As a final and last averaging step we always take the {\bf $c$-average} $\la \ldots \ra_c$ 
over all independent configurations $c$ (equilibrated using swap MC moves).
It is assumed that this ensemble is isotropic and achiral
and that the number $\Nc$ of configurations of this ensemble is as large as possible. 
We have sampled at least $\Nc=100$ independent configurations.
For $n=10000$ we have $\Nc=200$.

The {\bf $k$-average} 
$a(c) \equiv \la \ahat_{ck} \ra_k$ for some observable $\ahat_{ck}$ depending
on the state $k$ for the given independent configuration $c$ corresponds ideally to the 
standard thermodynamic average over all allowed states $k$  \cite{ChaikinBook}. 
For non-ergodic systems this average generally depends on $c$.
Naturally, in practice only a finite number $\Nk$ is sampled.
Since this is done by analyzing the $\Nk$ stored ``frames" $k$ of each configuration $c$,
we are limited to $\Nk=10000$ for $n=10000$ and $\Nk=1000$ for larger $n$.

We store for each $n$ and $c$ {\em four} time-series with $\Nk$
frames $k$ with equidistant time intervals $\tincr = 1, 10, 100$ and $1000$ MCS.
The total production time of each time-series is thus $\tsampmax=\Nk \tincr$,
e.g., $\tsampmax=10^6$ MCS for $n > 10000$ and $\tincr=1000$ MCS.
Storing these time-series allows the characterization of dynamical properties.
This is done here by analyzing {\bf $t$-averages}
\begin{equation}
\bar{a}(\tsamp) \equiv \frac{1}{\tsamp} \int_0^{\tsamp} \ddiff t \ \ahat(t)
\approx \frac{1}{\Nt} \sum_{i=1}^{\Nt} \ahat(t=i\tincr)
\label{eq_taver_def}
\end{equation}
of instantaneous $\ahat(t)$ over ``preaveraging times"
$\tsamp = \Nt \tincr \le \tsampmax$ with $\tincr$ being the discrete time increment
of the $\Nt \le \Nk$ equidistant measurements. 
As described in Appendix~\ref{corr},
$t$-averaged fields have $\tsamp$-dependent CFs which are for 
stationary stochastic TFs related to the more common CFs of these fields
with time-lag $t$. 

\subsection{Instantaneous TFs probed}
\label{instTFs}

The displacement field $u_{\alpha}(\qvec)$ in reciprocal space 
(for a given configuration $c$ and a given state $k$ or time $t$)
may in principle be defined by integrating the velocity field 
$v_{\alpha}(\qvec,t) = \dot{u}_{\alpha}(\qvec,t)$.
To avoid the arbitrary integration constant, it is imposed that
\begin{equation}
\la u_{\alpha}(\qvec,c,k) \ra_k = 0
\mbox{ for all } \qvec \mbox{ and } c,
\label{eq_udef_q}
\end{equation}
i.e. any measured $u_{\alpha}(\qvec,c,k)$ is shifted by its $k$-average.
This means physically and in numerical practice that we use in real space
as a reference position $\rref_{\alpha}^a$ for the displacement vector 
$u_{\alpha}^{a}=r_{\alpha}^a - \rref_{\alpha}^a$ of each particle $a$
at time $t$ the $k$-averaged monomer position
\begin{equation}
\rref_{\alpha}^a \equiv \la r_{\alpha}^a(k) \ra_k,
\mbox{ i.e.} \la u_{\alpha}^{a}(k) \ra_k \equiv 0,
\label{eq_udef_rref}
\end{equation}
with $r_{\alpha}^a(k)$ denoting the coordinates of particle $a$ 
of configuration $c$ and state $k$ \cite{spmP5b}.
We remind that the displacement field in real space may be defined by \cite{foot_urref}
\begin{equation}
u_{\alpha}(\rvec) \equiv \frac{1}{n/V} \sum_a u_{\alpha}^{a} \delta(\rvec-\rref_{\alpha}^a).
\label{eq_udef_r}
\end{equation}
It follows from Eq.~(\ref{eq_udef_rref}) that the $k$-average indeed vanishes. 
The linear strain TF is defined by \cite{LandauElasticity}
\begin{equation}
\epsab(\qvec) \equiv \frac{i}{2}
\left[ q_{\alpha} u_{\beta}(\qvec) + q_{\beta} u_{\alpha}(\qvec) \right]
\label{eq_epsab_def}
\end{equation}
as a symmetric second-order TF associated to $u_{\alpha}(\qvec)$.
(See Appendix~\ref{rotate} for the definition of related ``rotation TF" $\rotab(\qvec)$.)
By construction Eq.~(\ref{eq_udef_q}) and Eq.~(\ref{eq_epsab_def}) imply that 
the $k$-averaged strain field must also vanish.
That a displacement or strain TF is an instantaneously
taken phase function is often emphasized by carrets,
i.e. we write $\uhat_{\alpha}(\qvec)$ and $\epshatab(\qvec)$.
We store and manipulate these TFs using periodic square lattices 
with a lattice constant $\agrid \approx 0.1$.

\section{Static correlation functions}
\label{cstat}

\subsection{Displacement correlations in reciprocal space}
\label{cstat_u}
\label{cstat_uq}

The CFs $\cab(\qvec)$ of the instantaneous displacement TF $\hat{u}_{\alpha}(\qvec)$
in reciprocal space are defined by \cite{foot_intro_carrets}
\begin{eqnarray}
\cab(\qvec)   & \equiv & \la \cab(\qvec,c) \ra_c \mbox{ with } \label{eq_cab_def} \\
\cab(\qvec,c) & \equiv & \la \cab(\qvec,c,k) \ra_k \mbox{ and }  \nonumber \\
\cab(\qvec,c,k) & \equiv & \uhat_{\alpha}(\qvec,c,k) \uhat_{\beta}(-\qvec,c,k) 
\label{eq_cab_def_B} 
\end{eqnarray}
assuming by construction $\la \uhat_{\alpha}(\qvec,c,k) \ra_k =0$.
In agreement with Eq.~(\ref{eq_ten_sum_o2}) we may write
\begin{equation}
c_{\alpha\beta}(\qvec) = k_1(q) \delta_{\alpha\beta} + k_2(q) \qhat_{\alpha}\qhat_{\beta}.
\label{eq_cab_ITF}
\end{equation}
Using the first relation of Eq.~(\ref{eq_ten_sum_d3}) we express $k_1(q)$ and $k_2(q)$ 
in terms of the corresponding invariants in NRC 
\begin{eqnarray}
\kL(q) =
k_1(q)+ k_2(q) & = & \la c_{11}^{\circ}(\qvec) \ra_{\qhatvec} \mbox{ and } 
\label{eq_cab_NRC} \\ 
\kN(q) = k_1(q)         & = & 
\la c_{22}^{\circ}(\qvec) \ra_{\qhatvec} = \ldots = 
\la c_{dd}^{\circ}(\qvec) \ra_{\qhatvec}.
\nonumber
\end{eqnarray}
While for idealized iso\-tropic systems $c_{\alpha\beta}^{\circ}(\qvec)$ 
must only dependent on the wavenumber $q$ and not on the wavevector direction $\qhatvec$,
this is generally only approximatively true for real experiments or 
computer simulations for often merely statistical reasons.
It is then justified to additionally take the average $\langle \ldots \rangle_{\qhatvec}$ 
over all measured directions $\qhatvec$ \cite{spmP5a,spmP5b}.
Let us assume following Ref.~\cite{ChaikinBook} that
the displacements $\uhat_{\alpha}^{\circ}(\qvec)$ 
in NRC are  Gaussian variables following Maxwell-Boltzmann statistics.
All contributions to the free energy from different wavevectors and modes 
thus factorize and the partition function for a given $\qvec$ becomes an integral
over all $u_{\alpha}(\qvec)$ with a Gibbs weight 
$\exp[ -\beta V \ \delta f(\qvec)]$ set by the free energy density
\begin{eqnarray}
\delta f(\qvec) & \equiv & \frac{1}{2} \uhat_{\alpha}(\qvec) 
q^2 \Eab(\qvec)  \uhat_{\beta}(-\qvec)
\nonumber \\
& = & \frac{1}{2} \uhat_{\alpha}^{\circ}(\qvec) 
q^2 E_{\alpha\beta}^{\circ}(\qvec) \uhat_{\beta}^{\circ}(-\qvec)
\label{eq_cab_df}
\end{eqnarray}
where we have used in the second step that $\delta f(\qvec)$ is a scalar.
Here, $\Eab(\qvec)$ denotes a second-order symmetric ITF with invariants given by
\begin{eqnarray}
\kEL(q) = \kEone(q)+\kEtwo(q) & \equiv & L(q) \mbox{ and } \nonumber \\
\kEN(q) = \kEone(q) & \equiv & G(q)
\label{eq_cab_Eab_invariants}
\end{eqnarray} 
where we have introduced the $q$-dependent longitudinal modulus $L(q)$
and the $q$-dependent shear modulus $G(q)$.
Performing then the Gaussian integrals leads to
\begin{equation}
q^2\kL(q) = \frac{1}{\beta V L(q)} \mbox{ and }
q^2\kN(q) = \frac{1}{\beta V G(q)}
\label{eq_cab_LqGq} 
\end{equation}
in agreement with Ref.~\cite{ChaikinBook}.
Let us define by
\begin{equation}
E_{\alpha\gamma}(\qvec) K_{\gamma\beta}(\qvec) = \delta_{\alpha\beta}
\label{eq_Kab_def}
\end{equation}
a second-order TF $\Kab(\qvec)$ as the inverse with respect to $\Eab(\qvec)$.
Using Eq.~(\ref{eq_ten_sum_inv_invariants}) the invariants of $\Kab(\qvec)$ are
\begin{equation}
\kKL(q) = \frac{1}{L(q)} \mbox{ and } \kKN(q) = \frac{1}{G(q)}.
\label{eq_kKLkKN_q}
\end{equation}
We thus rewrite Eq.~(\ref{eq_cab_LqGq}) compactly as
\begin{equation}
q^2 \cab(\qvec) = \Kab(\qvec)/\beta V
\label{eq_cab_Kab}
\end{equation}
in ordinary coordinates.
Following Ref.~\cite{ChaikinBook} $\Eab(\qvec)$ may alternatively be defined by the 
contraction $\Eab(\qvec) \equiv \OPdown[\Eabcd(\qvec)]$ of the fourth-order  
static elasticity TF $\Eabcd(\qvec)$ being also an ITF.
This will be elaborated in Sec.~\ref{boltz} and Sec.~\ref{dyn_u}
within a more general context which includes the time-dependence.

\begin{figure}[t]
\centerline{\resizebox{.9\columnwidth}{!}{\includegraphics*{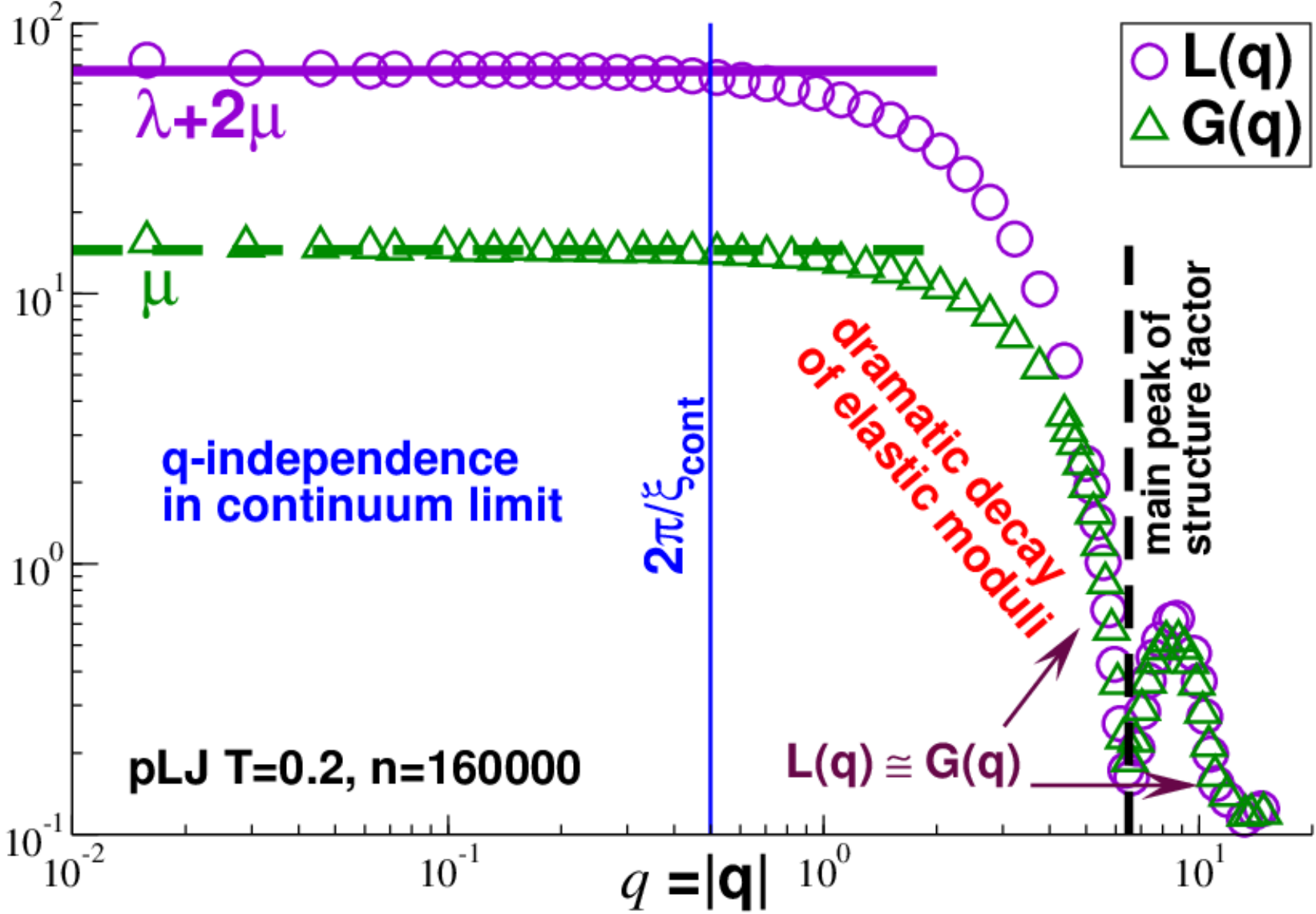}}}
\caption{Static microscopic elastic moduli $L(q)$ and $G(q)$ in reciprocal space obtained 
by rescaling according to Eq.~(\ref{eq_cab_LqGq}) the measured ICFs $\kL(q)$ and $\kN(q)$ 
of the displacement TFs of pLJ particles at $T=0.2$ for $n=160000$ particles. 
The horizontal lines indicate the macroscopic elastic moduli $\lambda + 2 \mu$ and $\mu$.
$L(q)$ and $G(q)$ are seen to strongly decrease for larger $q$ with 
a striking minimum at the position of the main peak of the coherent structure factor 
(dashed vertical line).
}
\label{fig_pdLJ_LqGq}
\end{figure}

The relations Eq.~(\ref{eq_cab_LqGq}) may be used to determine $L(q)$ and $G(q)$ from the 
fluctuations of displacement TFs. This is demonstrated in Fig.~\ref{fig_pdLJ_LqGq} for 
our pLJ particle model with $n=160000$.
As can be seen, $L(q)$ and $G(q)$ are constant below $q \approx 0.5$ (thin vertical line).
Using the known macroscopic moduli $\lambda \approx 38$ and $\mu \approx 14$ 
confirms
\begin{equation}
L(q) \to \lambda + 2\mu \mbox{ and } G(q) \to \mu \mbox{ for } q \ll 2\pi/\xicont.
\label{eq_cab_LqGq_qxicont}
\end{equation}
Depending somewhat on the criterion, the length scale $\xicont$ 
characterizing the breakdown of the elastic continuum assumption is thus about 
$\xicont \approx 10$ for the presented pLJ particle system. 
(The trivial prefactor $2\pi$ used here for the determination of $\xicont$ is 
often suppressed elsewhere for clarity.)
See Refs.~\cite{Klix12,Klix15} for related experimental work using 
Eq.~(\ref{eq_cab_LqGq}) to extract the shear modulus $\mu(T)$ 
as a function of temperature $T$ from the low-$q$ limit of $G(q,T)$
using the recorded positions of effectively two-dimensional colloidal systems.
Both generalized static moduli are seen to dramatically decay over two orders of magnitude 
for larger $q$ and the minimum of both moduli at $q \approx 6.5$ (dashed vertical line) 
coincides perfectly with the main peak of the coherent structure factor presented 
elsewhere \cite{spmP1}. 
Note also that $L(q)$ and $G(q)$ are surprisingly similar for the largest $q$.

\subsection{Displacement correlations in real space}
\label{cstat_ur}

According to Eq.~(\ref{eq_ten_q2r_ITFs}) 
the inverse FT $\cab(\rvec)=\Fcal[\cab(\qvec)]$ must also be an ITF,
i.e.
\begin{equation}
\cab(\rvec) = \ktild_1(r) \delta_{\alpha\beta} + \ktild_2(r) \rhat_{\alpha}\rhat_{\beta}
\label{eq_cab_cr_ITF}
\end{equation}
where the two invariants $\ktild_1(r)$ and $\ktild_2(r)$ are in principle given by 
$k_1(q) = \kN(q)$ and $k_2(q) = \kL(q)-\kN(q)$. 
Fortunately, both generalized moduli become constant in the continuum limit,
cf.~Fig.~\ref{fig_pdLJ_LqGq}, which suggests for sufficiently large systems 
the approximation $L(q) \approx \lambda + 2\mu$ and $G(q) \approx \mu$.
Let us introduce the convenient constants 
\begin{equation}
J_1 \equiv \frac{1}{\mu}-\frac{1}{\lambda+2\mu} = \frac{\lambda+\mu}{\mu (\lambda+2\mu)}
\mbox{ and }
J_2 \equiv \frac{2}{\lambda+2\mu}
\label{eq_J1J2_def}
\end{equation}
having the same units as inverse moduli.
Using Table~\ref{tab_eta} for the exponent $\eta=2$ then implies
\begin{equation}
\beta \ktild_1(r) = \frac{J_1+J_2}{8\pi r} \mbox{ and } 
\beta \ktild_2(r) = \frac{J_1}{8\pi r}  
\label{eq_cab_d3}
\end{equation}
for $d=3$ and a logarithmic behavior for $d=2$.
As shown in Sec.~\ref{resp}, these CFs determine 
the displacement response due to an applied perturbative force density.

\subsection{Strain correlations in reciprocal space}
\label{cstat_eq}

The CFs of the instantaneous strain TF, cf.~Eq.~(\ref{eq_epsab_def}), 
are given in reciprocal space by
\begin{eqnarray}
\cabcd(\qvec)   & \equiv & \la \cabcd(\qvec,c) \ra_c \mbox{ with } \label{eq_cabcd_def} \\
\cabcd(\qvec,c) & \equiv & \la \cabcd(\qvec,c,k) \ra_k \mbox{ and }  \nonumber \\
\cabcd(\qvec,c,k) & \equiv & \epshatab(\qvec,c,k) \epshatcd(-\qvec,c,k) \nonumber 
\end{eqnarray}
similarly as the CFs of the displacement TFs by Eq.~(\ref{eq_cab_def}).
For an achiral and isotropic system this TF must take the generic form given by 
Eq.~(\ref{eq_ten_sum_o4}). In fact, only two of the in general five invariants 
matter as may be seen by expressing the strain CFs using Eq.~(\ref{eq_epsab_def}) 
by the CF $\cab(\qvec)$ of the displacement fields. 
As readily seen,
\begin{equation}
\cabcd(\qvec) = q^2 \OPup[\cab(\qvec)]
\label{eq_cab2cabcd}
\end{equation}
as in Eq.~(\ref{eq_ten_sum_OPup_def}).
Using Eq.~(\ref{eq_ten_sum_OPup_invariants})
we get in NRC the two ICFs
\begin{eqnarray}
\cL(q)  & = & q^2 \kL(q) = \frac{1}{\beta V L(q)} \mbox{ and } \nonumber \\
4\cG(q) & = & q^2 \kN(q) =  \frac{1}{\beta V G(q)}
\label{eq_cabcd_invariants}
\end{eqnarray}
and $c_4(q)= k_2(q)$ and $4c_5(q)=k_1(q)$ in ordinary coordinates
while all other invariants vanish.
Note also that using $q_{\alpha}^{\circ}=q \delta_{\alpha 1}$ and Eq.~(\ref{eq_epsab_def})
it is seen that only 
\begin{eqnarray}
\hat{\varepsilon}_{11}^{\circ}(\qvec) & = & i q \uhat_1^{\circ}(\qvec) \mbox{ and } \nonumber \\
\hat{\varepsilon}_{1\beta}^{\circ}(\qvec) = \hat{\varepsilon}_{\beta 1}^{\circ}(\qvec) 
& = & \frac{i q}{2} \uhat_{\beta}^{\circ}(\qvec) \mbox{ for } \beta \ne 1
\label{eq_cabcd_NRC}
\end{eqnarray}
can be finite while $\epshatab^{\circ}(\qvec) = 0$ for all other strain components.
Only the longitudinal and shear ICFs in NRC
\begin{eqnarray}
\cL(q) & = & \la c_{1111}^{\circ}(\qvec) \ra_{\qhatvec} \mbox{ and }\nonumber \\
\cG(q) & = & \la c_{1212}^{\circ}(\qvec) \ra_{\qhatvec} =\ldots = 
             \la c_{1d1d}^{\circ}(\qvec) \ra_{\qhatvec}
\label{eq_cLcG}
\end{eqnarray}
can thus be finite.
This implies using Eq.~(\ref{eq_ten_sum_d3}) that only two invariants
in ordinary space can be finite:
\begin{equation}
c_4(q) = \cL(q)-4\cG(q) \mbox{ and } c_5(q) = \cG(q).
\label{eq_cLcG2cn}
\end{equation}

\subsection{Strain correlations in real space}
\label{cstat_er}

As we have seen, all ICFs in reciprocal space become constant for small $q$.
Hence, $\beta V c_4(q) \simeq -J_1$ and
$4\beta V c_5(q) \to J_1+J_2/2$ for $q\xicont \ll 1$
using the constants $J_1$ and $J_2$ of Eq.~(\ref{eq_J1J2_def}).
Let us thus define the two amplitudes $\chat_4=-J_1/\beta$ and 
$\chat_5=(J_1+J_2/2)/4\beta$ of dimension volume.
The inverse FT $\cabcd(\rvec)$ is then obtained using Table~\ref{tab_eta} for $\eta=0$
in terms of the five invariants $\ctild_n(r)$ in real space. For $d=3$ we get
\begin{eqnarray}
8\pi r^3 \ctild_1(r) & \simeq & \chat_4, \nonumber \\
8\pi r^3 \ctild_2(r) & \simeq & \chat_4+4\chat_5, \nonumber \\
8\pi r^3 \ctild_3(r) & \simeq & -3\chat_4, \nonumber \\
8\pi r^3 \ctild_4(r) & \simeq & 15 \chat_4 \mbox{ and } \nonumber \\
8\pi r^3 \ctild_5(r) & \simeq & -3\chat_4-6\chat_5
\label{eq_ctild_d3}
\end{eqnarray}
where ``$\simeq$" marks that this is the asymptotic limit
for large $r$ and sufficiently large systems.
For $d=2$ one may also use Table~\ref{tab_eta} and additionally
Eq.~(\ref{eq_ten_sum_d2_in_transform}). This leads to
\begin{eqnarray}
4\pi r^2 \ctild_1(r) & \simeq & 5\chat_4 + 8 \chat_5, \nonumber \\
4\pi r^2 \ctild_2(r) & \simeq & -\chat_4, \nonumber \\
4\pi r^2 \ctild_3(r) & \simeq & -6\chat_4 - 8 \chat_5, \nonumber \\
4\pi r^2 \ctild_4(r) & \simeq & 8\chat_4 \mbox{ and} \nonumber \\
\ctild_5(r) & = & 0. 
\label{eq_ctild_d2}
\end{eqnarray}
Note that this result is consistent with Ref.~\cite{spmP5b}.
Importantly, all invariants $\ctild_n(r)$ for $d=3$ decay asymptotically as $1/r^3$
and all for $d=2$ as $1/r^2$.
More generally, the strain CFs of isotropic elastic bodies 
in real space thus decay in any dimension $d$ analytically as $\cabcd(\rvec) \simeq 1/r^d$.

\begin{figure}[t]
\centerline{\resizebox{.9\columnwidth}{!}{\includegraphics*{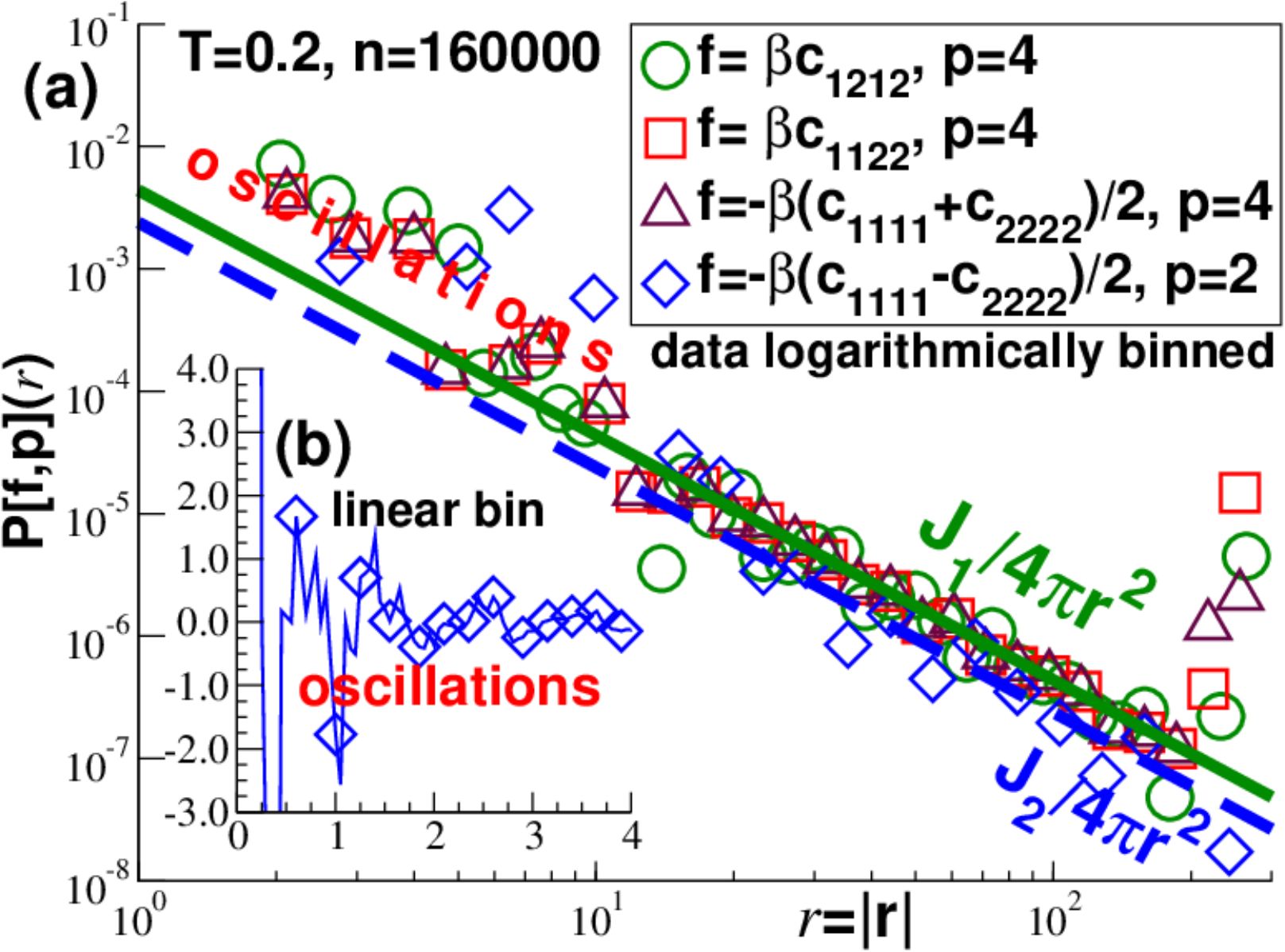}}}
\caption{$r$-dependence of strain CFs $\cabcd(\rvec)$ obtained for pLJ particles in $d=2$ 
for $n=160000$ and $T=0.2$. The expected $\theta$-dependence is projected out using 
Eq.~(\ref{eq_Pfp_def}) for $f(r,\theta)$ and $p$ as indicated in the legend.
{\bf (a)} 
Double-logarithmic representation for logarithmically binned data.
The two power-law slopes indicate the expected asymptotic behavior 
using the constants $J_1 \approx 0.07$ and $J_2 \approx 0.03$, cf.~Eq.~(\ref{eq_J1J2_def}).
{\bf (b)}
Linear representation for $f=-\beta (c_{1111}(\rvec)-c_{2222}(\rvec))/2$ and $p=2$
emphasizing the generic oscillatory behavior for $r \ll 10$.
}
\label{fig_pdLJ_cr}
\end{figure}

The latter relations for $d=2$ are put to the test in Fig.~\ref{fig_pdLJ_cr}
for $n=160000$ beads.
Following Ref.~\cite{spmP5b} we first compute the $\cabcd(\qvec)$ in reciprocal space  
and perform than numerically an inverse FT to obtain $\cabcd(\rvec)$.
The shear-strain CF $c_{1212}(\rvec)$ has already been presented in the last panel of 
Fig.~\ref{fig_intro}. It follows from Eq.~(\ref{eq_ctild_d2}) that
\begin{equation}
\beta c_{1212}(\rvec) \simeq \frac{J_1}{4\pi r^2} \cos(4\theta) \mbox{ for } r \gg 1.
\label{eq_cxyxy_rtheta}
\end{equation}
The same large-$r$ limit holds also for  $\beta c_{1122}(\rvec)$ and for
$-\beta (c_{1111}(\rvec)+c_{2222}(\rvec))/2$.
Moreover,
\begin{equation}
\beta (c_{1111}(\rvec)-c_{2222}(\rvec))/2 \simeq - \frac{J_2}{4\pi r^2} \cos(2\theta)
\label{eq_cdiff_rtheta}
\end{equation}
for $r \gg 1$.
Note that if the coordinate system is turned by an angle $\phi$,
as shown in the first panel of Fig.~\ref{fig_intro},
the above CFs turn with the coordinate system. 
To obtain a precise test of the expected $r$-dependences we project out in 
Fig.~\ref{fig_pdLJ_cr} the angular dependences using 
\begin{equation}
P[f,p](r) \equiv
2 \times \frac{1}{2\pi} \int_0^{2\pi} \ddiff \theta \ f(r,\theta) \cos(p \theta)
\label{eq_Pfp_def}
\end{equation}
for any function $f(r,\theta)$ of the polar coordinates $r$ and $\theta$
using (here) $p=2$ and $p=4$. 
Panel {\bf (a)} of Fig.~\ref{fig_pdLJ_cr} presents logarithmically averaged
data using a double-logarithmic representation.  In agreement with 
Eq.~(\ref{eq_cxyxy_rtheta}) the indicated first three cases collapse for $p=4$ 
and $r \gtrsim 20$ on $J_1/4\pi r^2$ (bold solid line). This confirms the octupolar symmetry 
\cite{foot_wiki_multipolar} of these (rescaled) CFs.
Confirming Eq.~(\ref{eq_cdiff_rtheta}) the last indicated case with
$f(\rvec)=-\beta (c_{1111}(\rvec)-c_{2222}(\rvec))/2$ collapses onto $J_2/4\pi r^2$ (dashed line).
$p=2$ is used here in agreement with the predicted quadrupolar symmetry.
We note that the logarithmic average used in panel {\bf (a)}
suppresses oscillatory behavior for $r \ll 10$ which ultimately
stems from the packing of the discrete particles.
This is emphasized in panel {\bf (b)} using a linear representation
for $f(\rvec)=-\beta (c_{1111}(\rvec)-c_{2222}(\rvec))/2$.
Similar results are obtained for other particle numbers $n$ \cite{spmP5b}.

\section{Test of Gaussianity}
\label{gauss}

\begin{figure}[t]
\centerline{\resizebox{.9\columnwidth}{!}{\includegraphics*{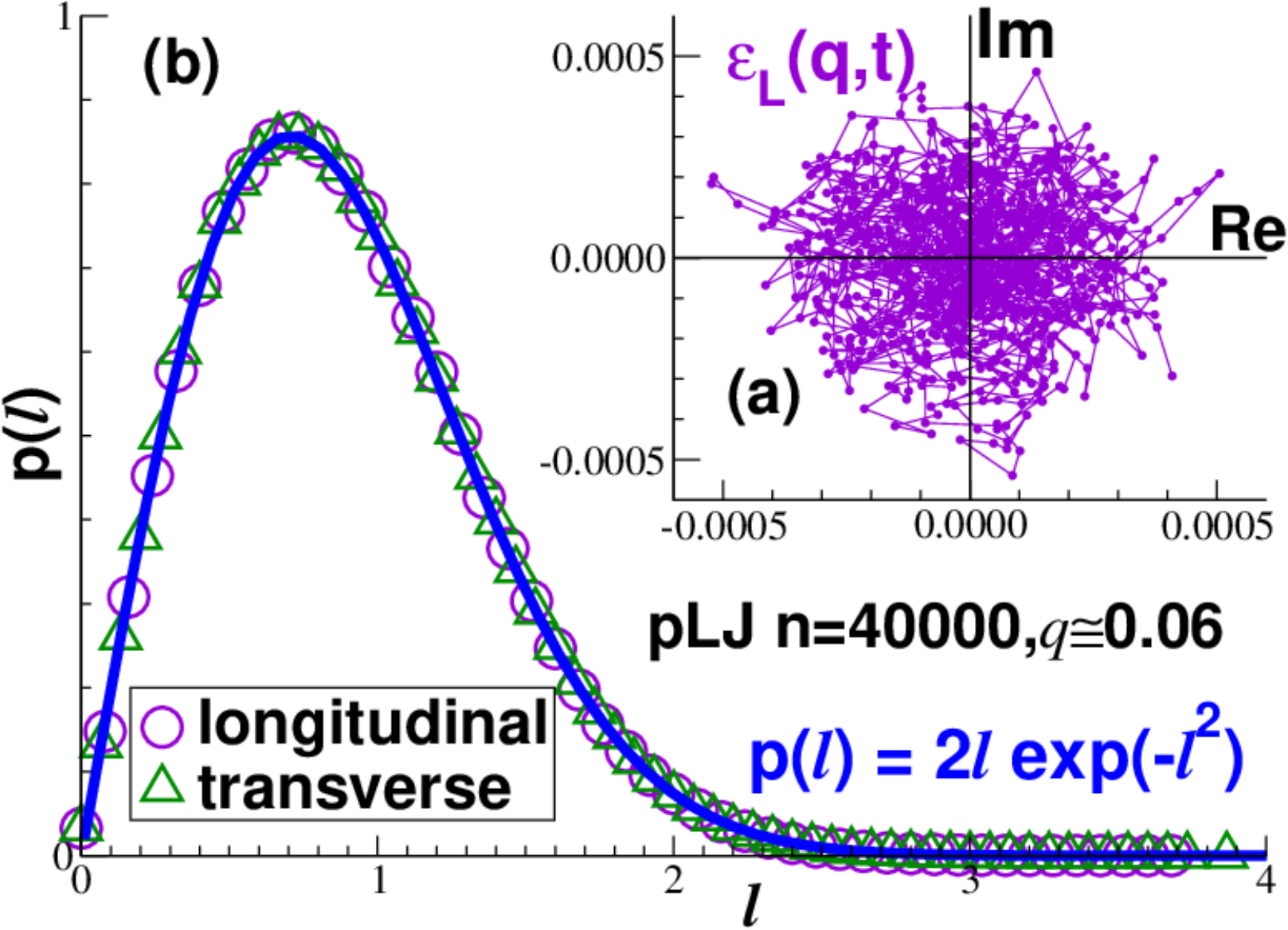}}}
\caption{Distribution of strain components $\epsL(\qvec)$ and $\epsN(\qvec)$ in NRC
for $n=40000$ pLJ particles:
{\bf (a)}
Trajectory of $\epsL(\qvec,t)$ in reciprocal space for one configuration and one wavevector with $q \ll 1$.
{\bf (b)}
Normalized distribution of the real-valued length $l$ defined by Eq.~(\ref{eq_gauss_e_def})
for $\epsL(\qvec)$ and $\epsN(\qvec)$.
The distribution confirms the expected Rayleigh distribution Eq.~(\ref{eq_gauss_pl_Gauss}). 
}
\label{fig_pdLJ_histo}
\end{figure}

We have assumed above 
that the displacements $\uhat_{\alpha}^{\circ}(\qvec)$ are distributed according to
a complex circularly-symmetric Gaussian distribution \cite{foot_wiki_complex_Gaussian}.
(The same applies to the strain components in NRC 
being merely rescaled displacement components.)
We show now that this assumption holds for sufficiently small wavenumbers $q$.
Let us introduce the convenient notation
\begin{eqnarray}
\epsL(\qvec) & \equiv & \hat{\varepsilon}_{11}^{\circ}(\qvec) = i q \uhat_1^{\circ}(\qvec) 
\mbox{ and } \label{eq_gauss_epsL} \\
\epsN(\qvec) & \equiv & \hat{\varepsilon}_{12}^{\circ}(\qvec) = \frac{i q}{2} \uhat_2^{\circ}(\qvec) 
\nonumber
\end{eqnarray}
for the longitudinal and normal (transverse) components in NRC. 
A scatter plot of $\epsL(\qvec)$ in the complex plane for one configuration and 
one wavevector $\qvec$ is shown in panel {\bf (a)} of Fig.~\ref{fig_pdLJ_histo}. 
The data appears to be indeed distributed symmetrically around the origin of the complex plane.
The goal is now to characterize this distribution. $\la \ldots \ra$ stands here for
the combined $c$- and $k$-average $\la \la \ldots \ra_k \ra_c$.

Let us rescale for later convenience $\epsL(\qvec)$ and $\epsN(\qvec)$
by the square root of their typical squared averages
\begin{equation}
\varepsilon(\qvec) \Rightarrow 
e(\qvec) \equiv \frac{\varepsilon(\qvec)}{\la \varepsilon(\qvec) \varepsilon(-\qvec) \ra^{1/2}}
\label{eq_gauss_phiLN}
\end{equation}
for both components ``L" and ``N".
Due to the Wiener-Khinchin theorem, cf. Eq.~(\ref{eq_FT_WKT}), 
both averages are real and positive and, moreover,
equivalent to the ICFs or the elastic moduli $L(q)$ and $G(q)$, cf.~Eq.~(\ref{eq_cab_LqGq}).
Let us write 
\begin{equation}
e(\qvec) = \eRe(\qvec) + i \eIm(\qvec) = l(\qvec) e^{i \phi(\qvec)}
\label{eq_gauss_e_def}
\end{equation}
with $\eRe$ and $\eIm$ being the real and the imaginary parts of $e$,
$l \ge 0$ its real-valued length and $\phi$ its phase angle.
($l$ and $\phi$ are the polar coordinates of $e(\qvec)$.)
Hence,
\begin{equation}
l^2(\qvec) = e(\qvec) e(-\qvec) = (\eRe)^2 + (\eIm)^2.
\label{eq_gauss_l2_def}
\end{equation}
Due to the above rescaling $\langle l^2(\qvec) \rangle = 1$ for both strain components.
Panel {\bf (b)} of Fig.~\ref{fig_pdLJ_histo} shows the normalized distributions $p(l)$ 
of the lengths $l$ of the reduced longitudinal (circles) and normal (squares) displacements.
As emphasized by the bold solid line, a Rayleigh distribution
\cite{foot_wiki_Rayleigh_distribution,Rayleigh1905}
with $p(l) = 2 l \exp(-l^2)$ is observed.
We also note that plotting $p(l)/2l$ as a function of $l^2$ in half-logarithmic coordinates
yields a purely exponential decay (not shown).
The observed distribution can be explained by reworking Maxwell's argument \cite{Maxwell60}
for the velocity distribution of an ideal gas for the (effectively two-dimensional) complex plane.
This assumes that the two components $\eRe$ and $\eIm$ are decorrelated, 
equivalent and isotropically distributed. 
This implies a random phase angle $\phi$ of uniform distribution 
and the factorization 
\begin{equation}
p_2(\eRe,\eIm) \ddiff \eRe \ddiff \eIm = p_1(\eRe)\ddiff \eRe \times p_1(\eIm)\ddiff \eIm
\label{eq_gauss_p2factor}
\end{equation}
of the probability for observing both $\eRe$ and $\eIm$
with $p_1(x)$ being the same distribution for each component.
Moreover, isotropy implies that $p_2(\eRe,\eIm)$ must be a function of the scalar $l^2$.
Following Maxwell this functional equation is solved by the normalized Gaussian distribution
\begin{equation}
p_2(\eRe,\eIm) \ddiff \eRe \ddiff \eIm = \frac{1}{\pi}\exp(-l^2) \ddiff \eRe \ddiff \eIm
\label{eq_gauss_p2}
\end{equation}
with $l^2 = (\eRe)^2+(\eIm)^2$.
Such a probability density of a complex random variable $e(\qvec)$
is called a ``complex circularly-symmetric Gaussian probability density function"
\cite{foot_wiki_complex_Gaussian}.
The probability $p(l)$ for observing a length $l$ is then the Rayleigh distribution
\cite{Rayleigh1905}
\begin{equation}
p(l) = 2\pi l \times p_2(\eRe,\eIm) = 2 l \exp(-l^2)
\label{eq_gauss_pl_Gauss}
\end{equation}
with $l=l(\qvec)$ being either the length of the rescaled longitudinal or normal strain component.

\begin{figure}[t]
\centerline{\resizebox{.9\columnwidth}{!}{\includegraphics*{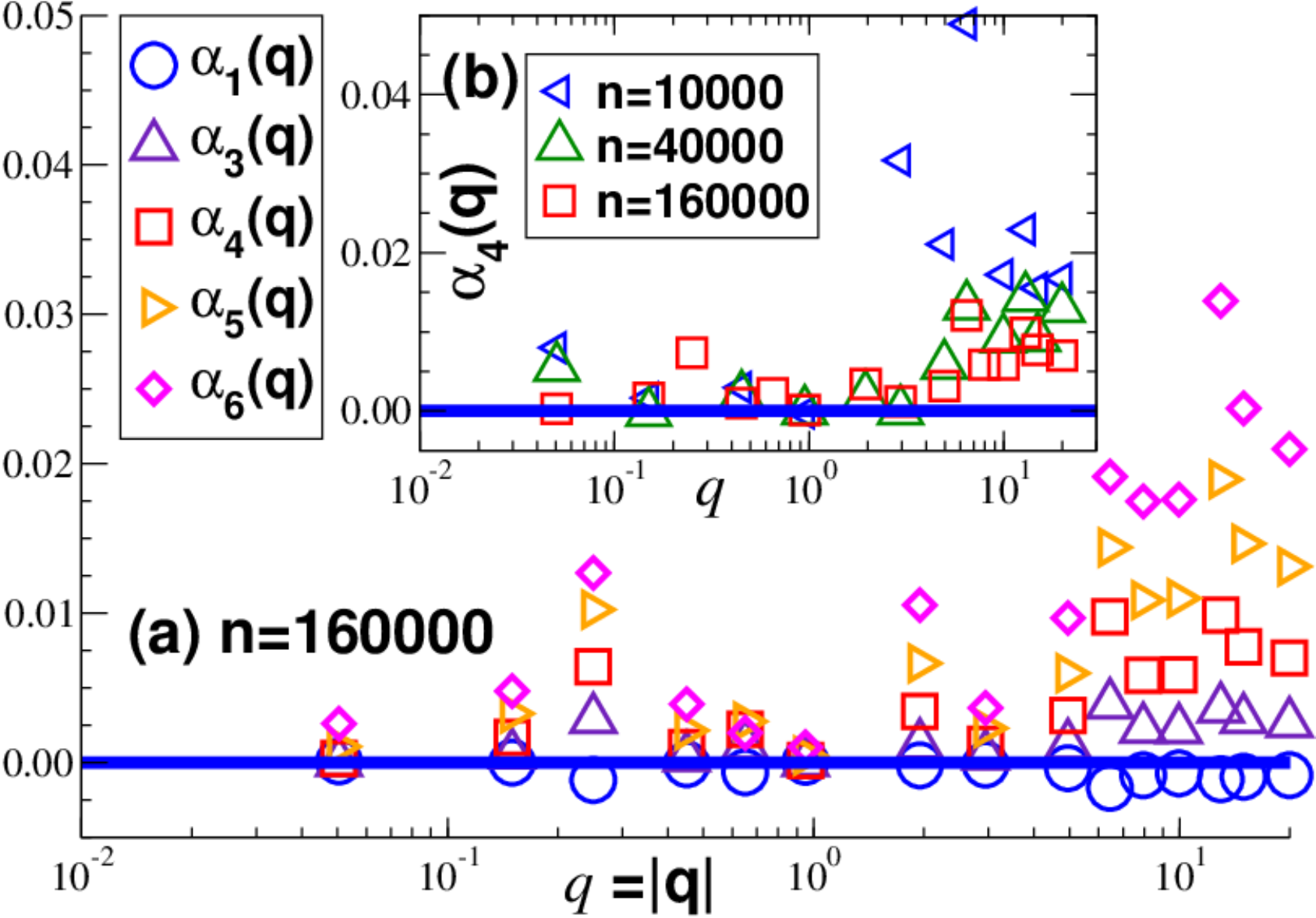}}}
\caption{Non-Gaussianity parameters $\alpha_m(q)$ for the rescaled length $l$ of the 
longitudinal strain components $\epsL(\qvec)$ in NRC obtained for our pLJ model at $T=0.2$:
{\bf (a)}
$\alpha_m(q)$ for $m=1,3,4,5$ and $6$ for $n=160000$.
{\bf (b)}
comparison of $\alpha_4(q)$ for three system sizes.
As seen, $\alpha_m(q)$ is tiny for all $m$, $q$ and $n$
but increases with $m$ and $q$ and decreases with $n$.
}
\label{fig_pdLJ_alpha}
\end{figure}

We have thus demonstrated in panel {\bf (b)} of Fig.~\ref{fig_pdLJ_histo} 
for one small wavevector that the distribution is a circularly-centered complex Gaussian.
Whether this still holds for larger $q$ is best tested
by computing for both strain components the non-Gaussianity parameters
\begin{eqnarray}
\alpha_1(q) & \equiv & \frac{2}{\sqrt{\pi}} \la l^1 \ra -1, \nonumber \\
\alpha_3(q) & \equiv & \frac{4}{3\sqrt{\pi}} \la l^3 \ra -1, \nonumber \\
\alpha_4(q) & \equiv & \frac{1}{2} \la l^4 \ra -1, \nonumber \\
\alpha_5(q) & \equiv & \frac{8}{15\sqrt{\pi}} \la l^5 \ra -1 \mbox{ and } \nonumber \\
\alpha_6(q) & \equiv & \frac{1}{6} \la l^6 \ra -1
\label{eq_gauss_alpha_def} 
\end{eqnarray}
which vanish for perfect Rayleigh distributions. 
(Note that $\alpha_2(q) \equiv 0$ by definition.)
The results obtained for the longitudinal strain components are given in 
Fig.~\ref{fig_pdLJ_alpha}. Panel {\bf (a)} presents for our largest system
with $n=160000$ particles the moments $m=1,3,4,5$ and $6$. 
Not surprisingly, the observed values systematically increase with $m$.
A comparison of $\alpha_4(q)$ for different system sizes is seen in panel {\bf (b)}.
We also note an increase with $q$, especially for the largest $q$.
Importantly, $|\alpha_m(q)| \ll 0.1$ for all measured $m$, $q$ and $n$.
Similar results have been observed for the transverse strain components.

\section{Test of isotropy}
\label{aniso}

We have estimated in Sec.~\ref{cstat_u} $\xicont \approx 10$ for the 
length characterizing the breakdown of the continuum assumption. Interestingly, 
the standard monomer structure factor, which measures the density fluctuations, 
only becomes constant for similar $q$ \cite{spmP5a} as the moduli $L(q)$ and $G(q)$. 
The inhomogeneity length $\xihom$ must thus be similar to $\xicont$.
Following Sec.~\ref{ten_aniso} the isotropy length $\xiiso$ is bounded from below by $\xihom$. 
This argument suggests that $\xicont$, $\xiiso$ and $\xihom$ are 
of similar order for the presented computer model.
We attempt here to characterize $\xiiso$ {\em directly} following Sec.~\ref{ten_aniso}
by characterizing the anisotropies of the lateral and transverse strain ICFs.
Naturally, the order of the averaging procedure is important here since $c$-averaged CFs
(assuming an arbitrarily large ensemble with $\Nc \to \infty$)
must by construction be perfectly isotropic as confirmed below. 
What is meant by $\xiiso$ is the (finally $c$-averaged) length 
characterizing the isotropy of each configuration $c$.
As defined by the second relation of Eq.~(\ref{eq_cabcd_def}),
we focus in this section on the $k$-averaged strain CFs $\cabcd(\qvec,c)$
and their longitudinal and transverse ICFs $\cL(\qvec,c)$ and $\cG(\qvec,c)$ in NRC. 
(The argument $\qvec$ emphasizes that we first do not average over $\qhatvec$.)
As in Sec.~\ref{gauss} the ICFs are normalized by their
 $c$- and $\qhatvec$-averaged means $\cL(q)$ and $\cG(q)$ presented in Fig.~\ref{fig_pdLJ_LqGq}.
These rescaled ICFs are called $\dL(\qvec,c)$ and $\dG(\qvec,c)$.
 
\begin{figure}[t]
\centerline{\resizebox{.9\columnwidth}{!}{\includegraphics*{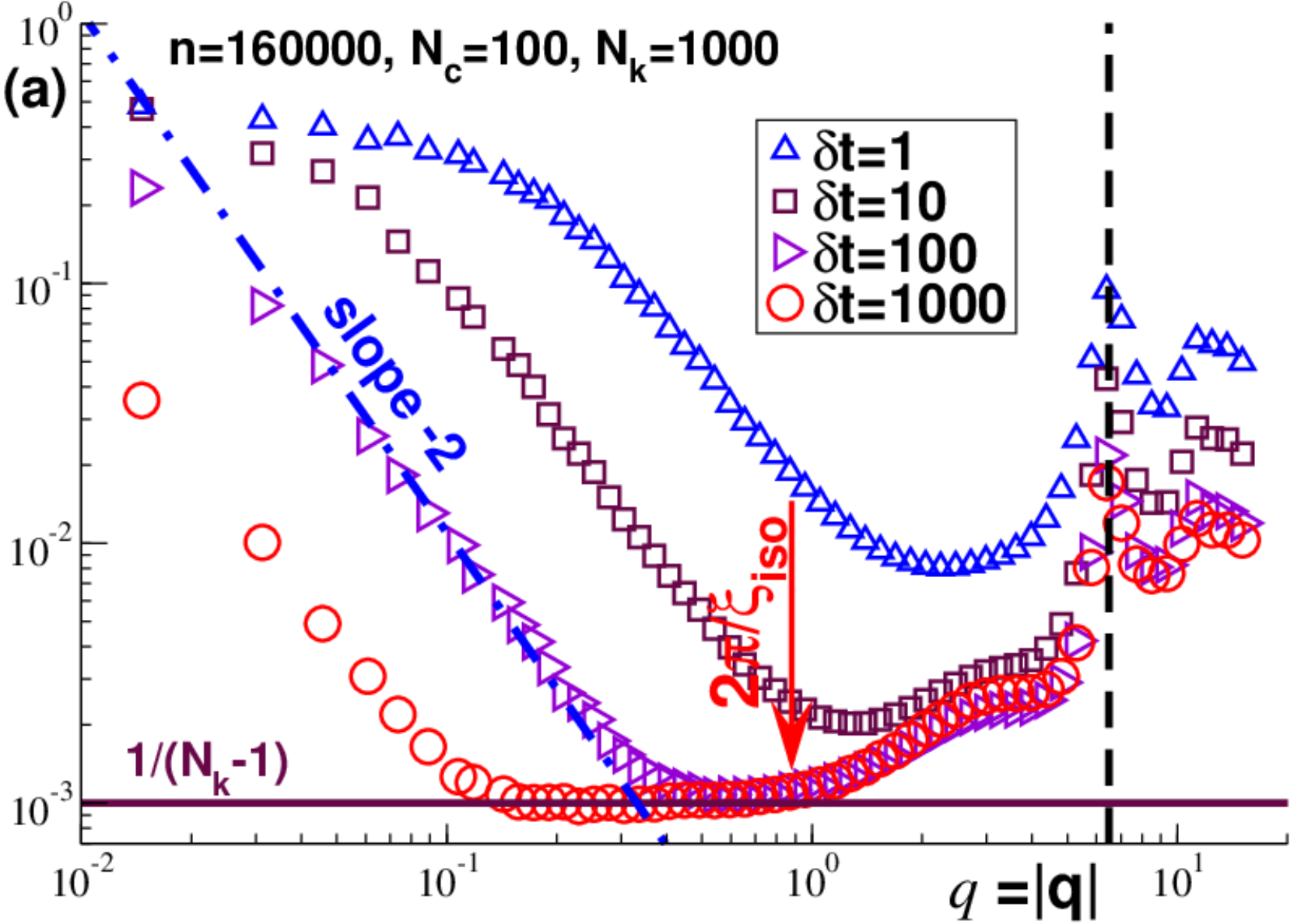}}}
\centerline{\resizebox{.9\columnwidth}{!}{\includegraphics*{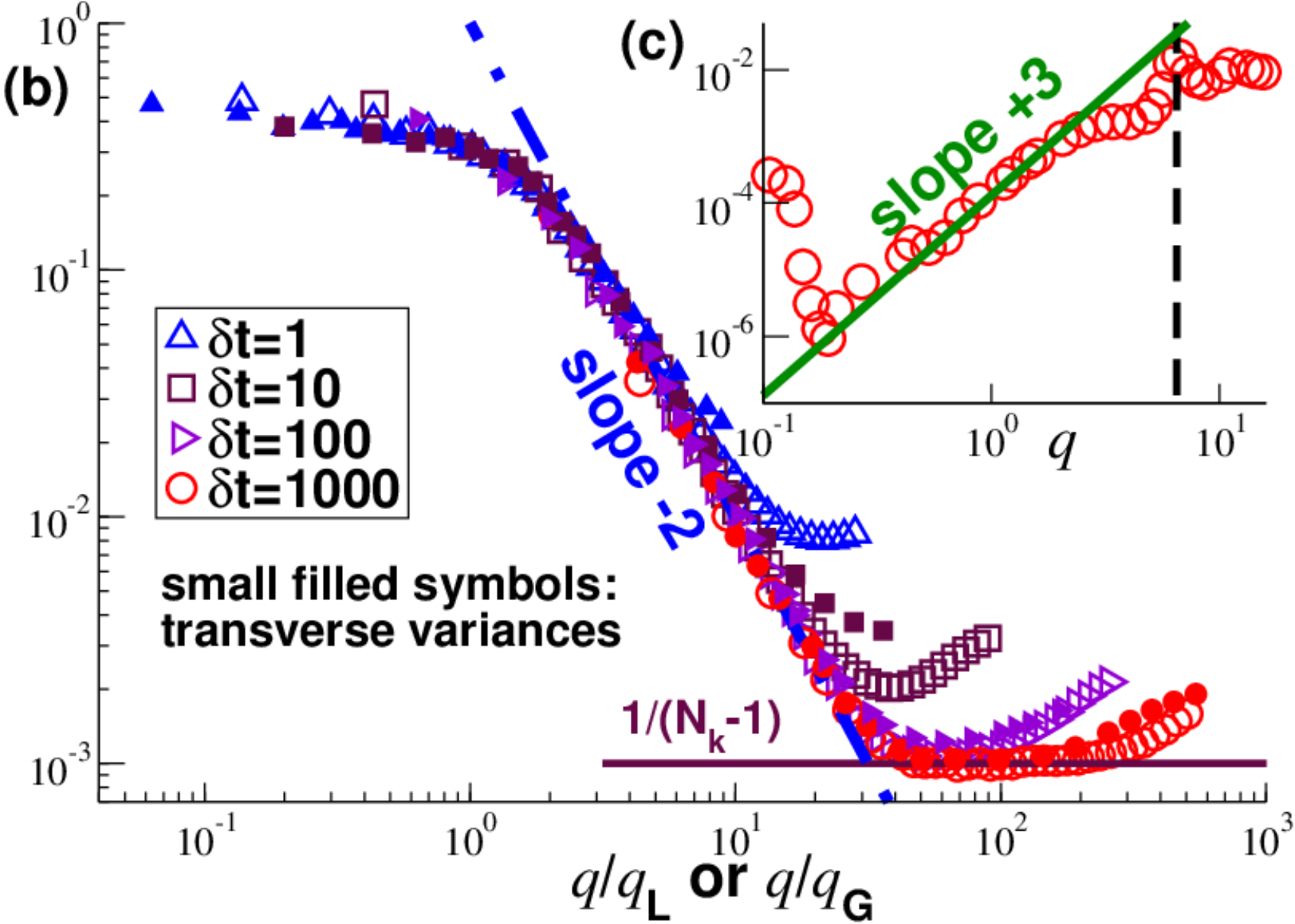}}}
\caption{$c$-variances of rescaled $k$-averaged ICFs 
for pLJ particles with $n=160000$:
{\bf (a)} 
Data for longitudinal ICFs $\dL(\qvec,c)$ for different $\tincr$. 
The power law with exponent $-2$ (dash-dotted line)
is a dynamical effect due to relaxation times $\tauL(q) \propto 1/q^2$,
the intermediate regime (bold horizontal line) is expected for uncorrelated
Gaussian fluctuations.
{\bf (b)}
Data collapse using $q/\qL$ for longitudinal ICFs (open symbols) and $q/\qG$ 
for transverse ICFs (filled symbols).
{\bf (c)} 
Longitudinal variances for $\tincr=1000$ after subtracting $1/(\Nk-1)$.
}
\label{fig_pdLJ_cvari}
\end{figure}

\begin{figure}[t]
\centerline{\resizebox{.9\columnwidth}{!}{\includegraphics*{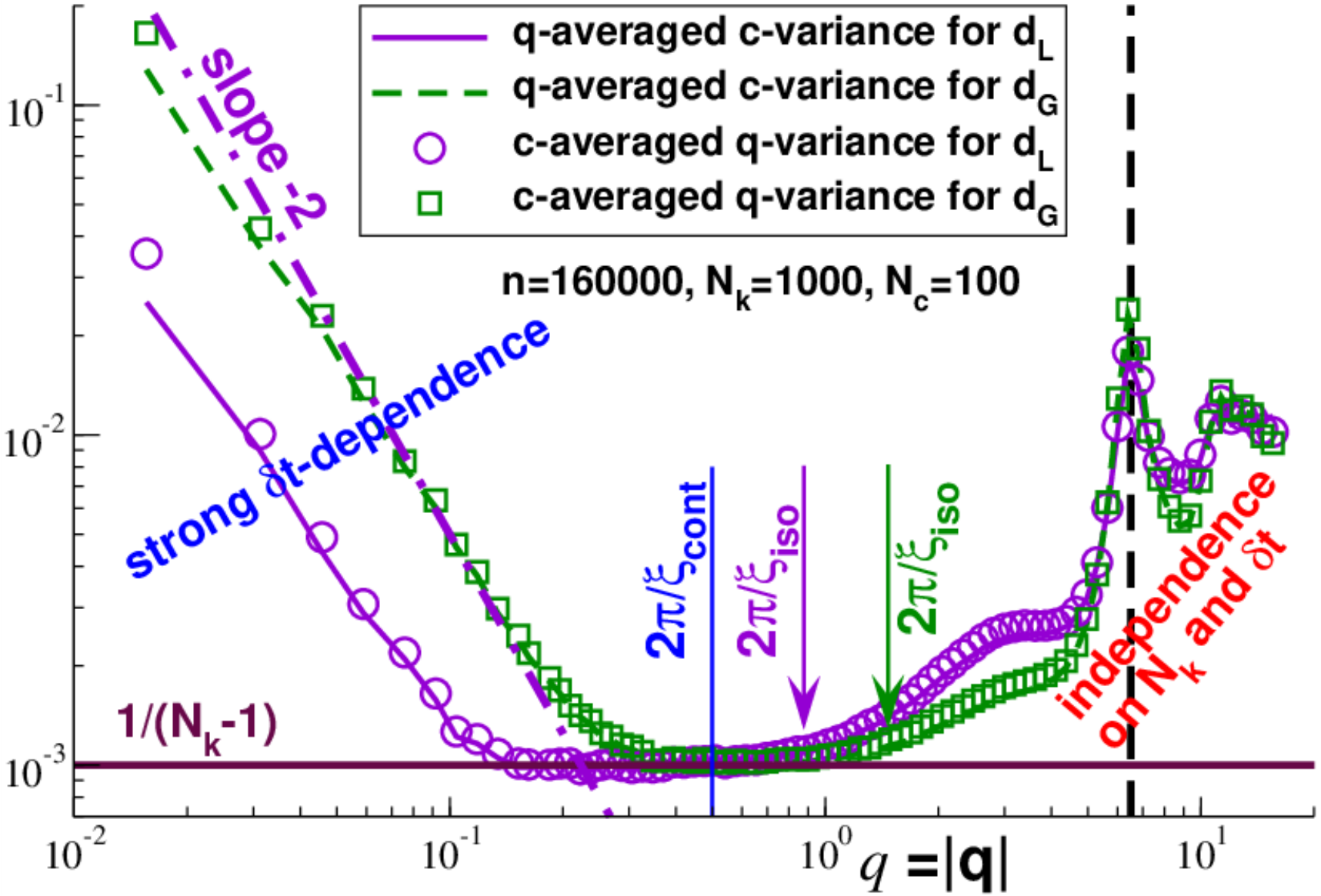}}}
\caption{Comparison of two types of variances for the longitudinal
and transverse ICFs $\dL(\qvec,c)$ and $\dG(\qvec,c)$ for pLJ particles with $n=160000$,
$\Nk=1000$ and $\tincr=1000$. 
The thin solid and dashed lines represent the finally $\qhatvec$-averaged $c$-variances,
the open symbols the fluctuations with $\qvec$ for a given $q$-bin being finally $c$-averaged. 
Similar non-monotonic behavior is obtained for both properties. 
}
\label{fig_pdLJ_aniso_qfirst}
\end{figure}

Let us first compute for $\dL(\qvec,c)$ and $\dG(\qvec,c)$ the $c$-averaged (empirical) variances 
\begin{equation}
\la \ \la d(\qvec,c)^2 \ra_c - \la d(\qvec,c) \ra_c^2 \ \ra_{\qhat}
\label{eq_aniso_cvari_A}
\end{equation}
for each wavevector $\qvec$ taking in a final step the $\qhatvec$-average 
over all $||\qvec||$ in a bin around $q$. 
The data for the longitudinal ICFs are shown in panel {\bf (a)} of Fig.~\ref{fig_pdLJ_cvari} 
for a broad range of time-increments $\tincr$ between always $\Nk=1000$ stored frames.
The thin solid and dashed lines presented in Fig.~\ref{fig_pdLJ_aniso_qfirst}
show the corresponding variances for both the longitudinal and the transverse ICFs
for $\tincr=1000$.
As can be seen from both representations, 
non-monotonic behavior with three different $q$-regimes is observed.
Let us begin with the most simple intermediate regime for $0.1 \ll q \ll 1$
where the variances approach $1/(\Nk-1)$ for large $\tincr$.
We recall from Fig.~\ref{fig_pdLJ_alpha} that the instantaneous strains
are Gaussian variables for $q \ll 1$. The $k$-averaged ICFs $\dL(\qvec,c)$ and 
$\dG(\qvec,c)$ must thus also be Gaussian and this with a variance
given by the number of {\em independent} contributions $k$. 
The horizontal lines 
thus correspond to the empirical standard deviation (without Bessel correction) 
if the relaxation times $\tauL(q)$ and $\tauT(q)$ for the corresponding longitudinal
and transverse relaxation modes are smaller than $\tincr$. 
One expects that the relaxation times strongly increase in the continuum limit 
with increasing wavelength. The contributions $k$ to the ICFs $\dL(\qvec,c)$ and $\dG(\qvec,c)$
thus become increasingly correlated. 
As will be shown in Sec.~\ref{dyn_e_simu}, 
the longitudinal and transverse relaxation processes are characterized
in the continuum limit by the relaxation times 
\begin{equation}
\tauL(q) \simeq \zeta/q^2(\lambda+2\mu) \mbox{ and } \tauT(q) \simeq \zeta/q^2\mu.
\label{eq_tauLT_def}
\end{equation}
($\zeta \approx 750$ will be determined in Sec.~\ref{dyn_e_simu}.)
This explains the strong power law decay with apparent exponent $-2$ indicated
by dot-dashed lines. As suggested by Eq.~(\ref{eq_tauLT_def}),
the small-$q$ data should thus collapse by rescaling the horizontal axis as
$q \to q/\qL$ and $q \to q/\qG$ for, respectively, the longitudinal and transverse
variances with 
\begin{equation}
\qL^2 \equiv \frac{\zeta}{(\lambda+2\mu)\tsamp} \mbox{ and }
\qG^2 \equiv \frac{\zeta}{\mu\tsamp}.
\label{eq_qLT_def}
\end{equation}
The expected scaling is confirmed in panel {\bf (b)} of Fig.~\ref{fig_pdLJ_cvari}.

More importantly, 
deviations from the plateau for uncorrelated frames (horizontal lines) are also seen
for large wavenumbers $q \gg 1$. Since we know already from Fig.~\ref{fig_pdLJ_alpha}
that non-Gaussianity becomes relevant in this limit, this is not unexpected.
This last regime must in fact become more striking if numerical data were available 
with both larger $\Nk$ and $\tincr \gg \tauLT(q)$ making the first two $\Nk$- and 
$\tincr$-dependent regimes decay more rapidly. This can be also seen by simply
subtracting the trivial limit $1/(\Nk-1)$ from the measured variances
as shown in panel {\bf (c)} of Fig.~\ref{fig_pdLJ_cvari}.
This subtraction makes manifest the strong increase of the reduced variances with increasing $q$. 
The solid line indicates an empirical power law. 

That non-Gaussian and anisotropic behavior are related in the large-$q$ limit may be better 
understood from the second type of variances indicated by the open symbols in 
Fig.~\ref{fig_pdLJ_aniso_qfirst}.
We sample here 
\begin{equation}
\la \ \la d(\qvec,c)^2 \ra_{\qhat} - \la d(\qvec,c) \ra_{\qhat}^2 \ \ra_c,
\label{eq_aniso_qvari_A}
\end{equation}
i.e. the (empirical) variances of the rescaled ICFs $\dL(\qvec,c)$ and $\dG(\qvec,c)$ 
are computed for all $\qvec$ in a given $q$-bin in a first step which is followed
by a final $c$-average.
As can be seen, the data are very similar to the $\qhatvec$-averaged $c$-variances discussed above.
Importantly, this shows that the deviations from $1/\Nk$ seen for $q \gg 1$ are due to 
anisotropic correlations. 
Consistently, the data become rapidly $\Nk$- and $\tincr$-independent in this $q$-limit.
As shown by the vertical arrows we may use the observed deviations from the bold horizontal line
to estimate $\xiiso$. The anisotropic effects are apparently slightly
different for longitudinal and transverse correlations, rising faster in the former case.
 
\begin{figure}[t]
\centerline{\resizebox{.9\columnwidth}{!}{\includegraphics*{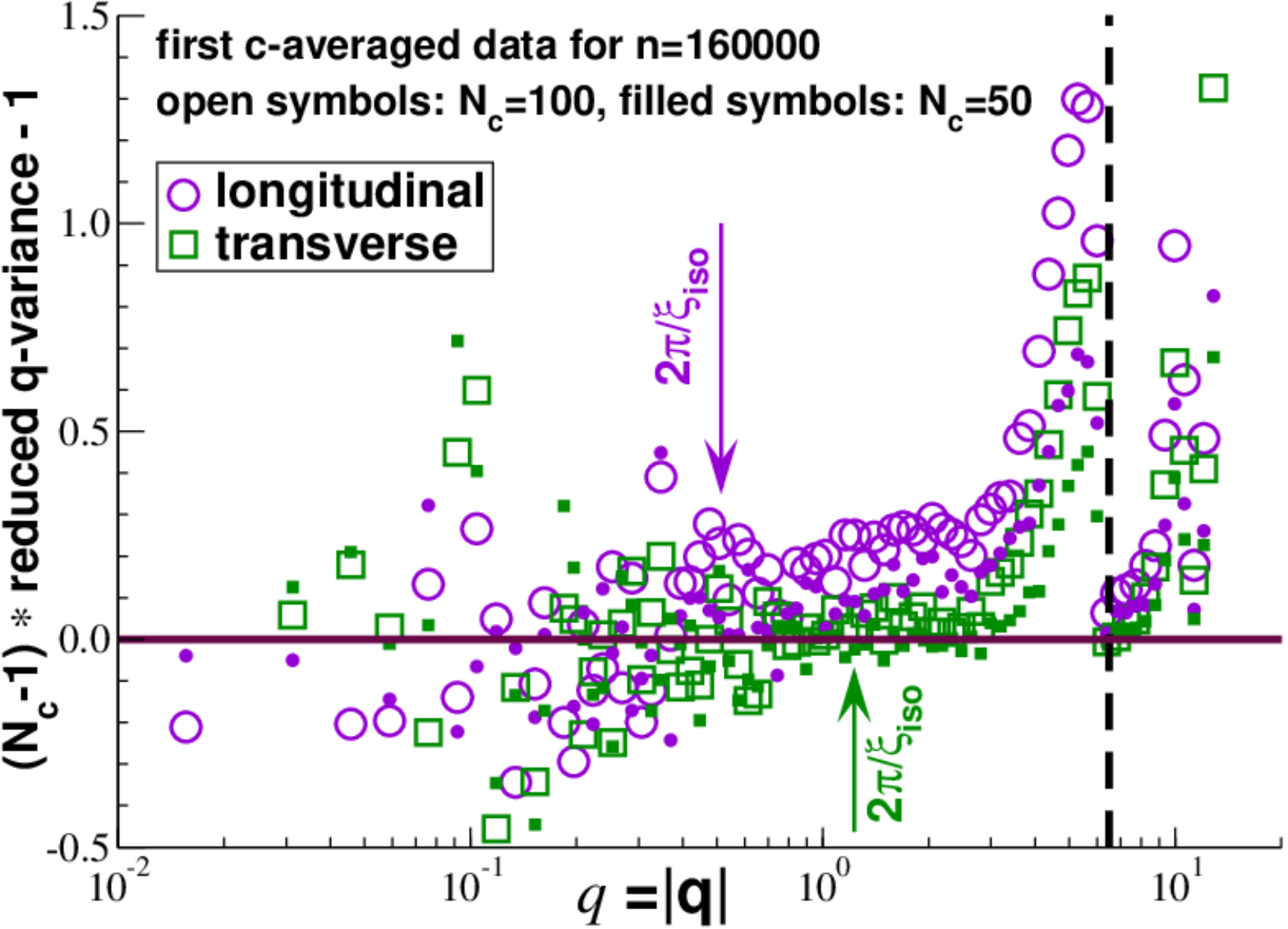}}}
\caption{Reduced dimensionless $q$-variances of $c$-averaged ICFs 
$\dL(\qvec)$ and $\dG(\qvec)$ for pLJ systems. 
Data collapse for $\Nc=100$ (open symbols) and $\Nc=50$ (filled symbols) is observed 
upon multiplying the vertical axis with $\Nc-1$.
Estimations of $\xiiso$ for both ICFs are indicated by arrows.}
\label{fig_pdLJ_aniso_cfirst}
\end{figure}

As already emphasized above, one expects that anisotropic effects become irrelevant for 
the variance
\begin{equation}
\la d(\qvec)^2 \ra_{\qhat} - \la d(\qvec) \ra_{\qhat}^2 \mbox{ with }
d(\qvec) = \la d(\qvec,c) \ra_c
\label{eq_aniso_cfirst_A}
\end{equation}
being the first $c$-averaged ICFs for each wavevector $\qvec$. 
This point is made in Fig.~\ref{fig_pdLJ_aniso_cfirst}.
To get rid of the dynamical effect in the small-$q$ limit
we divide these variances by the $\qhatvec$-averaged $c$-variances
shown by the thin solid and dashed lines in Fig.~\ref{fig_pdLJ_aniso_cfirst}.
Since all $c$ are by construction independent this ratio must decay a $1/(\Nc-1)$.
This is confirmed in Fig.~\ref{fig_pdLJ_aniso_cfirst} for data obtained for two $\Nc$
by rescaling the vertical axis by $\Nc-1$.
Interestingly, the anisotropic behavior for $q \gg 1$ remains visible for the rescaled data
(albeit with more scatter since a very small signal has been amplified).
This can be used to crosscheck the values of $\xiiso$ for the two ICFs (vertical arrows).

\section{Linear response of tensor fields}
\label{linres}

\subsection{Green and growth function fields}
\label{GFs}

TFs may be characterized by measuring within linear response the ``response field" (RF) 
due to a small external ``source field" (SF) perturbing the system.
Let us consider a vector field $\rhohata(\qvec,t)$ in reciprocal space. 
We assume that at $t=0$ a tiny SF $\Sa(\qvec) H(t)$ is 
switched with $H(t)$ denoting the Heaviside function.
To leading order this yields a first-order RF
\begin{eqnarray}
\Ra(\qvec,t) & \equiv & \la \rhohata(\qvec,t) \ra - \la \rhohata(\qvec,t<0) \ra \nonumber \\
& = & \Gab(\qvec,t) \Sb(\qvec) H(t)
\label{eq_GFs_Ra_q}
\end{eqnarray}
with $\Gab(\qvec,t)$ being a second-order TF depending in general also on $t$.
Due to the ``convolution theorem" of FTs, cf.~Eq.~(\ref{eq_FT_convolution}), 
this becomes
\begin{equation}
\Ra(\rvec,t) = \frac{1}{V} \int \ddiff \rvec' \Gab(\rvec-\rvec',t) \Sb(\rvec') H(t)
\label{eq_GFs_Ra_r}
\end{equation}
in real space. We call $\Gab(\qvec,t)$ a ``Green function field" since a localized SF 
$\Sa(\rvec) \propto \delta(\rvec)$ becomes a $\qvec$-independent tensor in reciprocal space. 
As stated above, it is assumed that the system is not driven
by an instantaneous $\delta(t)$-pulse but rather by a perturbation
constant in time. While $\Gab(\qvec,t)$ is thus a Green function with respect
to space it is strictly speaking not a Green function with respect to time
but a ``growth function" \cite{DoiEdwardsBook},
i.e. the time integral of the $\delta(t)$-response.
Hence, ``GF" denotes below ``Green function field" if the spatial aspects matter
and ``growth function field" otherwise.
For an isotropic system the GF must be an ITF, i.e.
\begin{equation}
\Gab(\qvec,t) = k_1(q,t) \delta_{\alpha\beta} + k_2(q,t) \qhat_{\alpha}\qhat_{\beta}
\label{eq_GFs_iso}
\end{equation}
must hold with $k_1(q,t)$ and $k_2(q,t)$ being two time-dependent invariants.
While the GF is an ITF, this does in general not apply for the SF,
even for a perfectly isotropic system. This may happen especially if the
SF for the perturbation is generated not by an external perturbation
but by an intrinsic instantaneous fluctuation of the system, e.g.,
due to a local plastic reorganization of an elastic body such as
the change of connectivity matrix of a polymer network \cite{RubinsteinBook}. 
While on average for isotropic systems such intrinsic fluctuating SFs must also be isotropic,
this does in general not hold for an individual event. 
Hence, although the GF is an ITF, SFs and RFs are in general not.
It is thus important to carefully distinguish between the three types of TFs
\cite{spmP5a,spmP5b}.

\subsection{Fluctuation dissipation theorem for TFs}
\label{fdt}

It is crucial that time-dependent CFs (as defined in Appendix~\ref{corr} 
with all external perturbations switched off) and GFs may be linearly related 
according to the ``Fluctuation Dissipation Theorem" (FDT) 
\cite{DoiEdwardsBook,ForsterBook,ChaikinBook}.
This implies that the perturbed TF and the SF must be thermodynamically conjugate,
i.e. their inner product yields a contribution to the (scalar) Hamiltonian.
Details depend now on whether the fluctuating TF is thermodynamically an {\em extensive} 
field and the perturbation TF an {\em intensive} field or the opposite. 
Only the former case is relevant here for which 
\cite{DoiEdwardsBook,spmP5b}
\begin{equation}
\Gab(\qvec,t) = \beta V \left[\cab(\qvec,t=0)-\cab(\qvec,t)\right] H(t) 
\label{eq_fdt_ex_A}
\end{equation}
holds with $\cab(\qvec,t)$ being defined by Eq.~(\ref{eq_corr_cab_def}). 
The RF $\Rab(\qvec,t)$ thus reveals a continuous growth behavior 
starting continuously at $\Rab(\qvec,0)=0$ \cite{DoiEdwardsBook}. 
Let us assume that we can {\em by construction} impose $\la \rhohata(\qvec,t) \ra =0$ 
as it is the case for the displacement TFs, cf.~Sec.~\ref{instTFs}. 
Using Eq.~(\ref{eq_corr_cab_tlarge}) we thus get
\begin{equation}
\Gab(\qvec,t) \to \beta V \cab(\qvec,t=0) \mbox{ for } t \to \infty.
\label{eq_fdt_ex_B}
\end{equation}
Dropping the time argument, 
we may write the FDT relation for the static limit concisely as 
\begin{equation}
\Gab(\qvec) = \beta V \cab(\qvec) = \beta V \la \rhohata(\qvec)\rhohatb(-\qvec) \ra.
\label{eq_fdt_ex_C}
\end{equation}
We will use Eq.~(\ref{eq_fdt_ex_A}) to establish below in Sec.~\ref{dyn} 
the general relations between the time-dependent CFs of displacement and strain fields 
with the longitudinal and transverse material functions $L(q,t)$ and $G(q,t)$.
Interestingly, Eq.~(\ref{eq_fdt_ex_C}) can be used to compute the static 
response of displacement fields under externally applied force fields
from the CFs computed in Sec.~\ref{cstat}.

\subsection{Linear displacement response}
\label{resp}

The displacements generated by a small force density $\gext_{\alpha}(\qvec)$ applied 
to a linear elastic body are given by 
\begin{equation}
u_{\alpha}(\qvec) = \Gab(\qvec) \ \gext_{\beta}(\qvec)
\label{eq_resp_intro_A}
\end{equation}
in reciprocal space in terms of the symmetric second-order GF TF $\Gab(\qvec)$.
For isotropic systems $\Gab(\qvec)$ must be an ITF.
Taking advantage of the FDT relation $\Gab(\qvec) = \beta V \cab(\qvec)$ 
for the static displacement CF $\cab(\qvec)$, the GFs are given by the CFs 
\cite{foot_resp_strain}.

As a first example let us consider an external perturbation 
given by localized and well separated point forces, i.e. by a force density 
$\gext_{\alpha}(\qvec) = \sum_a f_{\beta}^a/V$ in reciprocal space.
Using Eq.~(\ref{eq_fdt_ex_C}) the response in real space is thus
\begin{equation}
u_{\alpha}(\rvec) = \sum_a \beta \cab(\rvec-\rvec^a) f_{\alpha}^a.
\label{eq_resp_simpl}
\end{equation}
Importantly, since $\cab(\rvec) \sim 1/\beta$ the GF $\Gab(\rvec)$
does not depend explicitly on the temperature of the system.
Note also that $u_{\alpha}(\rvec)$ cannot be an ITF
since the applied force in reciprocal space is a finite first-order tensor,
i.e. according to Eq.~(\ref{eq_ten_ten_o1}) {\em not} an isotropic tensor.
One specific case may be worth noting for comparison.
Let us consider following W. Thomson (1848) one 
point force $\gext_{\alpha}(\rvec) = f_{\alpha} \delta(\rvec)$ at the origin.
Using Eq.~(\ref{eq_resp_simpl}) 
and Eq.~(\ref{eq_cab_d3}) one directly obtains the well-known linear response 
\cite{LandauElasticity}
\begin{equation}
u_{\alpha}(\rvec) = \frac{1}{8\pi r} \frac{1+\nu}{E (1-\nu)}
\left[ (3-4\nu)\delta_{\alpha\beta} + \rhat_{\alpha} \rhat_{\beta}
\right]
f_{\beta} 
\label{eq_resp_Thomson}
\end{equation}
with $E$ and $\nu$ being, respectively, the standard Young modulus and Poisson ratio 
for $d=3$ \cite{LandauElasticity}.

Let us next consider the force field
\begin{equation}
\gext_{\alpha}(\rvec) = f \delta_{\alpha \beta} 
\left[ \delta(\rvec- h  \evec_{\beta}) - \delta(\rvec+ h \evec_\beta) \right],
\label{eq_iso_source_r}
\end{equation}
being the sum of $d$ dipoles created by pairs of close point forces of same magnitude 
but opposite sign. Using Eq.~(\ref{eq_FT_G}) for sufficiently small $p=2h$ 
the source force density in reciprocal space is the ITF
$\gext_{\alpha}(\qvec) = -i \ p q_{\alpha} \ r/V$.
Since both $\cab(\qvec)$ and the source are ITFs, the same applies 
due to the product theorem for the RF. Hence,
$u_{\alpha}(\qvec) = l_1(q) \qhat_{\alpha}$.
Using the results found in Sec.~\ref{cstat_u} it is seen that $l_1(q) = -i p f / V L(q) q$ 
only depends on the longitudinal modulus $L(q)$ and not on the shear modulus $G(q)$
as one expects due to the imposed isotropic pressure.
Focusing on the continuum limit where $L(q) \approx \lambda + 2\mu$
and using Table~\ref{tab_eta} for $\eta=1$ we finally get the invariant in real space
\begin{eqnarray}
\ltild_1(r) & = & 
\frac{p f}{\lambda+2\mu} \frac{1}{2\pi r} \mbox{ for } d=2 \mbox{ and } \nonumber \\
\ltild_1(r) & = & 
\frac{p f}{\lambda+2\mu} \frac{1}{4\pi r^2} \mbox{ for } d=3.
\label{eq_iso_ltild}
\end{eqnarray}

\subsection{Boltzmann superposition relations}
\label{boltz}

Boltzmann superposition relations are generally formulated in terms of
second-order stress and strain fields \cite{ChaikinBook,RubinsteinBook,lyuda18}.
Using the convolution relations Eq.~(\ref{eq_FT_convolution}) and Eq.~(\ref{eq_LT_convolution})
these relations are best stated in Fourier-Laplace space, e.g.,
the stress increment $\sigab(\qvec,t)$ caused by a strain
$\epsab(\qvec,t)$ may be compactly written as \cite{lyuda18}
\begin{equation}
\sigab(\qvec,s) = \Eabcd(\qvec,s) \epscd(\qvec,s)
\label{eq_boltz_sEe}
\end{equation}
with $\Eabcd(\qvec,s)$ being the generalized elasticity TF characterizing
the viscoelastic material properties.
It is assumed here that the stress and strain increments both vanish
in the time domain for $t < 0$.
For $s=0$ and $\qvec=\bfzero$ Eq.~(\ref{eq_boltz_sEe}) 
reduces to the macroscopic static Hooke law with $\Eabcd(\bfzero,0)$ being
the macroscopic elasticity tensor $\Eabcd$, cf.~Eq.~(\ref{eq_moduli_Eabcd_iso}). 
For $\qvec=\bfzero$ but finite $s$ Eq.~(\ref{eq_boltz_sEe}) describes the 
macroscopic viscoelastic stress response \cite{RubinsteinBook}
while for $s=0$ and finite $\qvec$ the microscopic static (long-time) behavior
\cite{LandauElasticity}.
For isotropic systems
\begin{equation}
L(q,s) \equiv E_{1111}^{\circ}(\qvec,s) \mbox{ and } 
G(q,s) \equiv E_{1212}^{\circ}(\qvec,s) 
\label{eq_LqsGqs_def}
\end{equation}
denote the only two invariants in NRC of $\Eabcd(\qvec,s)$ relevant in the present work. 
We need in Sec.~\ref{dyn_u} the corresponding relation for the displacement fields 
caused by a force density. We demonstrate here that
\begin{equation}
u_{\alpha}(\qvec,s) = q^{-2} \Kab(\qvec,s) g_{\beta}(\qvec,s)
\label{eq_boltz_uKg}
\end{equation}
with $\Kab(\qvec,s) = \Lcal[\Kab(\qvec,t)]$ being the $s$-dependent generalization of the
static creep compliance $\Kab(\qvec,s=0)$ introduced above in Sec.~\ref{cstat_u}.
Using that the second-order ICFs $\Eab(\qvec,s)$ and $\Kab(\qvec,s)$
are defined to be inverse with respect to each other, i.e.
\begin{equation}
E_{\alpha\gamma}(\qvec,s) K_{\gamma\beta}(\qvec,s)=\delta_{\alpha\beta},
\label{eq_EabqsKabqs_def}
\end{equation}
we get by tracing Eq.~(\ref{eq_boltz_uKg}) with $\Eab(\qvec,s)$
the corresponding Boltzmann superposition relation
\begin{equation}
g_{\alpha}(\qvec,s) = q^2 \Eab(\qvec,s) u_{\beta}(\qvec,s)
\label{eq_boltz_gEu}
\end{equation}
expressing the force TF by the displacement TF.
Both relations are equivalent as may be seen by tracing the latter relation with $\Kab(\qvec,s)$.
Interestingly, Eq.~(\ref{eq_boltz_gEu}) can be directly obtained from the more familiar
stress-strain relation Eq.~(\ref{eq_boltz_sEe}) if we define $\Eab(\qvec,s)$ as the
contraction
\begin{equation}
\Eab(\qvec,s) \equiv \OPdown[\Eabcd(\qvec,s)].
\label{eq_Eab_qs}
\end{equation}
Let us also remind that
the displacement and strain fields are related by Eq.~(\ref{eq_epsab_def}) and
that the force density field is given by the contraction
\begin{equation}
g_{\alpha}(\qvec,s) \equiv - i q_{\beta} \sigab(\qvec,s)
\label{eq_boltz_s2g}
\end{equation}
of the stress field. Applying the linear operator $\OPdown$ to Eq.~(\ref{eq_boltz_sEe})
then yields Eq.~(\ref{eq_boltz_gEu}) and thus in turn Eq.~(\ref{eq_boltz_uKg}).
Using finally Eq.~(\ref{eq_ten_sum_OPdown_invariants}) and Eq.~(\ref{eq_ten_sum_inv_invariants}) 
the invariants of $\Eab(\qvec,s)$ and $\Kab(\qvec,s)$ are thus given in NRC by
\begin{eqnarray}
\kEL(q,s) = 1/\kKL(q,s) & = & L(q,s)  \nonumber \mbox{ and } \\
\kEN(q,s) = 1/\kKN(q,s) & = & G(q,s) \label{eq_kKLkKN_qs}
\end{eqnarray}
with the superscript indicating the respective ITF.

\section{Time-dependent correlations}
\label{dyn}

\subsection{Introduction}
\label{dyn_intro}

We present finally applications of the mathematical formalism for ITFs
for time-dependent GFs and CFs of the displacement and strain fields.
Taking advantage of the Boltzmann linear superposition
relation for the first-order displacement and force density TFs
formulated in Sec.~\ref{boltz}, we show in Sec.~\ref{dyn_u} 
how the GFs and CFs of displacement fields are 
related to the two longitudinal and transverse material functions $L(q,t)$ and $G(q,t)$.
The corresponding relations for strain GFs and CFs are investigated
in the subsequent subsections.

\subsection{Invariants of CFs of displacement fields}
\label{dyn_u}

Let us consider an isotropic elastic body 
at equilibrium for $t<0$ perturbed by an external force density 
$\gaimp(\qvec,t)=\gaimp(\qvec)H(t)$ being switched on at $t=0$
and kept constant for all $t\ge 0$.
Within linear response this will generate a time-dependent displacement field
\begin{equation}
u_{\alpha}(\qvec,t) = \Gab(\qvec,t) \gbimp(\qvec) H(t).
\label{eq_GF_displ}
\end{equation}
$\Gab(\qvec,t)$ must be a second-order ITF for $q\xiiso \ll 1$
characterized by two invariants $\kL(q,t)$ and $\kN(q,t)$ in NRC.
We show here that these invariants are given 
by the two material functions $L(q,s)$ and $G(q,s)$ defined by Eq.~(\ref{eq_LqsGqs_def}).
 
Let us first emphasize that there is a profound difference between applied
macroscopic ($\qvec=\bfzero$) and microscopic (finite $\qvec)$ forces.
While in the former case the total inner force $g_{\alpha}(t)=\gaimp(t)$ is imposed
and under direct experimental control, this is different for the inner force density 
$g_{\alpha}(\qvec,t)$ at finite $\qvec$ due to the internal degrees of freedom of the system.
These allow the system to respond by means of generated forces.
For an overdamped fluid the external force density is simply
diminished by the frictional force generated by the internal motion, 
i.e. the effective inner force density driving the system is 
\begin{eqnarray}
g_{\alpha}(\qvec,t) & = & \gaimp(\qvec) + \gagen(\qvec,t) \mbox{ with }\label{eq_dyn_u_A}\\
\gagen(\qvec,t) & = & - \zeta v_{\alpha}(\qvec,t) \nonumber  \label{eq_dyn_u_B}
\end{eqnarray}
and $v_{\alpha}(\qvec,t)$ being the velocity field in reciprocal space.
The friction is caused by the motion generated at $t > 0$ \cite{foot_phy_friction}.
Using that $v_{\alpha}(q,s) = s u_{\alpha}(\qvec,s)$ we may rewrite the
generated force density in Fourier-Laplace space as
\begin{equation}
\gagen(\qvec,s)=  - \frac{q^2}{w(q,s)} u_{\alpha}(\qvec,s) \mbox{ with } 
w(q,s) \equiv \frac{q^2}{\zeta s} \label{eq_wqs_def}
\end{equation}
standing for a convenient scalar.
As a next step we insert now Eq.~(\ref{eq_dyn_u_A}) and Eq.~(\ref{eq_wqs_def}) 
into the Boltzmann relation Eq.~(\ref{eq_boltz_uKg}). This leads to the recursion relation 
\begin{eqnarray}
u_{\alpha}(\qvec,s) & = & q^{-2} \Kab(\qvec,s) \gbimp(\qvec) \nonumber \\
& - & w(q,s)^{-1} \Kab(\qvec,s) u_{\beta}(\qvec,s)
\label{eq_dyn_u_E}
\end{eqnarray}
for the displacement field.
To obtain the invariants $\kL(q,s)$ and $\kN(q,s)$ one compares the invariants of 
Eq.~(\ref{eq_GF_displ}) and Eq.~(\ref{eq_dyn_u_E}) in NRC. Using Eq.~(\ref{eq_kKLkKN_qs}) we get
\begin{eqnarray}
q^2 \kL(q,s) & = & \frac{w(q,s)}{1+w(q,s) L(q,s)} \mbox{ and } \nonumber \\ 
q^2 \kN(q,s) & = & \frac{w(q,s)}{1+w(q,s) G(q,s)}. \label{eq_dyn_u_kLN} 
\end{eqnarray}
We use then in a final step Eq.~(\ref{eq_fdt_ex_A}) to relate the GFs via
\begin{equation}
\Gab(\qvec,t) = \beta V \left[\cab(\qvec,t=0)-\cab(\qvec,t) \right]
\label{eq_dyn_u_F}
\end{equation}
to the time-dependent displacement CFs $\cab(\qvec,t)$ defined by Eq.~(\ref{eq_corr_cab_def}).
Using Eq.~(\ref{eq_dyn_u_kLN}) one readily obtains the ICFs in Fourier-Laplace for $\cab(\qvec,t)$.
We shall state this below for the essentially equivalent strain CFs.

\subsection{Corresponding invariants for strain fields}
\label{dyn_e}

The linear strain increment $\epsab(\qvec,t)$ in reciprocal space caused 
by a weak constant external stress $\sigabext(\qvec)$ applied at $t=0$ is given by
\begin{eqnarray}
\epsab(\qvec,t) & = & \Gabcd(\qvec,t) \sigcdext(\qvec) H(t) \mbox{ with }
\label{eq_GF_strain}\\
\Gabcd(\qvec,t) & = & q^2 \OPup[\Gab(\qvec,t)]
\label{eq_GF_displ2strain}
\end{eqnarray}
being the outer product of the displacement GFs $\Gab(\qvec,t)$ discussed 
in the preceding paragraph. 
To show this we have used Eq.~(\ref{eq_GF_displ}) and Eq.~(\ref{eq_epsab_def})
and that the imposed external force density is a contraction  
\begin{equation}
\gext_{\alpha}(\qvec,s) \equiv - i q_{\beta} \sigabext(\qvec,s)
\label{eq_sigabext_def}
\end{equation}
of the externally imposed stress $\sigabext(\qvec)$.
Using Eq.~(\ref{eq_ten_sum_OPup_invariants}) the only two finite invariants 
of $\Gabcd(\qvec,s)$ are thus
\begin{eqnarray}
\gL(q,s) & = & q^2 \kL(q,s) = \frac{w(q,s)}{1+w(q,s) L(q,s)} \mbox{ and } \nonumber \\ 
4\gG(q,s) & = & q^2 \kN(q,s) = \frac{w(q,s)}{1+w(q,s) G(q,s)} \label{eq_dyn_e_gLG}
\end{eqnarray}
with $\kL(q,s)$ and $\kN(q,s)$ being given by Eq.~(\ref{eq_dyn_u_kLN}).
While for large $s$ (small $t$) both $\gL(q,s)$ and $\gG(q,s)$ vanish,
i.e. the GFs of the strain TF are continuous at $t=0$,
they naturally approach for small $s$ (large $t$) 
\begin{equation}
\gL(q,s) \to \frac{1}{L(q)} \mbox{ and }
4\gL(q,s) \to \frac{1}{G(q)} 
\label{eq_dyn_e_GFs_A}
\end{equation}
with $L(q,s) \to L(q)$ and $G(q,s) \to G(q)$ for $s \to 0$.

According to the FDT relation Eq.~(\ref{eq_fdt_ex_A}) the strain GFs derived above 
are related to the time-dependent CFs of instantaneous strain TFs by
\begin{equation}
\Gabcd(\qvec,t) = \beta V \left[\cabcd(\qvec,t=0)-\cabcd(\qvec,t) \right].
\label{eq_dyn_e_CFs_A}
\end{equation}
This leads in Fourier-Laplace space to 
\begin{equation}
\beta V\cabcd(\qvec,s) = \beta V \cabcd(\qvec,t=0)-\Gabcd(\qvec,s).
\label{eq_dyn_e_CFs_B}
\end{equation}
Consistently with Eq.~(\ref{eq_cabcd_invariants}) and Eq.~(\ref{eq_dyn_e_gLG})
the invariants $\cL(q,s)$ and $\cG(q,s)$ of the fourth-order CF TF thus are 
\begin{eqnarray}
\beta V \cL(q,s)  & = & \frac{1}{L(q)}- \frac{w(q,s)}{1+w(q,s) L(q,s)} 
\mbox{ and } \nonumber \\ 
\beta V 4\cG(q,s) & = & \frac{1}{G(q)} - \frac{w(q,s)}{1+w(q,s) G(q,s)} \label{eq_dyn_e_cLG}
\end{eqnarray}
which is the most important result of this work.
The special limit for $s \to \infty$ ($t \to 0$) has already been considered 
in Sec.~\ref{cstat_u}. 
In the opposite small-$s$ (large-$t$) limit $w(q,s)$ becomes very large and thus cancels out 
and both ICFs thus vanish as expected.

\subsection{Large-time limit}
\label{dyn_e_tlarge}

The leading corrections for the small-$s$ (large-$t$) limit can be readily obtained 
supposing that $L(q,s) \simeq L(q)$ and $G(q,s) \simeq G(q)$,
i.e. the time-dependence of the material function is {\em assumed} to be negligible. 
It is convenient to introduce by 
\begin{equation}
\tauL(q) = \zeta/q^2L(q) \mbox{ and } \tauT(q) = \zeta/q^2G(q)
\label{eq_tauLT_defB}
\end{equation}
two time scales characterizing the overdamped strain relaxation. 
(This definition reduces in the continuum limit to Eq.~(\ref{eq_tauLT_def}).)
Using Eq.~(\ref{eq_dyn_e_cLG}) this implies
\begin{eqnarray}
\beta V \cL(q,s)  & \simeq & \frac{1}{L(q)} \frac{s}{s+1/\tauL(q)} \mbox{ and } \nonumber \\
	\beta V 4\cG(q,s)  & \simeq & \frac{1}{G(q)} \frac{s}{s+1/\tauT(q)}
\label{eq_dyn_e_tlarge_A}
\end{eqnarray}
for $s \to 0$. Using Eq.~(\ref{eq_LT_exponential}) an exponential decay 
\begin{eqnarray}
L(q) \beta V \cL(q,t)  & \simeq & \exp(-t/\tauL(q)) \mbox{ and } \nonumber \\
G(q) \beta V 4\cG(q,t)  & \simeq & \exp(-t/\tauT(q)) 
\label{eq_dyn_e_tlarge_B}
\end{eqnarray}
is obtained for large $t$ as expected for overdamped motion.
Instead of the standard time-dependent CFs we shall investigate below the CFs of $t$-averaged 
strain TFs, i.e. we switch from a Green-Kubo representation to an Einstein 
representation as described in Appendix~\ref{corr}.
Using Eq.~(\ref{eq_taver_Debye}) the above results can be reformulated as
\begin{eqnarray}
L(q) \beta V \cLbar(q,\tsamp)  & \simeq & D(\tsamp/\tauL(q)) \mbox{ and } \nonumber \\
G(q) \beta V 4\cGbar(q,\tsamp)  & \simeq & D(\tsamp/\tauT(q)) 
\label{eq_dyn_e_tlarge_C}
\end{eqnarray}
with $\cLbar(q,\tsamp)$ and $\cGbar(q,\tsamp)$ being the ICFs of the CFs of the $t$-averaged strain,
cf.~Eq.~(\ref{eq_taver_cbarabcd_def}),
and $D(x)$ denoting Debye's function, cf.~Eq.~(\ref{eq_taver_Debye}).
We have assumed above that the material functions can be assumed to be time-independent.
This approximation is sufficient for the continuum limit considered below.
More care is needed in general as shown in Appendix~\ref{tauterm}.

\subsection{Simulation results}
\label{dyn_e_simu}

\begin{figure}[t]
\centerline{\resizebox{.9\columnwidth}{!}{\includegraphics*{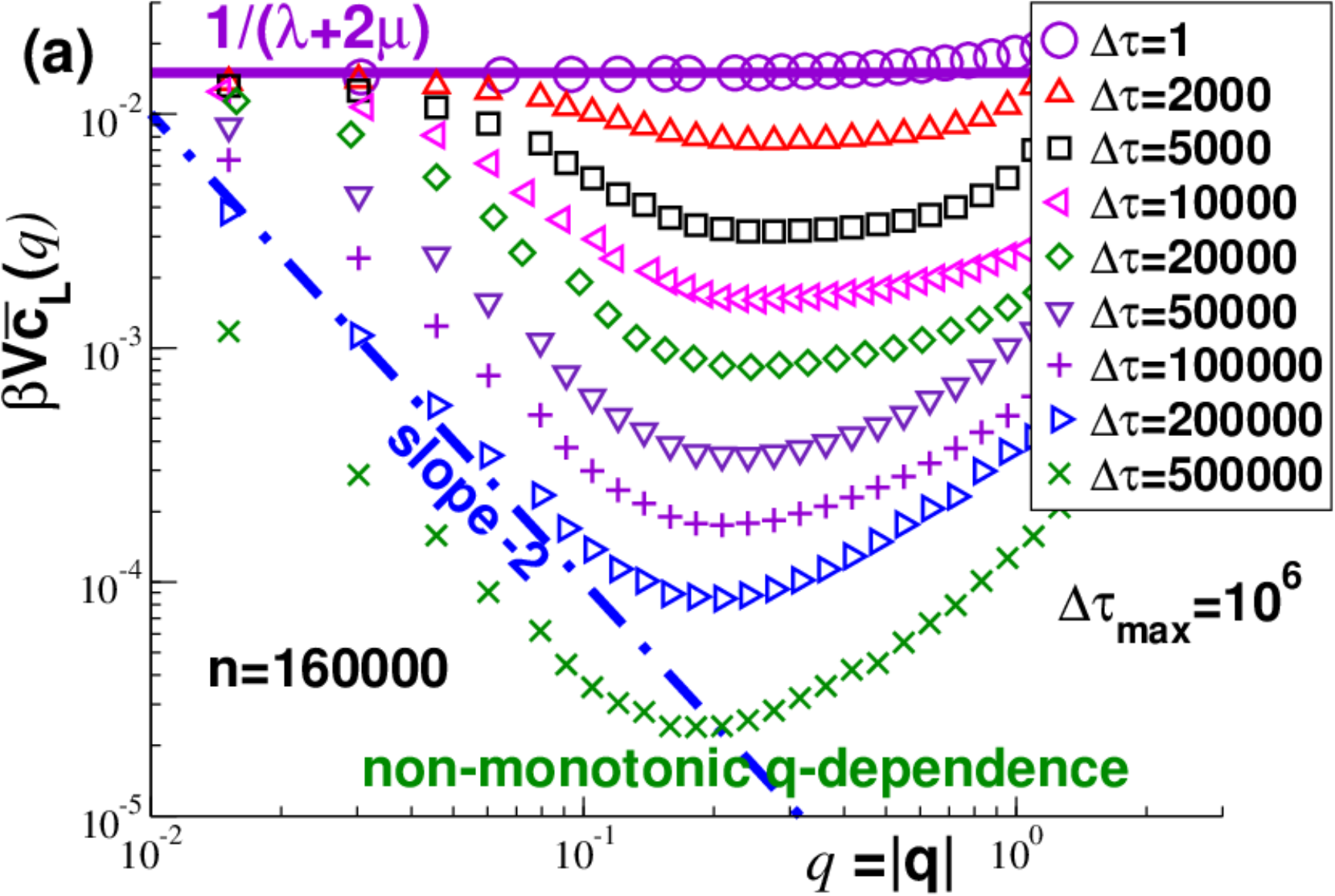}}}
\centerline{\resizebox{.9\columnwidth}{!}{\includegraphics*{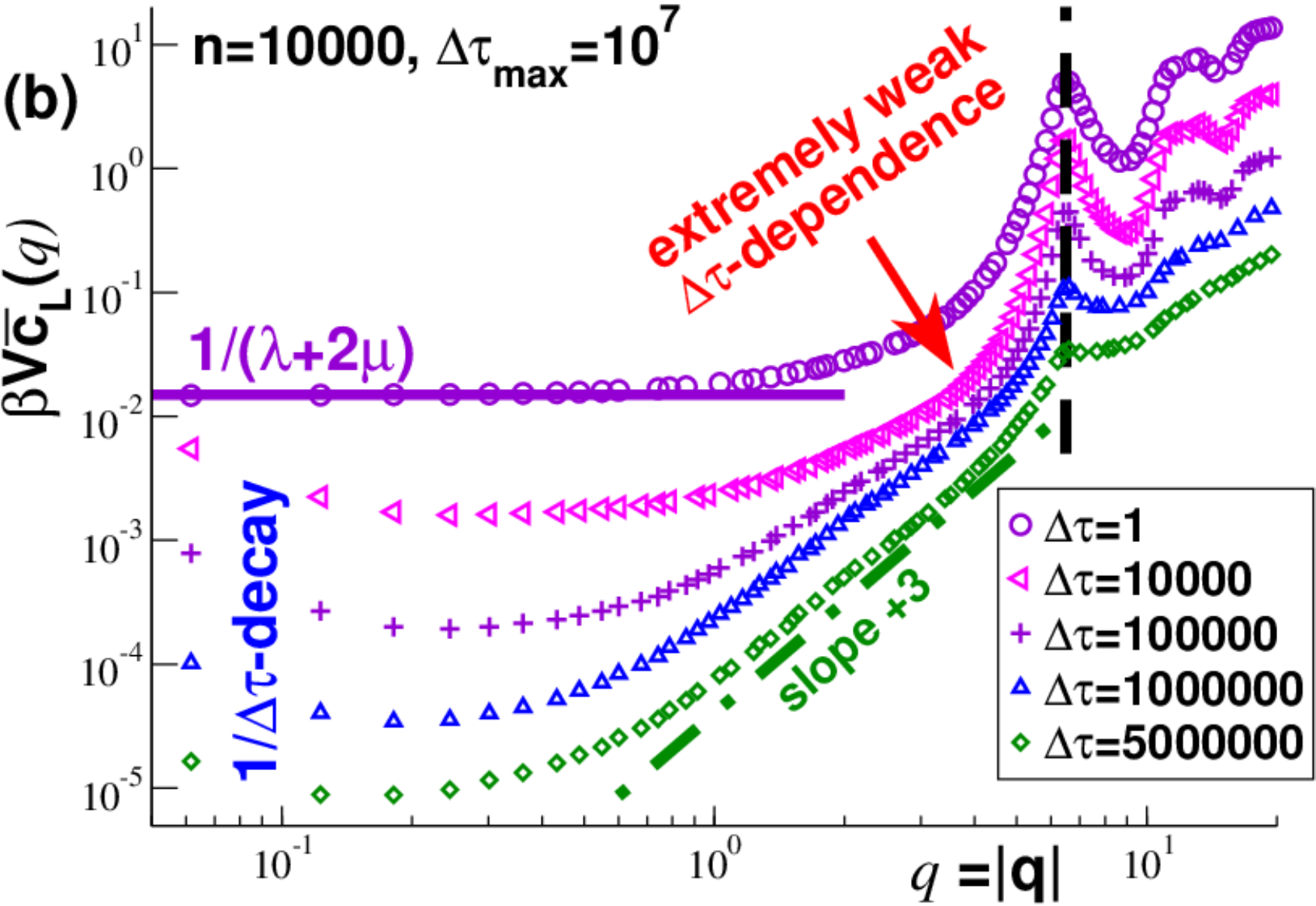}}}
\caption{Longitudinal ICF $\beta V \cLbar(q,\tsamp)$ for pLJ systems 
for a broad range of preaveraging times $\tsamp \ll \tsampmax$:
{\bf (a)} 
$n=160000$ with $\tsampmax=10^6$ focusing on $q \le 1$ 
showing that $\beta V \cLbar(q,\tsamp) \to 1/L(q)$ for $\tsamp \to 1$
and that all data decay systematically with increasing $\tsamp$ for all $q$
but are non-monotonic for constant $\tsamp$ with respect to $q$.
{\bf (b)}
$n=10000$ for $\tsampmax=10^7$ focusing on large $\tsamp$ and $q$
demonstrating a strong increase with $q$ (dashed-dotted line) 
and a weak dependency on $\tsamp$ especially 
for $q$ slightly below the maximimum of the structure factor (dashed vertical line). 
}
\label{fig_pdLJ_dyn_e_q}
\end{figure}

Using our pLJ model system we have computed the ICFs $\cLbar(q)$ 
and $\cGbar(q)$ for $t$-averaged longitudinal and transverse strain fields in NRC.
Results obtained for $\beta V \cLbar(q,\tsamp)$ are plotted in Fig.~\ref{fig_pdLJ_dyn_e_q} 
as a function of $q$ for different preaveraging times $\tsamp$ and for two system sizes.
Panel {\bf (a)} focuses on data for $q \le 1$ obtained for $n=160000$ and $\tsampmax=10^6$,
panel {\bf (b)} on the large-$q$ and large-$\tsamp$ behavior for the smaller systems
with $n=10000$ and $\tsampmax=10^7$. 
Since the total time series is used to construct the displacement/strain fields 
(cf.~Sec.~\ref{instTFs}) 
the $t$-averaged strain fields trivially vanish for $\tsamp \to \tsampmax$
and therefore the corresponding CFs.
Only data with $\tsamp \ll \tsampmax$ is thus useful.
Note that data for $\tsamp=1$ (circles) corresponds to the CF of the instantaneous
longitudinal strain already shown in Fig.~\ref{fig_pdLJ_LqGq} and used for the 
determination of the static longitudinal modulus $L(q)$. From this upper limit 
the ICFs are seen to monotonically decrease with $\tsamp$ and this for all $q$.
As can be better seen from panel {\bf (a)}, the data are {\em strongly non-monotonic} 
with respect to $q$ decreasing first in the continuum limit to a minimum roughly located 
at $q \approx 2\pi/\xicont$ (shifting to lower $q$ with increasing $\tsamp$).

\begin{figure}[t]
\centerline{\resizebox{.9\columnwidth}{!}{\includegraphics*{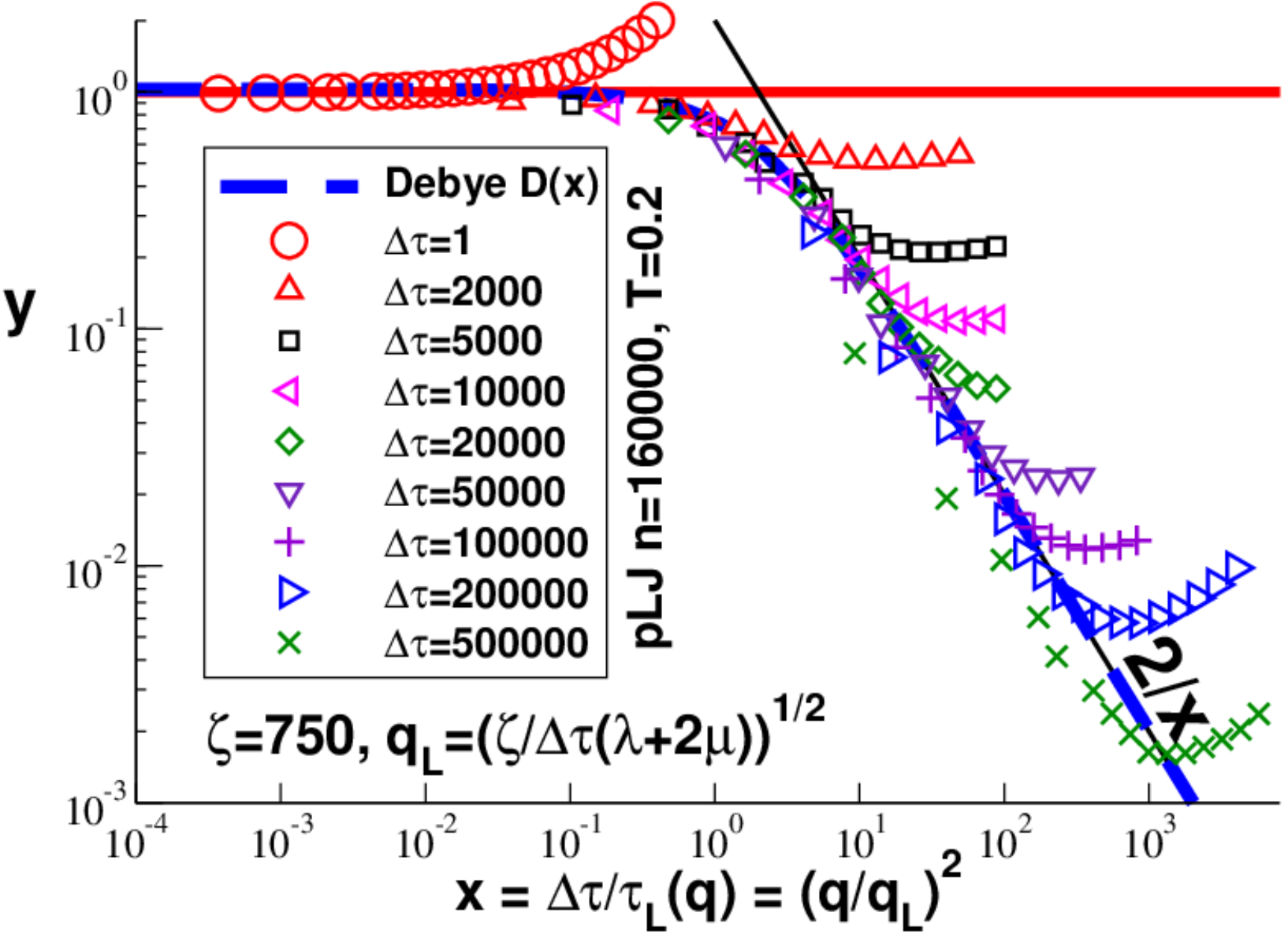}}}
\caption{Rescaled longitudinal ICF $y = (\lambda+2\mu)\beta V \cLbar(q,\tsamp)$
for $n=160000$ as a function of $x=\tsamp/\tauL(q) = (q/\qL(\tsamp))^2$ as suggested by
Eq.~(\ref{eq_dyn_e_tlarge_C}) for sufficiently small $q$ and large $\tsamp$. 
The bold dashed line indicates the Debye function $D(x)$,
the thin solid line the asymptotic power law $D(x) \simeq 2/x$ for $x \gg 1$.
The same friction constant $\zeta \approx 750$ is used for all $\tsamp$.
}
\label{fig_pdLJ_dyn_e_qlow}
\end{figure}

The $1/q^2$-decay seen in panel {\bf (a)} of Fig.~\ref{fig_pdLJ_dyn_e_q} (dash-dotted line) 
is expected in the continuum limit from Eq.~(\ref{eq_dyn_e_tlarge_C}) for reduced times 
$x=\tsamp/\tauL(q) \gg 1$.
Focusing on the data for $q \ll 1/\xicont$ and large $\tsamp$ we present
in Fig.~\ref{fig_pdLJ_dyn_e_qlow} a scaling plot suggested by Sec.~\ref{dyn_e_tlarge}
and using $L(q) \simeq \lambda + 2\mu$. 
We focus again on the longitudinal ICFs computed for different $\tsamp$.
Following the first relation given in Eq.~(\ref{eq_dyn_e_tlarge_C})
we plot $y = (\lambda + 2\mu) \beta V \cLbar(q,\tsamp)$ as a function of
the scaling variable $x=\tsamp/\tauL(q) = (q/\qL(\tsamp))^2$
with the $\tsamp$-dependent characteristic wavevector
$\qL(\tsamp)=(\zeta/\tsamp (\lambda+2\mu))^{1/2}$.
The bold dashed line indicates the Debye function $D(x)$,
the thin solid line its large-$x$ limit $D(x) \simeq 2/x$.
Note that in the latter limit the ICFs asymptotically decay as $1/\tsamp q^2$. 
Since $\lambda \approx 38$ and $\mu \approx 14$ are known, there is only one
fitting parameter, namely the effective friction coefficient $\zeta = 750$.
This value was determined by horizontally shifting the data sets
for $10^4 \le \tsamp \le 10^5$ onto the Debye function.
Naturally, data for $q \gg 1/\xicont$ deviates from $D(x)$ as expected from 
the unscaled data in Fig.~\ref{fig_pdLJ_dyn_e_q}. Note also that the data
for $\tsamp=500000$ are slightly too small. This deviation can be explained from the
fact that all data for $\tsamp \simeq \tsampmax$ must vanish due to the 
numerical construction of the strain field, cf.~Eq.~(\ref{eq_udef_q}).
Similar behavior has been observed for $\cGbar(q,\tsamp)$ 
using the {\em same} friction coefficient (not shown).
 
It is unfortunately not possible for our pLJ systems to determine
the terminal relaxation times for $q \gg 1/\xicont$ 
since we are unable to reach the final $1/\tsamp$-decay of the ICFs
for the available $\tsampmax$.
This can be better seen from the data presented in panel {\bf (b)} of Fig.~\ref{fig_pdLJ_dyn_e_q} 
for $n=10000$ and $\tsampmax=10^7$ MSD. As emphasized by the arrow, 
much larger production times $\tsampmax$ are warranted to get $\tauLT(q)$.
More details on the large-$q$ relaxation may be found in Appendix~\ref{tauterm}.

\section{Conclusion}
\label{conc}

\subsection{Summary}
\label{conc_summary}

Focusing on displacement and strain TFs in amorphous solids
we have addressed various aspects of ITFs relevant more generally
for isotropic and achiral condensed matter systems.
It was shown that a generic mathematical structure in terms of a small number of invariants 
is expected, cf. Sec.~\ref{ten} and Appendix~\ref{ten_field}.
We emphasized that theoretical and numerical studies should focus on these invariants and 
this especially in reciprocal space where the results can be formulated in a $d$-independent manner. 
Generic ITFs contain in general terms depending on the direction of the field vector,
e.g., in reciprocal space on the components $\qhat_{\alpha}$ of the wavevector,
and for this reason components of ITFs may superficially appear to be ``anisotropic" 
(cf.~Fig.~\ref{fig_intro}).
Importantly, any true anisotropy should be characterized in terms of proper invariants 
(cf.~Sec.~\ref{ten_aniso} and Sec.~\ref{aniso}). 
Under the additional assumptions stated in Sec.~\ref{ten_sum},
the most general fourth-order ITFs are given by {\em five} invariants,
cf.~Eq.~(\ref{eq_ten_sum_o4}).  Using the important transformation 
Eq.~(\ref{eq_ten_sum_d2_in_transform}) one may reduce for $d=2$
the number of independent invariants of fourth-order ITFs from five to four.
This allows to simplify the general results for $d \ge 2$ for two-dimensional systems.
We also stressed the advantages to analyze ITFs using NRC
by means of an alternative set of equivalent invariants, cf.~Eq.~(\ref{eq_ten_sum_d3}).
Using the general formalism of ITFs we have been able to compute analytically
for $d=2$ and $d=3$ for several cases the invariants in real space 
from those in reciprocal space (cf.~Table~\ref{tab_eta}).

Several predictions and numerical procedures have been illustrated using a pLJ particle model
in strictly {\em two} dimensions and sampled using an overdamped MC dynamics.
We have investigated in Sec.~\ref{cstat} static correlations of displacement and strain fields 
in linear elastic bodies taking advantage of well-known relations 
from statistical physics, cf.~Eq.~(\ref{eq_cab_df}) \cite{ChaikinBook}.
The wavenumber-dependent static elastic moduli $L(q)$ and $G(q)$ of the pLJ particle system 
used in this study have been determined, cf.~Fig.~\ref{fig_pdLJ_LqGq}.
$L(q)$ and $G(q)$ were observed to become very small for large $q$, 
especially around the position of the main peak of the structure factor.
We have explicitly checked in Sec.~\ref{gauss} that the strain components 
in NRC are complex circularly-symmetric Gaussian variables for small wavenumbers $q \ll 1$. 
Deviations become relevant for larger $q$, cf.~Fig.~\ref{fig_pdLJ_alpha}.
As shown in Sec.~\ref{aniso}, 
it is possible by $k$-averaging the strongly fluctuating instantaneous strain ICFs to obtain 
a precise characterization of anisotropic effects. These were shown to become again noticible
above $q \approx 1$. The three length scales $\xicont$, $\xiiso$ and $\xihom$ 
characterizing an amorphous elastic body were argued to be of same order, 
cf.~Fig.~\ref{fig_pdLJ_aniso_qfirst}.
 
Taking advantage of the general FDT relation Eq.~(\ref{eq_fdt_ex_A}) and 
the reduced form Eq.~(\ref{eq_boltz_uKg}) of the well-known Boltzmann superposition 
relation Eq.~(\ref{eq_boltz_sEe}), we have derived in Sec.~\ref{dyn} 
(focusing on systems with overdamped motion)
the general relations between the time-dependent CFs 
of displacement and strain fields with the viscoelastic material functions $L(q,t)$ 
and $G(q,t)$, i.e. two invariants in NRC of the fourth-order elasticity TF $\Eabcd(\qvec,t)$.
By analyzing the ICFs $\cLbar(q)$ and $\cGbar(q)$ of $t$-averaged strain fields in NRC 
(defined in Appendix~\ref{corr})
for a broad range of preaveraging times $\tsamp$ 
we have confirmed that the terminal relaxation times decay as $\tauLT(q) \propto 1/q^2$ 
in the continuum limit for our overdamped systems (cf.~Fig.~\ref{fig_pdLJ_dyn_e_qlow}). 
The determination of $\tauLT(q)$ was unfortunately not possible beyond the continuum
limit due to the strong slowing-down shown in the second panel of Fig.~\ref{fig_pdLJ_dyn_e_q}
and further discussed in Appendix~\ref{tauterm}.

\subsection{Outlook}
\label{conc_outlook}

We assumed overdamped dynamics in Sec.~\ref{dyn}.
One reason for considering this case was simply that local MC hopping moves were used
to obtain our numerical results. Interestingly, the presented arguments can readily be 
reformulated for momentum-conserving molecular dynamics simulations. 
One merely has to replace 
the friction coefficient $\zeta$ in Eq.~(\ref{eq_dyn_u_B}) by $\rho \frac{\ddiff}{\ddiff t}$ 
with $\rho$ being the mass density \cite{lyuda18}. The recursion relation Eq.~(\ref{eq_dyn_u_E}) 
still holds but now in terms of the scalar $w(q,s) \equiv q^2/\rho s^2$. One verifies that with this
rewriting Eq.~(\ref{eq_dyn_u_kLN}) and Eq.~(\ref{eq_dyn_e_cLG}) are equivalent to the 
relations obtained using the Mori-Zwanzig formalism in Ref.~\cite{Fuchs18b}.
  
Moreover, 
we simulated two-dimensional systems simply since for a feasible particle number 
$n  \ll 10^6$ larger linear system sizes $L \propto n^{1/d}$ can be simulated and thus smaller 
wavenumbers $q$ sampled than for $d=3$. This allowed us to probe the continuum limit 
as shown by Fig.~\ref{fig_pdLJ_LqGq}.
Since our theoretical results are relevant for all $d \ge 2$, 
the presented numerical methodology focusing on ICFs in NRC carries over to more realistic 
three-dimensional systems which should be addressed in the future.
Especially we predict long-range correlations for strain TFs decaying as $1/r^d$ 
in any viscoelastic system with a broad intermediate elastic plateau regime.
Such long-range correlations should also be relevant, e.g., for
bulks of entangled polymer melts for times below the reptation time \cite{RubinsteinBook,DoiEdwardsBook}. 
The presented study suggests that the typical ``tube diameter" of the 
reptation model and the continuum limit length $\xicont$ must be physically strongly related 
and numerically similar. Work in this direction is currently underway.



\vspace*{0.1cm}
\paragraph*{Acknowledgments.}
We are indebted to A.N. Semenov (Strasbourg) and H.~Xu (Metz) for helpful discussions.
 
\vspace*{0.1cm}
\paragraph*{Data availability statement.}
Data sets are available from the corresponding author on reasonable request.

\appendix
\section{Useful general transformations}
\label{trans}

We investigate in this work tensors depending on time $t$ and TFs depending 
additionally on the spatial position $\rvec$.
Due to the various convolution and correlation relations it is useful to move 
to Fourier space with $\qvec$ being the wavevector and 
to Laplace space with $s$ being the Laplace variable. 
We define the ``Fourier transformation" (FT) $f(\qvec) = \Fcal[f(\rvec)]$ and 
the ``Laplace transformation" (LT) $f(s) = \Lcal[f(t)]$ 
such that the original functions and their transformations have the {\em same} units. 
This makes it easier to dimensionally check the relations.
For notational simplicity the function names remain unchanged by the transform.
Which space is meant is indicated by the argument.
Some well-known properties of these transformations 
are summarized here for convenience \cite{foot_trans_fqs}. 

We consider real-valued functions $f(\rvec)$ of a $d$-dimensional spatial 
``position vector" $\rvec$. 
Following Refs.~\cite{lyuda22a,spmP5a,spmP5b} 
we define the FT from ``real space" (variable $\rvec$) to ``reciprocal space" (variable $\qvec)$ by
\begin{equation}
f(\qvec) \equiv \frac{1}{V} \int \ddiff \rvec \ f(\rvec) \exp(-i \qvec \cdot \rvec)
\label{eq_FT_def}
\end{equation}
with $V$ being the $d$-dimensional volume of the system
\cite{foot_trans_qdiscrete}. 
We note the FTs
\begin{eqnarray}
\partial_{\alpha} f(\rvec) & \stackrel{\Fcal}{\Leftrightarrow} & 
i q_{\alpha} f(\qvec) \mbox{ and } \label{eq_FT_A}\\
V \delta(\rvec) & \stackrel{\Fcal}{\Leftrightarrow} & 1 \label{eq_FT_C}
\end{eqnarray}
with $\partial_{\alpha} \equiv \partial/\partial  r_{\alpha}$ for the partial derivative
in the $\alpha$-direction in real space
and $\delta(\rvec)$ being Dirac's delta function. 
It follows from Eq.~(\ref{eq_FT_A}) and Eq.~(\ref{eq_FT_C}) that
\begin{equation}
\partial_{\alpha} V \delta(\rvec) \stackrel{\Fcal}{\Leftrightarrow} 
i q_{\alpha}.
\label{eq_FT_F}
\end{equation}
Using $\Fcal[f(\rvec-\vvec)]=f(q) \exp(-i\qvec\cdot \vvec)$ for a constant vector $\vvec$ we have
\begin{eqnarray}
V\delta(\rvec-\vvec) & \stackrel{\Fcal}{\Leftrightarrow}&
\exp\left[-i \qvec \cdot \vvec \right] \label{eq_FT_CC}.
\end{eqnarray}
This implies consistently with Eq.~(\ref{eq_FT_F}) for a ``dipole distribution" of Dirac functions
that
\begin{equation}
V \Fcal\left[\delta(\rvec+\vvec/2) - \delta(\rvec-\vvec/2)\right] \simeq i \qvec\cdot \vvec 
\mbox{ for } |\qvec \cdot \vvec| \ll 1.
\label{eq_FT_G}
\end{equation}
Let us next consider the ``correlation function" (CF)
\begin{equation}
c(\rvec) = \frac{1}{V} \int \ddiff \rvec' g(\rvec+\rvec') h(\rvec')
\label{eq_FT_correlation_r} 
\end{equation}
with real-valued fields $g(\rvec)$ and $h(\rvec)$.
According to the ``correlation theorem" \cite{numrec} we get in reciprocal space
\begin{equation}
c(\qvec) = \Fcal[c(\rvec)]  
=  g(\qvec) h(-\qvec)
\label{eq_FT_correlation_q} 
\end{equation}
with $g(\qvec)=\Fcal[g(\rvec)]$, $h(\qvec)=\Fcal[h(\rvec)]$. 
For ``auto-correlation functions" (ACFs), i.e. for $g(\rvec)=h(\rvec)$,
this simplifies to (``Wiener-Khinchin theorem") 
\begin{equation}
c(\qvec)  = g(\qvec) g(-\qvec) = |g(\qvec)|^2 \ge 0.
\label{eq_FT_WKT} 
\end{equation}
All CFs of this work are assumed to be even,
\begin{equation}
c(\rvec)=c(-\rvec) \stackrel{\Fcal}{\Leftrightarrow} c(\qvec)=c(-\qvec),
\label{eq_FT_even}
\end{equation}
i.e. all $c(\qvec)$ are real functions.
We also remind that according to the ``convolution theorem" of FTs \cite{numrec}
\begin{equation}
\frac{1}{V} \int \ddiff \rvec' g(\rvec-\rvec') h(\rvec')
\stackrel{\Fcal}{\Leftrightarrow} g(\qvec) h(\qvec).
\label{eq_FT_convolution}
\end{equation}

As in our previous work \cite{lyuda22a,spmP5a,spmP5b} we use the ``Laplace-Carson transform" 
\begin{equation}
f(s) = \Lcal[f(t] = s \ \lim_{\epsilon\to 0} \int_{t=-\epsilon}^{\infty} \ddiff t \ f(t) \exp(-s t).
\label{eq_LT_def}
\end{equation}
Due the prefactor $s$ of the transform $f(t)$ and $f(s)$ have the same dimension.
We note that
\begin{eqnarray}
a H(t) &  \stackrel{\Lcal}{\Leftrightarrow} & a \label{eq_LT_ut} \\
a \delta(t) & \stackrel{\Lcal}{\Leftrightarrow} & a s \label{eq_LT_delta} \mbox{ and } \\
\exp(-t/\tau) & \stackrel{\Lcal}{\Leftrightarrow} &\frac{s}{s+1/\tau} \label{eq_LT_exponential}
\end{eqnarray}
with $a$ and $\tau$ being some finite constants, 
$H(t)$ the Heaviside function (unit step) 
and $\delta(t) = \dot{H}(t)$ Dirac's delta function \cite{abramowitz}.
Following Newton a dot marks a derivative with respect to time.
We finally remind
\begin{eqnarray}
\int_0^t f(t) \ddiff t 
& \stackrel{\Lcal}{\Leftrightarrow} & 
f(s)/s \label{eq_LT_int} \mbox{ and } \\
\int_0^t g(t-t') h(t') \ddiff t' 
& \stackrel{\Lcal}{\Leftrightarrow} & \frac{g(s) h(s)}{s} \label{eq_LT_convolution} 
\end{eqnarray}
and the initial and final value theorems
\begin{equation}
\lim_{s\to \infty} f(s) = \lim_{t\to 0} f(t) \mbox{ and }
\lim_{s\to 0} f(s) = \lim_{t\to \infty} f(t)
\label{eq_LT_FVT}
\end{equation}
for reasonably behaved functions \cite{foot_trans_fqs}.

\section{More on isotropic tensor fields}
\label{tenmore}

\subsection{Macroscopic isotropic tensors}
\label{ten_ten}

It is important to distinguish ``macroscopic tensors" $T_{\alpha\beta\ldots}$ 
(without argument \cite{foot_trans_fqs})
and the more general TFs $T_{\alpha\beta\ldots}(\qvec)$. 
(TFs may reduce to macroscopic tensors for $\qvec \to \bfzero$ assuming this limit to exist.)
For tensors the general isotropy condition Eq.~(\ref{eq_ten_sum_iso}) becomes
\begin{equation}
T_{\alpha_1\ldots\alpha_o}^{\Tgen} = T_{\alpha_1\ldots\alpha_o},
\label{eq_ten_ten_iso}
\end{equation}
i.e. all tensor components are unchanged under any orthogonal transform.
Macroscopic isotropic tensors of different order are discussed, e.g., 
in Sec.~2.5.6 of Ref.~\cite{TadmorCMTBook}. 
With $k_1$, $i_1$ and $i_2$ being invariant scalars we have 
\begin{eqnarray}
T_{\alpha} & = & 0, \label{eq_ten_ten_o1} \\
T_{\alpha\beta} & = & k_1 \IDab, \label{eq_ten_ten_o2} \\
T_{\alpha\beta\gamma} & = & 0 \mbox{ and } \label{eq_ten_ten_o3} \\
T_{\alpha\beta\gamma\delta} & = & i_1 \IDab \IDcd 
+ i_2 \left(\IDac \IDbd + \IDad \IDbc\right) \label{eq_ten_ten_o4}
\end{eqnarray}
with $\IDab$ denoting the Kronecker symbol. Importantly, 
all tensors of odd order must vanish \cite{TadmorCMTBook} and
\begin{equation}
T_{12}= T_{1112} = T_{1222} = T_{1234} = T_{1344} = 0
\label{eq_ten_ten_Eabcd_B}
\end{equation}
These are consequences of a general property of 
macroscopic isotropic tensors \cite{spmP5a}:
the sign of tensor components change for a reflection of one axis if the number of 
indices equal to the inverted axis is {\em odd}.
Consistency with Eq.~(\ref{eq_ten_ten_iso}) implies then that
components of macroscopic isotropic tensors
with an odd number of equal indices vanish.
As emphasized below (cf. Sec.~\ref{ten_NRC}), 
this is different for the more general ITFs.

\subsection{Constructing isotropic tensor fields}
\label{ten_product}

Let us consider the TF $\Cvec(\qvec) = \Avec(\qvec) \otimes \Bvec(\qvec)$ 
being the product of two ITFs $\Avec(\qvec)$ and $\Bvec(\qvec)$ 
and $\otimes$ standing either for an outer product, 
e.g. $C_{\alpha\beta\gamma\delta}(\qvec)=A_{\alpha\beta}(\qvec)B_{\gamma\delta}(\qvec)$,
or an inner product, 
e.g. $C_{\alpha\beta\gamma\delta}(\qvec)=A_{\alpha\beta\gamma\nu}(\qvec)B_{\nu\delta}(\qvec)$.
Hence,
\begin{eqnarray}
\Cvec^{\Tgen}(\qvec) & = & \left(\Avec(\qvec) \otimes \Bvec(\qvec)\right)^{\Tgen} 
= \Avec^{\Tgen}(\qvec) \otimes \Bvec^{\Tgen}(\qvec) \nonumber \\
& = & \Avec(\qvec^{\Tgen}) \otimes \Bvec(\qvec^{\Tgen}) = \Cvec(\qvec^{\Tgen})
\label{eq_ten_product_A}
\end{eqnarray}
using in the second step a general property of TF products
and in the third step Eq.~(\ref{eq_ten_sum_iso}) for $\Avec(\qvec)$ and $\Bvec(\qvec)$
where $\qvec^{\Tgen}$ stands for the ``actively" transformed field position.
We have thus demonstrated that $\Cvec(\qvec)$ is also an ITF
(``product theorem for ITFs").
Similarly, one can show that the sum of two ITFs must also be an ITF.
The above relation Eq.~(\ref{eq_ten_product_A}) can also be used to state a
``quotient theorem for ITFs" similar to the 
well-known general quotient theorem for TFs \cite{McConnell}:  
If $\Cvec(\qvec)$ and, say, $\Avec(\qvec)$ are known to be ITFs,
Eq.~(\ref{eq_ten_product_A}) implies under mild and obvious conditions that $\Bvec(\qvec)$ must 
also be an ITF. 
Similarly, $C(\qvec)=A(\qvec)+B(\qvec)$ for ITFs $C(\qvec)$
and $A(\qvec)$ implies that $B(\qvec)$ must also be an ITF.
   
We note that the Kronecker symbol $\IDab$ \cite{Schultz_Piszachich} 
and also each component $q_{\alpha}$ of the field vector 
are isotropic according to Eq.~(\ref{eq_ten_sum_iso}), i.e.
$(\IDab)^{\Tgen} = \IDab$ and $(q_{\alpha})^{\Tgen} = q_{\alpha}^{\Tgen}$.
The above theorems allow quite generally the construction of ITFs from known ITFs. 
For instance, assuming $l(q)$, $k(q)$, $j(q)$ and $i(q)$ to be scalar invariants
any product of these terms, e.g.,
$l(q) q_{\alpha}$, $k(q) q_{\alpha} q_{\beta}$ or $i(q) q_{\alpha} q_{\beta} q_{\gamma} q_{\delta}$
must be an ITF and the same applies to sums of such terms. 
Albeit being legitimate ITFs, such sums may thus depend 
on the direction of the wavevector $\qvec$ and on the orientation of the coordinate system.

\subsection{Generic structure of isotropic tensor fields}
\label{ten_field}

Let us state the most general ITFs for $1 \le o \le 4$ and any dimension $d>1$
compatible with the symmetries stated in Sec.~\ref{ten_sum}.
With $l_n(q)$, $k_n(q)$  and $i_n(q)$ being invariant scalar functions we have
\begin{eqnarray}
T_{\alpha}(\qvec) & = & l_1(q) \ q_{\alpha}, 
\label{eq_ten_field_o1} \\
T_{\alpha\beta}(\qvec) & = & 
k_1(q) \ \delta_{\alpha\beta} + k_2(q) \ q_{\alpha}q_{\beta},
\label{eq_ten_field_o2_old} \\
T_{\alpha\beta\gamma\delta}(\qvec) & = &
i_1(q) \ \delta_{\alpha\beta} \delta_{\gamma\delta} \label{eq_ten_field_o4_old} \\
& + & i_2(q) \ \left(
\delta_{\alpha\gamma} \delta_{\beta\delta} + \delta_{\alpha\delta} \delta_{\beta\gamma}
\right) \nonumber \\
& + & i_3(q) \ \left(
q_{\alpha} q_{\beta}\delta_{\gamma\delta} + q_{\gamma}q_{\delta}\delta_{\alpha\beta} 
\right) \nonumber \\
& + & i_4(q) \ q_{\alpha} q_{\beta} q_{\gamma} q_{\delta} \nonumber\\
& + & 
i_5(q)  \left( 
q_{\alpha}q_{\gamma} \delta_{\beta\delta}+  
q_{\alpha}q_{\delta} \delta_{\beta\gamma}+\right. \nonumber \\
& & \hspace*{.90cm}\left. q_{\beta}q_{\gamma} \delta_{\alpha\delta}+  
q_{\beta}q_{\delta} \delta_{\alpha\gamma}  
\right) \nonumber
\end{eqnarray}
with $q_{\alpha} \equiv \qvec \cdot \evec_{\alpha}$.
One way to obtain these results is to construct ITFs using all possible 
``multilinear forms" \cite{Schultz_Piszachich,TadmorCMTBook} of order $o$ 
for additive terms of scalars of inner and triple products \cite{Schultz_Piszachich} 
and to eliminate then in a second step all terms not compatible with the 
additionally assumed symmetries 
\cite{Schultz_Piszachich,Lemaitre15,Lemaitre18,spmP5a,spmP5b}.
If the invariants are smooth non-singular functions
all relations reduce for $\qvec \to \bfzero$ to the isotropic tensors stated 
in Appendix~\ref{ten_ten}.

We have noted in Appendix~\ref{ten_ten} that all components of isotropic tensors
with an odd number of equal indices do vanish. 
Apparently, this does not hold for ITFs since ITFs of odd order may be finite, 
cf. Eq.~(\ref{eq_ten_field_o1}). 
The reader may also verify that while the isotropic tensor component $T_{1112}=0$ vanishes 
the component $T_{1112}(\qvec)$ is finite in general.
The reason for this is that the condition Eq.~(\ref{eq_ten_sum_iso}) 
for ITFs is less restrictive than Eq.~(\ref{eq_ten_ten_iso}).
Fortunately, there are convenient coordinates,
called ``Natural Rotated Coordinates" (NRC), where the nice symmetry Eq.~(\ref{eq_ten_ten_Eabcd_B}) 
for isotropic tensors can be also used for TFs of even order. 
We return to this in Appendix~\ref{ten_NRC}.

For dimensional reasons it is useful to rewrite for $\qvec \ne \bfzero$ 
the above ITFs in terms of the normalized components
$\qhat_{\alpha} = \qhatvec \cdot \evec_{\alpha}$.
It is thus convenient to bring in factors of $q$ and to redefine
$l_1(q) \to l_1(q)/q$,
$k_2(q) \to k_2(q)/q^2$,
$i_3(q) \to i_3(q)/q^2$, $i_4(q) \to i_4(q)/q^4$ and $i_5(q) \to i_5(q)/q^2$.
This rescaling leads to Eq.~(\ref{eq_ten_sum_o1}), Eq.~(\ref{eq_ten_sum_o2}) and 
Eq.~(\ref{eq_ten_sum_o4}) stated in Sec.~\ref{ten_sum}.
Now {\em all} invariants of each ITF of order $o$
have the {\em same} physical units.

For two-dimensional systems it is possible and useful to rewrite Eq.~(\ref{eq_ten_sum_o4}) 
more compactly in terms of the first four invariants $i_1(q)$, $i_2(q)$, $i_3(q)$ and $i_4(q)$.
One way to see this is to rewrite the last parenthesis of Eq.~(\ref{eq_ten_sum_o4}) as
\begin{eqnarray}
\qhat_{\alpha} \qhat_{\gamma} \delta_{\beta\delta} + 
\qhat_{\alpha} \qhat_{\delta} \delta_{\beta\gamma} + 
\ldots
& = & -2 \left[ \delta_{\alpha\beta} \delta_{\gamma\delta} \right]
\label{eq_d2_rewriting} \\
& + & \left[\delta_{\alpha\gamma} \delta_{\beta\delta} + 
\delta_{\alpha\delta} \delta_{\beta\gamma} \right] \nonumber \\
& + & 2 \left[\qhat_{\alpha} \qhat_{\beta}\delta_{\gamma\delta} + 
\qhat_{\gamma}\qhat_{\delta}\delta_{\alpha\beta} \right] \nonumber
\end{eqnarray}
as may be verified by using that $\qhat_1^2+\qhat_2^2=1$ in $d=2$
and comparing all possible cases for $(\alpha,\beta,\gamma,\delta)$,
e.g., $(1,1,1,1)$, $(2,2,2,2)$, $(1,1,2,2)$, $(1,2,1,2)$ or $(1,1,2,2)$.
Hence, if one has obtained by means of a numerical or theoretical argument a representation 
of fourth-order ITFs with a finite value for the invariant $i_5(q)$, this result may be 
rewritten by means of the transformation Eq.~(\ref{eq_ten_sum_d2_in_transform}) 
given in Sec.~\ref{ten_sum} in terms of only four finite invariants.

\subsection{Natural Rotated Coordinates}
\label{ten_NRC}

That {\em four} invariants for fourth-order ITFs are sufficient in $d=2$ can also be seen using NRC.
Following Refs.~\cite{lyuda18,spmP5a,spmP5b} let us rotate the coordinate system 
such that the $1$-axis points into the direction of $\qvec$.
We mark coordinates in NRC by ``$\circ$".  Let us define 
\begin{eqnarray}
\kL(q)  & \equiv & T^{\circ}_{11}(\qvec), \iL(q) \equiv T^{\circ}_{1111}(\qvec), 
\label{eq_ten_NRC_invariants} \\ 
\kN(q)  & \equiv & T^{\circ}_{22}(\qvec), \iN(q) \equiv T^{\circ}_{2222}(\qvec), \nonumber \\
\iM(q)  & \equiv & T^{\circ}_{1122}(\qvec) \mbox{ and } \nonumber \\
\iG(q)  & \equiv & T^{\circ}_{1212}(\qvec) \nonumber 
\end{eqnarray}
for second- and fourth-order ITFs in NRC.
Since the system is isotropic these functions depend on the scalar $q$,
i.e. they are {\em invariant} under rotation and they do not change either 
if one of the coordinate axes is inversed.
$\kL(q)$ and $\iL(q)$ are called ``longitudinal invariants",
$\kN(q)$ and $\iN(q)$ ``normal invariants",
$\iM(q)$ ``mixed invariant" and
$\iG(q)$ ``transverse" (or ``shear") invariant.
All other components $T^{\circ}_{\alpha\beta}(\qvec)$ and 
$T^{\circ}_{\alpha\beta\gamma\delta}(\qvec)$ are 
due to the assumed index symmetries identical to these invariants or
must vanish for an {\em odd number of equal indices} \cite{spmP5a},
behaving thus in NRC as isotropic tensors, cf.~Eq.~(\ref{eq_ten_ten_Eabcd_B}).

The fields $T_{\alpha\beta}(\qvec)$ and $T_{\alpha\beta\gamma\delta}(\qvec)$ 
in the original frame may then be obtained by an inverse rotation. 
Both sets of invariants are thus related by
\begin{eqnarray}
\kL(q) & = & k_1(q)+k_2(q), \kN(q) = k_1(q), \label{eq_ten_NRC_n2A} \\ 
\iL(q) & = & i_1(q) + 2i_2(q) + 2i_3(q) + i_4(q), \nonumber \\
\iG(q) & = & i_2(q), \nonumber \\
\iM(q) & = & i_1(q) + i_3(q) \mbox{ and } \nonumber \\
\iN(q) & = & i_1(q) + 2i_2(q).  \nonumber 
\end{eqnarray}

We can extend the above identification of the two sets of invariants
for two-dimensional systems to higher dimensions. Let us first consider $d=3$.
As above we turn the $1$-axis into the direction of the wavevector $\qvec$.
This rotation is, of course, now not unique. 
However, any rotation around $\qvec$ is equivalent due to isotropy. 
This implies, e.g., that $T^{\circ}_{2222}(\qvec)=T^{\circ}_{3333}(\qvec)$ or
$T^{\circ}_{1122}(\qvec)=T^{\circ}_{1133}(\qvec)$.
We use again the same invariants as in Eq.~(\ref{eq_ten_NRC_invariants}). 
While the two invariants $\kL(q)$ and $\kN(q)$ for second-order ITFs are sufficient for $d=3$,
the four invariants $\iL(q)$, $\iG(q)$, $\iM(q)$ and $\iN(q)$ for fourth-order ITFs
must be supplemented by {\em one} additional invariant.  Hence, either
\begin{equation}
\iS(q) \equiv T^{\circ}_{2233}(\qvec) \mbox{ or } \iT(q)  \equiv T^{\circ}_{2323}(\qvec). 
\label{eq_ten_d3_A}
\end{equation}
Both functions are clearly also invariants; since only five invariants $i_n(q)$ 
are needed for Eq.~(\ref{eq_ten_sum_o4}), both cannot be {\em independent}.
In any case, as in $d=2$ this completely determines
$T^{\circ}_{\alpha\beta}(\qvec)$ and $T^{\circ}_{\alpha\beta\gamma\delta}(\qvec)$ 
and we may rotate back to the original coordinate frame. 
If we choose, without restricting the generality of the argument,
a wavevector $\qvec = \evec_1$, this yields 
Eq.~(\ref{eq_ten_sum_d3}) stated in Sec.~\ref{ten_sum}.
Upon inversion this implies in turn
\begin{eqnarray}
i_1(q) & = & \iS(q),                               \label{eq_ten_d3_C} \\
i_2(q) & = & \iT(q) = (\iN(q)-\iS(q))/2,                       \nonumber \\ 
i_3(q) & = & \iM(q)-\iS(q),                                    \nonumber \\ 
i_4(q) & = & \iL(q)+\iN(q)-2\iM(q)-4\iG(q) \mbox{ and }    \nonumber \\ 
i_5(q) & = & \iG(q)-(\iN(q)-\iS(q))/2 = \iG(q)-\iT(q).   \nonumber 
\end{eqnarray}
The above five invariants (in either ordinary coordinates or NRC) 
are in fact also sufficient for higher dimensions $d$ since by symmetry
\begin{eqnarray}
\kN(q) & = & T^{\circ}_{22}(\qvec) = \ldots = T^{\circ}_{dd}(\qvec), 
\label{eq_ten_d3_E} \\
\iG(q) & = & T^{\circ}_{1212}(\qvec) = \ldots = T^{\circ}_{1d1d}(\qvec), \nonumber \\ 
\iM(q) & = & T^{\circ}_{1122}(\qvec) = \ldots = T^{\circ}_{11dd}(\qvec), \nonumber \\ 
\iN(q) & = & T^{\circ}_{2222}(\qvec) = \ldots = T^{\circ}_{dddd}(\qvec), \nonumber \\ 
\iS(q) & = & T^{\circ}_{2233}(\qvec) = T^{\circ}_{2244}(\qvec) =
T^{\circ}_{3344}(\qvec) =  \ldots \mbox{ and }  \nonumber \\
\iT(q) & = & T^{\circ}_{2323}(\qvec) = T^{\circ}_{2424}(\qvec) = T^{\circ}_{3535}(\qvec) 
= \ldots \nonumber
\end{eqnarray}
where in any case $\iT(q) = (\iN(q)-\iS(q))/2$ must hold.

\section{Rotation tensor field}
\label{rotate}

The gradient $\gamab(\rvec) \equiv \partial_{\alpha} u_{\beta}(\rvec)$ of the 
displacement field $u_{\alpha}(\rvec)$ is well-known to be uniquely decomposable 
in terms of the symmetric linear strain TF $\epsab(\rvec)$, 
defined above by Eq.~(\ref{eq_epsab_def}),
and an antisymmetric ``rotation TF" $\rotab(\rvec)$ \cite{LandauElasticity}.
The latter second-order TF is defined by
\begin{equation}
\rotab(\qvec) = \Fcal[\rotab(\rvec)] = \frac{i}{2}
\left[ q_{\alpha} u_{\beta}(\qvec) - q_{\beta} u_{\alpha}(\qvec) \right]
\label{eq_rotab_def}
\end{equation}
using the displacement field $u_{\alpha}(\qvec)$ in reciprocal space.
Importantly, $\rotab(\qvec)$ can not be assumed to vanish 
by construction for finite $\qvec$ as normally admitted for the 
macroscopic rotation tensor ($\qvec=0$).
It follows from Eq.~(\ref{eq_FT_A}), Eq.~(\ref{eq_epsab_def}) and 
Eq.~(\ref{eq_rotab_def}) that
\begin{equation}
\gamab(\qvec) 
= i q_{\alpha} u_{\beta}(\qvec) = \epsab(\qvec) + \rotab(\qvec),
\label{eq_rot_gamab}
\end{equation}
i.e. $\gamab(\qvec)$ is neither symmetric nor antisymmetric.
As for the strain TF, cf.~Eq.~(\ref{eq_cabcd_def}), 
one may compute now in reciprocal space the CFs $\cabcd(\qvec)$
for the instantaneous rotation TFs $\rotabhat(\qvec)$.
As in Eq.~(\ref{eq_cab2cabcd}) for the strain CFs,  
one can express the rotation CFs in terms of the displacement CFs $\cab(\qvec)$.
This yields
\begin{eqnarray}
\frac{4}{q^2} \cabcd(\qvec) & = &
\qhat_{\alpha}\qhat_{\gamma} c_{\beta\delta}(\qvec) 
+ \qhat_{\beta}\qhat_{\delta} c_{\alpha\gamma}(\qvec) 
  \nonumber \\
        & - &  
\qhat_{\alpha}\qhat_{\delta} c_{\beta\gamma}(\qvec)  
-\qhat_{\beta}\qhat_{\gamma} c_{\alpha\delta}(\qvec). 
\label{eq_rot_cab2cabcd}
\end{eqnarray}
While $\cabcd(\qvec)$ is still an even and isotropic TF with major index symmetry 
($\alpha\beta \leftrightarrow \gamma\delta$),
the minor index symmetries ($\alpha \leftrightarrow \beta, \gamma \leftrightarrow \delta$)
for fourth-order ITF do {\em not} hold due to $\rotabhat(\qvec)=-\rotbahat(\qvec)$.
We have instead
\begin{equation}
\cabcd = - \cbacd = - \cabdc = \cbadc
\label{eq_rot_indexsym}
\end{equation}
(dropping the argument). 
$\cabcd(\qvec)$ is thus not described by the relations for fourth-order ITFs
given in Sec.~\ref{ten_sum} and Appendix~\ref{tenmore}.
Fortunately, its description in NRC is very simple.
Using Eq.~(\ref{eq_rot_cab2cabcd}) and Eq.~(\ref{eq_cab_LqGq}) we get
\begin{equation}
4 c_{1212}^{\circ}(q) = 4 c_{1313}^{\circ}(\qvec) = q^2 \kN(q) = \frac{1}{\beta V G(q)}
\label{eq_rot_cshear_stat}
\end{equation}
while all other components either vanish or are given by symmetry,
cf.~Eq.~(\ref{eq_rot_indexsym}). 
Note that Eq.~(\ref{eq_rot_cshear_stat}) contains the same information
as the second relation in Eq.~(\ref{eq_cabcd_invariants}).
As one expects, the generalized shear modulus $G(q)$, cf.~Fig.~\ref{fig_pdLJ_LqGq}, 
determines the size of the typical rotation fluctuations. 
For the time-dependent CFs $\cabcd(\qvec,t)$ of the instantaneous rotation TF
Eq.~(\ref{eq_rot_cshear_stat}) can be generalized into the time domain.
Just as in Eq.~(\ref{eq_dyn_e_cLG}) for the shear strain in NRC 
this yields in Fourier-Laplace space
\begin{equation}
\beta V 4 c_{1212}^{\circ}(q,s) = \frac{1}{G(q)} - \frac{w(q,s)}{1+w(q,s) G(q,s)}.
\label{eq_rot_cshear_dyn}
\end{equation}

\section{Different types of time-dependent correlation functions for stationary tensor fields}
\label{corr}

All instantaneous TFs  \cite{foot_intro_carrets} are assumed in this work 
to be {\em stationary stochastic} TFs (including time-reversal symmetry).
Let us focus first on one independent configuration $c$ 
and on a vector field $\rhohata(\rvec,t)$. 
The time-dependent CFs of this TF are defined by
\begin{equation}
\cab(\qvec,t) \equiv \la \rhohata(\qvec,t)\rhohatb(-\qvec,t=0) \ra 
\label{eq_corr_cab_def}
\end{equation}
in reciprocal space with $t$ being the ``time lag" \cite{AllenTildesleyBook}.
$\la \ldots \ra$ stands here and in the next two subsections 
for the standard thermal average for the given configuration.
Taking advantage of the assumed stationarity, the statistics is commonly improved
by means of a ``gliding average" \cite{AllenTildesleyBook}
over all pairs of time $t'$ and $t''$ with $t=|t''-t'|$.
For large times $t$ the fields at $t=0$ and $t$ decorrelate and we get
\begin{equation}
\cab(\qvec,t) \to \la \rhohata(\qvec) \ra \la \rhohatb(-\qvec) \ra
\label{eq_corr_cab_tlarge} 
\end{equation}
for $t \to \infty$. 
For $\la \rhohata(\qvec,t) \ra =0$ this implies $\cab(\qvec,t) \to 0$.
For non-ergodic systems the above functions depend also explicitly on the configuration $c$. 
The ensemble average is obtained by $c$-averaging, i.e.,
$\cab(\qvec,t) \equiv \la \cab(\qvec,t,c) \ra_c$.
Only this final averaging step over an isotropic ensemble may guarantee that 
$\cab(\qvec,t)$ is an ITF. 
Similarly, we also consider fourth-order TFs $\cabcd(\qvec,t)$
characterizing second-order instantaneous TFs $\rhohatab(\qvec,t,c)$.

In many cases instantaneous stochastic TFs are strongly fluctuating.
It is thus useful to systematically project out irrelevant fluctuations by preaveraging 
the field by means of a $t$-average as defined by Eq.~(\ref{eq_taver_def}).
Let us focus on $t$-averaged second-order TFs $\rhobarab(\qvec,\tsamp)$. 
The corresponding fourth-order CFs are defined by
\begin{eqnarray}
\cbarabcd(\qvec,\tsamp) & = & \la \cbarabcd(\qvec,\tsamp,c) \ra_c \mbox{ with }  
\label{eq_taver_cbarabcd_def} \\
\cbarabcd(\qvec,\tsamp,c) & = &\rhobarab(\qvec,\tsamp,c) \rhobarcd(-\qvec,\tsamp,c) 
\nonumber
\end{eqnarray}
Dropping the TF indices and the $\qvec$-argument of the CFs let us write more compactly 
$c(t)$ for the standard CF with time lag $t$ and $\cbar(\tsamp)$ for the CF of 
the $t$-averaged stochastic TF.
Assuming a {\em stationary} stochastic process,
the $\tsamp$-dependence of $\cbar(\tsamp)$ can be traced back via \cite{spmP1}
\begin{equation}
\cbar(\tsamp) = \frac{2}{\tsamp^2} \int_0^{\tsamp} \ddiff t \ (\tsamp-t) \ c(t)
\label{eq_taver_ct2Ct}
\end{equation}
to the time-dependent corresponding CF $c(t)$.
Note that Eq.~(\ref{eq_taver_ct2Ct}) is closely related to the general equivalence 
\cite{HansenBook,AllenTildesleyBook,spmP1} for transport coefficients of Einstein relations, 
corresponding to $\cbar(\tsamp)$, and Green-Kubo relations, corresponding here to $c(t)$.
$\cbar(\tsamp)$ is a natural smoothing function of $c(t)$.
Eq.~(\ref{eq_taver_ct2Ct}) implies that $c(t)$ is constant {\em iff} 
$\cbar(\tsamp)$ is constant 
and for $c(t) = \chat_p \exp(-t/\tau_p)$ that \cite{spmP1}
\begin{eqnarray}
\cbar(\tsamp) & = & \chat_p D(\tsamp/\tau_p) \mbox{ with } \nonumber \\
D(x) &  = & 2 [\exp(-x)-1+x]/x^2 
\label{eq_taver_Debye}
\end{eqnarray}
being the ``Debye function" well known in polymer science \cite{DoiEdwardsBook,RubinsteinBook}.
Note that $D(x) \to 2/x$ for $x \gg 1$.
For systems with overdamped dynamics 
the relaxation dynamics can be efficiently described by a linear superposition
of a small number of such Maxwell modes \cite{RubinsteinBook,FerryBook}.

\section{Relaxation times for strain CFs}
\label{tauterm}

As shown in Sec.~\ref{dyn_e}, the ICFs $\cL(q,t)$ and $\cG(q,t)$ are related to the 
material functions $L(q,t)$ and $G(q,t)$ in Fourier-Laplace space by Eq.~(\ref{eq_dyn_e_cLG})
using for overdamped systems the scalar $w(q,s)=q^2/\zeta s$ 
with $\zeta$ being the effective friction constant.
We focus again on the simple limit for large $t$ (small $s)$.
It was stated in Sec.~\ref{dyn_e_tlarge} that (not surprisingly) the ICFs 
must asymptotically decay exponentially for large $t$ with relaxation times given by
Eq.~(\ref{eq_tauLT_defB}) in terms of the static generalized elastic moduli 
$L(q) \equiv \lim_{s\to 0} L(q,s)$ and  $G(q) \equiv \lim_{s\to 0} G(q,s)$.
These relations {\em assume} that the $s$-dependence of the material functions 
becomes irrelevant for $s \to 0$. 
While this leads indeed to useful relations for $q\xicont \ll 1$, 
higher order contributions may in general contribute.
To see this let us consider the low-$s$ expansion
\begin{equation} 
L(q,s) = L(q) [1+ a_1 s + \frac{1}{2} a_2 s^2 + \ldots]
\label{eq_dyn_e_tlarge_Lexpansion}
\end{equation}
with constants $a_n$ depending {\em apriori} on $q$. 
Inserting this into the relation for the longitudinal ICF, cf.~Eq.~(\ref{eq_dyn_e_cLG}), 
we find to leading order
\begin{equation}
L(q) \beta V \cL(q,s) \simeq \frac{s}{s+1/(\zeta/q^2L(q)+a_1)},	
\label{eq_dyn_e_tlarge_D}
\end{equation}
i.e. the linear term in Eq.~(\ref{eq_dyn_e_tlarge_Lexpansion}) cannot be neglected.

\begin{figure}[t]
\centerline{\resizebox{.9\columnwidth}{!}{\includegraphics*{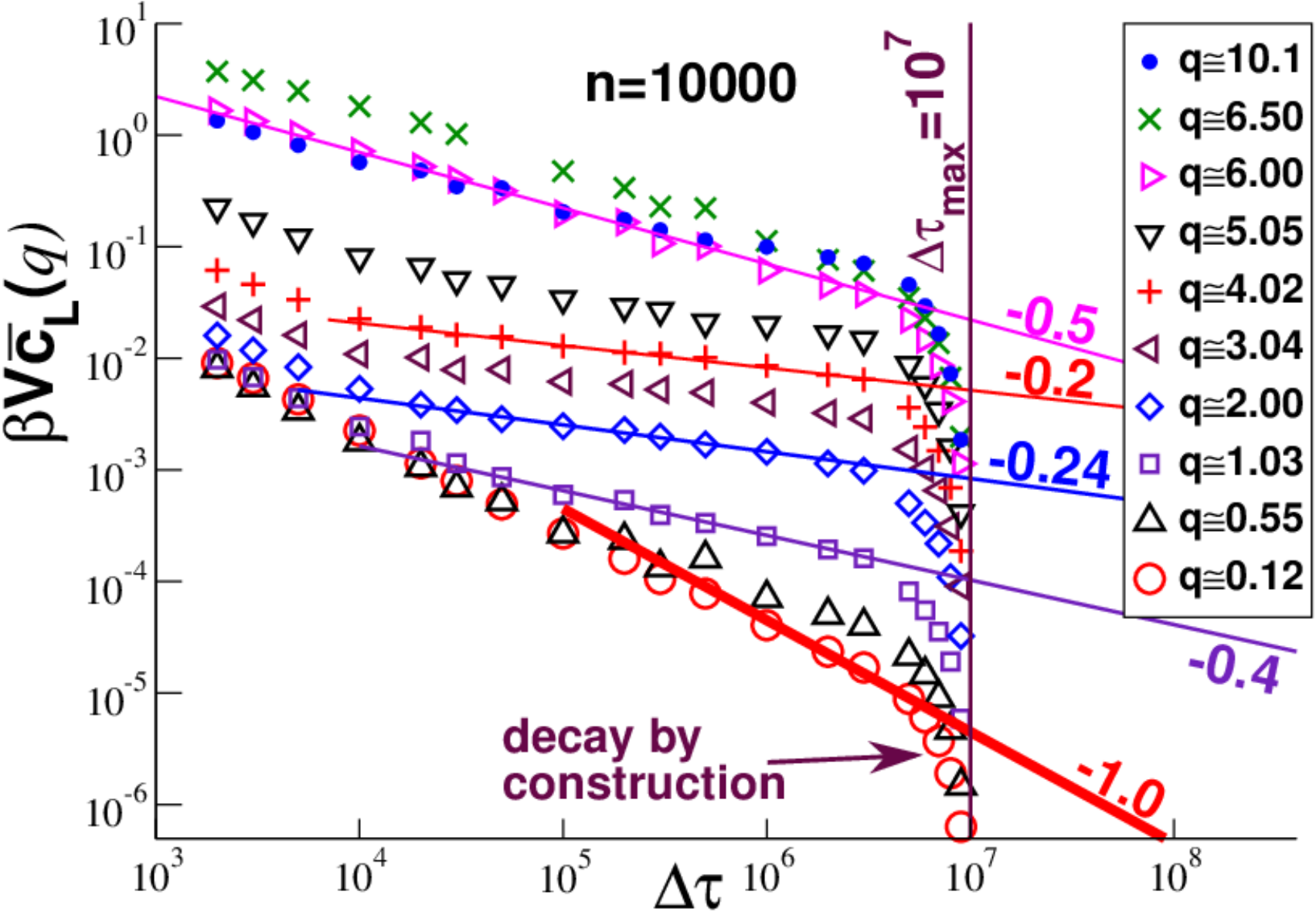}}}
\caption{Longitudinal ICF $\beta V \cLbar(q,\tsamp)$ for pLJ systems 
for several $q$ as a function of $\tsamp$.
As indicated by the vertical line $\tsampmax=10^7$ for $n=10000$,
i.e. the final strong decay of the data is expected due to the construction of the displacement field. 
The power-law slope with exponent $-1$ (bold solid line) expected for
asymptotically large $\tsamp \ll \tsampmax$ is barely consistent with the smallest $q$ 
indicated and much smaller effective exponents are visible for larger $q$ (thin solid lines). 
}
\label{fig_pdLJ_dyn_e_t}
\end{figure}

What is the physical meaning of $a_1$?
We note first that according to Eq.~(\ref{eq_LT_delta}) 
this term corresponds in the time domain to a generalized longitudinal modulus 
\begin{equation}
L(q,t) \approx L(q) [1+a_1 \delta(t)]
\label{eq_dyn_e_tlarge_E}
\end{equation}
with $a_1$ being an effective time scale characterizing the 
short-time behavior of $L(q,t)$.
Let us introduce by 
\begin{eqnarray}
\etaL(q,t) & \equiv & \int_0^t \ddiff t \ [L(q,t)-L(q)] \nonumber \\
\stackrel{\Lcal}{\Leftrightarrow} \etaL(q,s) & = & [L(q,s)-L(q)]/s.
\label{eq_etaL_def}
\end{eqnarray}
a generalize viscocity associated to the longitudinal material function $L(q,t)$.
Using the expansion Eq.~(\ref{eq_dyn_e_tlarge_Lexpansion}) and the final value theorem
of the LT, Eq.~(\ref{eq_LT_FVT}), we get \cite{foot_trans_fqs}
\begin{equation}
a_1 L(q) = \lim_{t \to \infty} \etaL(q,t) = \lim_{s \to 0} \etaL(q,s) = \etaL(q),
\label{eq_etaL2a1}
\end{equation}
i.e. $a_1$ characterizes the $q$-dependent longitudinal viscosity $\etaL(q)$.
Since $a_1$ is a time scale we call it from now $\tauLzero(q)$.
We introduce similarly $\tauTzero(q)$ and $\etaG(q)$ for the corresponding time scale 
and the generalized viscosity of the transverse ICF $\beta V \cG(q,s)$.
(The superscript ``v" marks that these times characterize generalized viscosities.)
Using again Eq.~(\ref{eq_LT_exponential}) 
one confirms that the above relations Eq.~(\ref{eq_dyn_e_tlarge_B}) 
are still applicable, however, in terms of the generalized relaxation times
\begin{eqnarray}
\tauL(q) & = & (\zeta/q^2+\etaL(q))/L(q) \mbox{ and } \nonumber \\
\tauT(q) & = & (\zeta/q^2+\etaG(q))/G(q). \label{eq_tauLT_defC}
\end{eqnarray}
The additional phenomenological time scales 
\begin{equation}
\tauLzero(q) \equiv \etaL(q)/L(q) \mbox{ and }  \tauTzero(q) \equiv \etaG(q)/G(q),
\label{eq_tauLGzero_def}
\end{equation}
characterizing the short-time behavior of both material functions,
are expected to strongly increase for large $q$.
One important reason is that $L(q)$ and $G(q)$ become extremely small for $q \xicont \gg 1$ 
as shown in Fig.~\ref{fig_pdLJ_LqGq}. 

Confirming the above statements, 
the dynamics for our two-dimensional overdamped model glass systems
has been found to dramatically slow down for $q \xicont \gg 1$. We have thus unfortunately been
unable to reach the predicted exponential decay for the possible production times $\tsampmax$.
This is shown in Fig.~\ref{fig_pdLJ_dyn_e_t} for the ICF $\cLbar(q,\tsamp)$ of the 
$t$-averaged longitudinal strain fields. We plot here ICFs for several 
wavenumbers $q$ as a function of the preaveraging time $\tsamp$ for our smaller systems 
with $n=10000$ particles and a total production time $\tsampmax=10^7$ (vertical line). 
The bold solid line indicates the
power-law exponent $-1$ corresponding for $\cLbar(q,\tsamp)$ 
to the exponential decay for $\cL(q,t)$.
Apparently, much larger $\tsampmax$ are warranted to get a
reasonable estimation of $\tauL(q)$ and $\tauT(q)$ for $q\xicont \gg 1$.
We have thus been unable to measure in this limit $\tauL(q)$ and $\tauT(q)$ 
and thus the respective viscosity contributions $\tauLzero(q)$ and $\tauTzero(q)$.
%



\begin{thebibliography}{49}
\bibitem[{foo({\natexlab{a}})}]{foot_intro_carrets}
\bibinfo{note}{We use carrets for the normalized components $\rhat_{\alpha}$
  and $\qhat_{\alpha}$ of the field vectors in real and reciprocal space and
  often also to emphasize instantaneous observables.}
\bibitem[{foo({\natexlab{b}})}]{foot_wiki_multipolar}
\bibinfo{note}{See the wikipedia entries on quadrupoles and multipolar
  expansion. For the (not normalized) basis functions $\cos(p \theta)$ or
  $\sin(p \theta)$ relevant for the $12$-plane a monopole corresponds to $p=0$,
  a dipole to $p=1$, a quadrupole to $p=2$ and an octupole to $p=4$.}
\bibitem[{foo({\natexlab{d}})}]{foot_urref}
\bibinfo{note}{It is also possible to use instead the current particle position
  $r_{\alpha}^a(t)$ in the $\delta(\rvec)$-function to construct the
  displacement field in real space. Both operational definitions are equivalent
  for wavenumbers $q$ smaller than the inverse typical (root-mean-squared)
  displacements \cite{spmP5b}.}
\bibitem[{foo({\natexlab{g}})}]{foot_resp_strain}
\bibinfo{note}{Similarily one may obtain the strain response $\epsab(\qvec) =
  \Gabcd(\qvec) \sigcdext(\qvec)$ due to an imposed stress $\sigabext(\qvec)$.
  Using Eq.~(\ref{eq_ten_sum_OPup_def}), Eq.~(\ref{eq_epsab_def}) and
  Eq.~(\ref{eq_sigabext_def}) this may be also done using $\Gabcd(\qvec) = q^2
  \beta V \OPup[\cab(\qvec)]$}.
\bibitem[{foo({\natexlab{h}})}]{foot_phy_friction}
\bibinfo{note}{The friction coefficient $\zeta$ is defined
  here with respect to a force per volume, i.e. the dimension $[\zeta]$ of
  $\zeta$ differs from the one used elsewhere
  \cite{DoiEdwardsBook,RubinsteinBook}. In our notation we have $[\zeta] =
  [\mbox{modulus}]/[\mbox{acceleration}]$.}
\bibitem[{foo({\natexlab{i}})}]{foot_trans_fqs}
\bibinfo{note}{Let $f(\qvec,s)$ be a function in Fourier-Laplace space with
  $\qvec$ being the wavevector and $s$ the Laplace variable. We often drop an
  argument if it vanishes, i.e. we write $f(s) \equiv f(\qvec=0,s)$, $f(\qvec)
  \equiv f(\qvec,s=0)$ and $f \equiv f(\qvec=0,s=0)$.}
\bibitem[{foo({\natexlab{j}})}]{foot_trans_qdiscrete}
\bibinfo{note}{The wavevectors $\qvec$ for finite real systems as in
  computational studies must obviously be discrete (``quantized") and
  commensurate with the boundary conditions reflecting the size and shape of
  the system \cite{numrec,spmP5a}.}
\end{thebibliography}
\end{document}